\documentclass{aa}
\usepackage{graphicx}
\usepackage{array}
\usepackage{subcaption}
\usepackage{arydshln}
\usepackage{upgreek}
\usepackage{txfonts}
\usepackage{hyperref}
\usepackage{float}
\usepackage{placeins}
\hypersetup{colorlinks=true, linkcolor=blue, urlcolor=blue, filecolor=blue, citecolor=blue}

\begin{document} 

   \title{The HARPS search for southern extra-solar planets}
   \subtitle{XLVII. Five Jupiter-mass planets in long-period orbits, one highly irradiated Neptune, one brown dwarf, and five stellar binaries.
   \thanks{Based on observations made with the HARPS instrument on the ESO 3.6-m telescope at La Silla Observatory (Chile), under GTO programme ID 072.C-0488, and its continuation programmes ID 085.C-0019, 087.C-0831, 089.C-0732, 090.C-0421, 091.C-0034, 092.C-0721, 093.C-0409, 095.C-0551, 096.C-0460, 098.C-0366, 099.C-0458, 0100.C-0097, 0101.C-0379, 0102.C-0558, 0103.C-0432 106.21R4.001, 108.222V.001, 183.C-0972, 192.C-0852 and 196.C-1006.}}

   \author{Y.G.C. Frensch\inst{1,2}, G. Lo Curto\inst{1}, F. Bouchy\inst{2}, M. Mayor\inst{2}, G. H\'ebrard\inst{3,4}, C. Lovis\inst{2},
          C. Moutou \inst{5}, F. A. Pepe\inst{2}, D. Queloz \inst{6}, N. Santos \inst{7,8}, D. Segransan\inst{2}, S. Udry \inst{2}
          \and N. Unger\inst{2}}

   \institute{European Southern Observatory, Karl-Schwarzschild-Strasse 3, 85748 Garching,   
            Germany\\ 
            email: \url{yolanda.frensch@eso.org}
         \and Observatoire de Genève, 51 Ch. des Maillettes, 1290 Sauverny, Switzerland
        \and Institut d'astrophysique de Paris, UMR7095 CNRS, Universit\'e Pierre \& Marie Curie, 98bis boulevard Arago, 75014 Paris, France
        \and Observatoire de Haute-Provence, CNRS, Universit\'e d'Aix-Marseille, 04870 Saint-Michel-l'Observatoire, France
        \and Universit\'e de Toulouse, UPS-OMP/CNRS, IRAP,
            14 avenue E. Belin, Toulouse, F-31400, France
        \and ETH Zurich, Department of Physics, Wolfgang-Pauli-Strasse 2, CH-8093 Zurich, Switzerland
        \and Instituto de Astrofisica e Ciencias do Espaco, Universidade do Porto, CAUP, Rua das Estrelas, 4150-762 Porto, Portugal
        \and Departamento de Fisica e Astronomia, Faculdade de Ciencias, Universidade do Porto, Rua do Campo Alegre, 4169-007 Porto, Portugal
             }
             
\date{Received 21 February 2023; accepted 15 May 2023}

  \abstract
   {The long-term ongoing HARPS radial velocity survey of extra-solar planets initiated in 2003 provides a unique data set with a 19 year baseline that allows the detection of long-period exoplanets, brown dwarfs, and low-mass binaries.}
   {Our aim is to detect and characterise long-period companions around main sequence stars (spectral types late F to early M). Only 6\% of the planets discovered so far have periods longer than 3 years; we are probing this still largely unknown population.}
   {We use the radial velocity method to search for exoplanets around stars. The radial velocity variations are measured with HARPS at the ESO 3.6 metre telescope. Difficulties in characterising long-period exoplanets arise from the entanglement of the radial velocity with the stellar magnetic cycle. We thoroughly examined the stellar activity indicators to rule out magnetic cycles as the source of the observed variation. The true
mass and inclination of our heavier companions are provided by astrometry, for which we use proper motions from Hipparcos and Gaia.}
    {Five Jupiter-mass exoplanets are reported to orbit HIP54597, BD-210397 ($\times 2$), HD74698, and HD94771 with 8.9 yr, 5.2 yr, 17.4 yr, 9.4 yr, and 5.9 yr orbits, and to have minimum masses of $2.01\pm 0.03$,
    $0.7 \pm 0.1$, $2.4^{+1.5}_{-0.2}$, $0.40 \pm 0.06,$ and $0.53 \pm 0.03 M_J$ respectively. HD74698 also hosts a highly irradiated Neptune in a 15 day orbit with a minimum mass of $0.07\pm 0.01$ $M_J$. The mass and inclination of the exoplanets cannot yet be well constrained by astrometric measurements. Only HIP54597 b, HD74698 c, and BD-210397 c have weak constraints. The mass of HIP54597 b can maximally increase by $10\%-30\%$, the minimum mass of HD74698 c is likely equal to its true mass, and BD-210397 c has a mass of ${2.66}_{-0.32}^{+0.63}$ $M_J$.
    HD62364 hosts a brown dwarf with a true mass of ${18.77}_{-0.63}^{+0.66} M_J$ in an orbit of 14 yr. The mass of HD62364 b is around the limit of the masses of brown dwarfs, but its orbit is highly eccentric ($e = 0.607 \pm 0.005$), which is more common among brown dwarfs than exoplanets. 
    HD56380B, HD221638B, and HD33473C have minimum masses within the brown dwarf limits, in orbits of 8.9 yr, 16.6 yr, and 50 yr respectively; however, astrometric measurements reveal them to be stellar binaries, with masses of ${375.3}_{-8.4}^{+8.6}$, ${110.0}_{-3.7}^{+3.9}$, and ${271.0}_{-3.8}^{+3.9} M_J$. The orbits of the stellar binaries HD11938 and HD61383 are incomplete.
    The preliminary result for HD61383 is a 0.190 $M_{\odot}$ binary in a 39 yr orbit. The secondary of the binary system HD11938 has a mass of 0.33 $M_{\odot}$ ---which is confirmed by a secondary peak in the cross-correlation function (CCF)--- and a preliminary period of 35 yr. The origin of the 3.0 yr radial velocity (RV) signal of HD3964 is uncertain as it shows entanglement with the magnetic cycle of the star. We finally report one more star, HD11608, with a magnetic cycle that mimics a planetary signal.}
   {We present the discovery of six exoplanets, one uncertain exoplanet candidate, one brown dwarf, and five stellar binaries around main sequence stars. We also improve the orbital solution of the stellar binary HD33473C thanks to long-term monitoring.}


    \authorrunning{Y.G.C. Frensch et al.}
    \titlerunning{The HARPS search for southern extra-solar planets. XLVII.}
   \maketitle
%

\section{Introduction}
Of the currently more than 5300 known exoplanets, only 6\% have periods of longer than 3 years, and less than 3\% have periods longer than 10 years\footnote{See The Extrasolar Planets Encyclopaedia, \url{http://exoplanet.eu}}. The radial velocity (RV) method, one of the leading methods in providing long-period exoplanet detections, was responsible for most of them. Long-term RV surveys using precise spectrographs made this accomplishment possible, including but not limited to: the over 12 year historical ELODIE program initiated in 1994 \citep{Mayor1995, Naef2005}, the SOPHIE large program started in 2006 \citep{Bouchy2009}, the Anglo-Australian Planet Search launched in 1998 \citep{Tinney2001}, and the HARPS\footnote{High Accuracy Radial velocity Planet Searcher.} volume-limited survey, from which the observations in this paper originate. From these different RV surveys, Jupiter analogues and long-period companions were discovered and their properties scrutinised (e.g. \citeauthor{JA_Boisse2012} \citeyear{JA_Boisse2012}, \citeauthor{JA_Moutou2014} \citeyear{JA_Moutou2014}, \citeauthor{JA_Wittenmyer2016} \citeyear{JA_Wittenmyer2016}, \citeauthor{JA_Rey2017} \citeyear{JA_Rey2017}, \citeauthor{JA_Bryan2019} \citeyear{JA_Bryan2019}, \citeauthor{JA_Feng2019} \citeyear{JA_Feng2019}, \citeauthor{LP_Kiefer2019} \citeyear{LP_Kiefer2019}, \citeauthor{LP_Dalal2021} \citeyear{LP_Dalal2021}, \citeauthor{LP_Demangeon2021} \citeyear{LP_Demangeon2021}, \citeauthor{JA_Sreenivas2022} \citeyear{JA_Sreenivas2022}). The HARPS volume-limited survey (up to 57.5pc) started as part of the HARPS Guaranteed Time Observations (GTO) in 2003 \citep{HARPS} and is targeting low-activity, solar-type dwarf stars with spectral types from late F to early M \citep{LoCurto2010}. The baseline of the survey, which is over 19 years, allows the detection of long-period signals, including gas giants beyond the ice line. The detection of lighter planets and other long-period companions such as brown dwarfs is also possible, although these targets are not the direct aim of the program. 

The detection of long-period planets, and in particular of `cold-Jupiters' ($M >0.3 M_{J}$, $a>1 \mathrm{AU}$), strongly correlates with the presence of inner super-Earths; more specifically, \cite{ZhuWu2018} quote a conditional probability of super-Earths of $90\% \pm 20\%$. \cite{Rosenthal2022} are probing a different class of outer planets with semi-major axis $a$ in the range $0.3$ to $3$ AU and in their recent publication present a value of $32^{+24}_{-16}\%$; as they are looking at different samples, this is effectively not in contrast.

In any case, the host stars to our detected long-period planets are potential sources of yet-undetected super-Earths. Our survey is not designed to detect these super-Earths, and our precision and sampling are not optimal. Outer giant planets might stabilise planetary systems, as for example shown by the Nice model \citep{Morbidelli2018} in which the gas giants are believed to shield the inner planets from collisions; Jupiter and Saturn for instance could be responsible for the existence of Earth and the other inner planets in their current orbits. 

Hot giant planets tend to orbit metal-rich host stars, a correlation known as the giant planet--metallicity correlation \citep{Fischer2005, Osborn2020}. As the metallicity of the host star is representative of the metallicity content in the protoplanetary disc, this result is an important piece of observational evidence in support of the core-accretion scenario \citep{Pollack1996}. In this formation scenario, a large metal core, which is expected to be more easily formed in a metal-rich disc (i.e. more planetesimals), efficiently accretes gas to form a gas giant. While hot Jupiters are more frequent around metal-rich stars, for longer-period giant planets this is not necessarily the case. \cite{Adibekyan2013} showed that planets (from $\sim 0.03 M_J$ to $\sim 4 M_J$) formed in metal-poor systems generally have longer periods than those formed in metal-rich systems. These authors suggest a metal-poor disc may form the giant planets further out and/or formation may start later, which would mean the planets undergo less migration. Short- and long-period giant planets may have the same formation but different migration histories \citep{Osborn2020}.  More data are needed to further constrain the formation and evolution of long-period giant planets. With this research, we are probing this still largely unknown population of exoplanets. 

The RV method observes the stellar wobble induced by the gravitational pull of the companion orbiting its host star. Stellar activity can mimic this wobble and create false-positive detections. As our program focuses on finding Jupiter-mass long-period companions, it does not require high precision and is not hampered by the short-period ($\sim$minutes) p-mode oscillations. 
For long-period planets, the greatest difficulties stem from the entanglement of the RV with the stellar magnetic cycle and the RV-induced signal from binary stars (or multiple-star systems). Although the magnetic cycle can inconveniently dominate the RV variation, the signal induced by the magnetic cycle correlates well with the stellar activity indices. The correlation can therefore be used to determine whether or not there is entanglement with the magnetic cycle and then be fitted and removed if the long-term activity index evolves smoothly \citep{Dumusque2011c}.

In search of gas giants, we discuss 12 stars in this paper. Each signal is thoroughly examined to ensure the variation is not caused by stellar activity. Section \ref{sec:Observations} contains information on the observations. In Section \ref{sec:Stars} we present the stellar parameters of the host stars. In Sect. \ref{sec:RV_obsol} we discuss one magnetic cycle and derive the orbital solutions for each star from the RV observations. In Sect. \ref{sec:Orvara} we present the combination of astrometric measurements from Hipparcos and Gaia with RVs in order to constrain the mass and inclination of the heaviest companions. Our results are discussed in Sect. \ref{sec:Conclusions}. 

\section{Observations}
\label{sec:Observations}
The observations used in the present study were carried out using the HARPS spectrograph at the ESO 3.6 meter telescope at La Silla Observatory (Chile) \citep{HARPS}. The presented signals are part of a long-term ongoing exoplanet survey that started in 2003 as part of the HARPS GTO program (Mayor 2003) and is still continuing today (Lo Curto 2010). In May 2015, the fibre link of HARPS was upgraded \citep{LoCurto2015}. With the upgrade, the instrumental profile was changed significantly. As this might induce an offset, we considered the data before and after the fibre upgrade as coming from two independent instruments. The magnitude of the offset varies for different stars and affects the CCF-FWHM as well.

The RVs result from version 3.5 of the HARPS data reduction software (DRS), which cross-correlates each spectrum with a numerical stellar template, matching its spectral type. The pipeline derives several cross-correlation function parameters: the RVs, the full width at half maximum (FWHM), the bisector span, and the contrast. Other results are the Mount Wilson S-index, the chromospheric emission ratio $\log \mathrm{R'_{HK}}$, and the activity indices based on the Na, Ca, and H$\mathrm{\alpha}$ lines.

The number of measurements and the basic characteristics of the measurements are summarised in Table \ref{tab:Observations}. Our program is targeting mainly Jupiter-mass planets and therefore requires only a moderate RV precision of $\sim 3$ m s$^{-1}$, which corresponds to a signal-to-noise ratio (S/N) of about 40. As this goal is not always obtained, we include observations with an S/N of larger than 25. This limit is used as additional quality control (QC); because the exposure times are defined for an S/N of 40, values below 25 indicate a poor observation (bad seeing, bad weather, etc.). The total number of measurements $N_{\mathrm{meas}}$ excludes the observations with a S/N below 25 and observations that did not pass the DRS QC. The DRS QC checks for the reliability and accuracy of the data obtained by verifying the instrumental stability, data-reduction process, and consistency with known calibration sources \citep{HARPS_DRS24}. The public HIRES \citep{HIRES} RV data are added to the analysis of BD-210397, and CORALIE observations \citep{CORALIE} are added to the study of HD61383. 

\begin{table*}
    \centering
    \caption{Characteristics of the observations of the target stars.}
    \label{tab:Observations}
    \begin{tabular}{l c c c c c c c c}
    \hline \hline
        & $N_{\mathrm{meas}}$ & $T_{\mathrm{min}}$ & $T_{\mathrm{max}}$ & Span & $\langle t_{exp} \rangle\,^{(c)}$ & $\langle \sigma_{\mathrm{RV}}\rangle$ & $\langle S/N \rangle_{550 \mathrm{nm}}\,^{(c)}$ \\
        & & & & [yr] & [s] & [m s$^{-1}$] \\
        \hline
    HD11608 & 42 & 2004-09-23 & 2022-10-13 & 18.1 & 448 & 1.6 & 52\\
    HD3964 & 65 & 2004-09-23 & 2022-12-08 & 18.2 & 250 & 2.0 & 54 \\
    HD94771 & 125 & 2004-02-15 & 2023-01-07 & 18.9 & 180 & 1.6 & 67 \\
    HIP54597 & 64 & 2004-02-22 & 2022-03-14 & 18.1 & 575 & 2.0 & 46 \\
    BD-210397 & 101$^{(a)}$ & 2005-01-03 & 2022-12-07 & 17.9 & 554 & 1.9 & 52 \\
    HD74698 & 131 & 2004-02-12 & 2023-01-08 & 18.9 & 182 & 2.0 & 56 \\
    HD62364 & 62 & 2004-02-12 & 2023-01-05 & 18.9 & 178 & 2.9 & 71 \\
    HD56380 & 92 & 2004-01-04 & 2022-11-13 & 18.9 & 364 & 2.1 & 50 \\
    HD221638 & 60 & 2005-11-21 & 2022-12-07 & 17.0 & 223 & 4.0 & 68 \\
    HD33473A & 57 & 2004-01-11 & 2023-01-05 & 19.0 & 181 & 1.7 & 81 \\
    HD11938 & 25 & 2004-07-16 & 2022-12-13 & 18.4 & 538 & 2.1 & 44\\
    HD61383 & 86$^{(b)}$ & 2001-02-17 & 2022-12-30 & 21.9$^\mathrm{(b)}$ & 179 & 3.1 &  68 \\
    \hline 
    \end{tabular}\begin{flushleft}{\footnotesize \textbf{Notes:} $^{(a)}$ Including public HIRES observations. $^{(b)}$ Including CORALIE data. $^{(c)}$ Excluding non-HARPS data.}\end{flushleft}
\end{table*}

\section{Stellar properties}
\label{sec:Stars}
All stars presented in this paper are main sequence stars, like our Sun. Table \ref{table:stellar_parameters} summarises the parameters of the stars analysed in the present paper. The spectral type, $V,$ and $B-V$ values originate from the Hipparcos catalogue \citep{Hipparcos}, where the magnitudes are dereddened by \cite{GomesdaSilva2021}. The parallax with the derived distance is taken from Gaia DR3 \citep{GaiaDR3}. The spectroscopic analysis of \cite{DelgadoMena2017} provides $\log g$ (corrected by \citeauthor{GomesdaSilva2021} \citeyear{GomesdaSilva2021}), $T_{\mathrm{eff}}$, and $[\mathrm{Fe/H}]$ for all stars, except BD-210397 \citep{Yee2017} and HIP54597 \citep{Sousa2011}, which are too cold for the parameter estimation by \cite{DelgadoMena2017}. The parameters $M_V$, $L,$ and $R_{\star}$ are determined from the above values using the bolometric correction from \cite{Flower1996}. The age and mass estimates are obtained by \cite{DelgadoMena2019}, apart from HIP54597 and BD-210397, which are calculated via theoretical isochrones \citep{Bressan2012}.

The average FWHM is obtained from the HARPS data from before 2015 and cross-correlated with G2 masks, as the $v \sin i$ approximation adapted from \cite{Santos2002} is calibrated on the same data period and spectral type. The FWHM uncertainty is the standard deviation and is therefore an indicator of activity in the RV signal. The average chromosphere emission ratio $\log \mathrm{R'_{HK}}$ results from the \cite{Noyes1984} method. The rotational period is averaged and estimated from the activity-Rossby relations described by \cite{Mamajek2008}, implementing the convective overturn time from \cite{Noyes1984}. These relations are defined for stars with ($\log\mathrm{R'_{HK}} > -5.0$). Here, we use the same method for relatively quiet stars to get an approximation of the rotational period. Long-term companions have periods that do not coincide with the rotational period of the  stars and we therefore do not require a very precise value. We use the standard deviations of $\log\mathrm{R'_{HK}}$ and $P_{\mathrm{rot}}$ as an estimate of their uncertainties.

\begin{table*}
\caption{Stellar parameters of the stars presented here.}\label{table:stellar_parameters}
\centering
\begin{tabular}{l l c c c c c c c c c c c} 
\hline\hline       
\multicolumn{2}{l}{Parameter} & HD11608 & HD3964 & HD94771 & HIP54597 & BD-210397 &  HD74698 \\ \hline
SpType & & K1/K2V & G5V & G3/G5V & K5V & K7V & G5V \\
$V$ & [mag] & $9.307 \pm 0.010$ & $8.383 \pm 0.010$ & $7.361 \pm 0.010$ & $9.836 \pm 0.010$ & $9.84 \pm 0.015^{(a)}$ & $7.760 \pm 0.010$ \\
$B-V$ & [mag] & $0.988 \pm 0.006$ & $0.675 \pm 0.002$ & $0.749 \pm 0.015$ & $1.072 \pm 0.015$ & $1.346 \pm 0.007^{(a)}$ & $0.658 \pm 0.010$ \\
$\pi$ & [mas] & $23.200 \pm 0.020$ & $20.541 \pm 0.025$ & $17.356 \pm 0.018$ & $25.286 \pm 0.015$ & $42.123 \pm 0.017$ & $19.285 \pm 0.014$ \\
$d$ & [pc] & $43.103 \pm 0.038$ & $48.684 \pm 0.060$ & $57.618 \pm 0.061^{(b)}$ & $39.547 \pm 0.023$ & $23.740 \pm 0.009$ & $51.854 \pm 0.039$ \\
$M_V$ & [mag] & $6.13 \pm 0.01$ & $4.95 \pm 0.01$ & $3.56 \pm 0.01$ & $6.85 \pm 0.01$ & $7.96 \pm 0.01$ & $4.19 \pm 0.01$  \\
$B.C.$ & [mag] & -0.413 & -0.089 & -0.109 & -0.407 & -1.050 & -0.079 \\
$L$ & [$L_{\odot}$] & $0.405 \pm 0.005$ & $0.89 \pm 0.01$ & $3.28 \pm 0.04$ & $0.208 \pm 0.002$ & $0.135 \pm 0.002$ & $1.79 \pm 0.02$ \\
$T_{\mathrm{eff}}$ & [K] & $4788 \pm 50.8$ & $5729 \pm 19.0$ & $5631 \pm 21.0$ & $4799 \pm 90$ & $4051 \pm 239$ & $5783 \pm 19.0$ \\
$R_{\star}$ & [$R_{\odot}$] & $0.925 \pm 0.021$ & $0.962 \pm 0.009$ & $1.903 \pm 0.019$ & $0.660 \pm 0.025$ & $0.747 \pm 0.088$ & $1.333 \pm 0.012$ \\
$\log(g)$ & [cm s$^{-2}$] & $4.39 \pm 0.18$ & $4.37 \pm 0.04$ & $3.94 \pm 0.03$ & $4.43 \pm 0.18$ & $4.67 \pm 0.06$ & $4.12 \pm 0.02$ \\
$\mathrm{[Fe/H]}$ & [dex] & $0.24 \pm 0.04$ & $0.05 \pm 0.01 $ & $0.22 \pm 0.02$ & $-0.22 \pm 0.06$ & $0.12 \pm 0.10$ & $0.07 \pm 0.02$ \\
$M_{\star}$ & [$M_{\odot}$] & $0.833 \pm 0.021$ & $1.005 \pm 0.032$ & $1.200 \pm 0.024$ & $0.69 \pm 0.02^{(c)}$ & $0.679 \pm 0.017^{(c)}$ & $1.039 \pm 0.025$ \\
Age & [Gyr] & $8.334 \pm 4.559$ & $2.501 \pm 2.175$ & $5.733 \pm 0.391$ & $6.9 \pm 4.5^{(c)}$ & $6.2 \pm 4.7^{(c)}$ & $7.847 \pm 1.191$ \\
FWHM & [km s$^{-1}$] & $7.004 \pm 0.014$ & $6.896 \pm 0.013$ & $7.624 \pm 0.009$ & $6.937 \pm 0.023$ & $7.203 \pm 0.022$ & $7.222 \pm 0.011$ \\
$\log \mathrm{R'_{HK}}$ & & $-4.98 \pm 0.07$ & $-4.87 \pm 0.11$ & $-5.20 \pm 0.09$ & $-4.89 \pm 0.07$ & - & $-5.03 \pm 0.09$  \\
$v\sin i$ & [km s$^{-1}$] & $1.5 \pm 0.5$ & $1.5 \pm 0.5$ & $3.2 \pm 0.5$ & $<1$ & $<1$ & $2.2 \pm 0.5$ \\
$P_\mathrm{rot}$ & [d] & $50 \pm 5$ & $24 \pm 4$ & $49 \pm 6$ & $45 \pm 5$ & - & $30 \pm 3$ \\
\hline
\hline 
\multicolumn{2}{l}{Parameter} & HD62364 & HD56380 & HD221638 &  HD33473A & HD11938 & HD61383\\
\hline 
SpType & & F7V & G8V & F6V & G3V & K4/K5V & G3V \\
$V$ & [mag] & $7.293 \pm 0.010$ & $9.180 \pm 0.010$ & $7.530 \pm 0.010$ & $6.743 \pm 0.010$ &  $9.854 \pm 0.010$ & $7.572 \pm 0.010$ \\
$B-V$ & [mag] & $0.529 \pm 0.004$ & $0.733 \pm 0.002$ & $0.503 \pm 0.006$ & $0.659 \pm 0.006$ & $1.117 \pm 0.009$ & $0.601 \pm 0.010$ \\
$\pi$ & [mas] & $18.881 \pm 0.023$ & $18.822 \pm 0.087$ & $19.160 \pm 0.027$ & $18.188 \pm 0.024$ & $24.760 \pm 0.080$ & $18.822 \pm 0.024$ \\
$d$ & [pc] & $52.963 \pm 0.064$ & $53.128 \pm 0.244$ & $52.192 \pm 0.073$ &  $54.981 \pm 0.074$ & $40.388 \pm 0.131$ & $53.130 \pm 0.068$  \\
$M_V$ & [mag] & $3.67 \pm 0.01$ & $5.55 \pm 0.02$ & $3.94 \pm 0.02$ & $3.04 \pm 0.01$ & $6.82 \pm 0.02$ & $3.95 \pm 0.01$ \\
$B.C.$ & [mag] & -0.015 & -0.189 & -0.005 & -0.087 & -0.568 & -0.091 \\
$L$ & [$L_{\odot}$] & $2.71 \pm 0.04$ & $0.56 \pm 0.01$ & $2.09 \pm 0.03$ & $5.18 \pm 0.07$ & $0.248 \pm 0.005$ & $2.26 \pm 0.03$ \\
$T_{\mathrm{eff}}$ & [K] & $6255 \pm 26.0$ & $5317 \pm 22.0$ & $6360 \pm 50.0$ & $5740 \pm 13.0$ & $4547 \pm 109.1$ & $5716 \pm 14.0$ \\
$R_{\star}$ & [$R_{\odot}$] & $1.401 \pm 0.015$ & $0.884 \pm 0.012$ & $1.192 \pm 0.020$ & $2.300 \pm 0.019$ & $0.802 \pm 0.039$ & $1.533 \pm 0.013$ \\
$\log(g)$ & [cm s$^{-2}$] & $4.14 \pm 0.04$ & $4.38 \pm 0.04$ & $4.16 \pm 0.08$ & $3.84 \pm 0.02$ & $4.41 \pm 0.32$ & $4.08 \pm 0.02$ \\
$\mathrm{[Fe/H]}$ & [dex] & $-0.11 \pm 0.02$ & $-0.42 \pm 0.02$ & $-0.21 \pm 0.04$ & $-0.13 \pm 0.01$ & $0.08 \pm 0.06$ & $-0.49 \pm 0.01$ \\
$M_{\star}$ & [$M_{\odot}$] & $1.157 \pm 0.028$ & $0.768 \pm 0.011$ & $1.123 \pm 0.040$ & $1.248 \pm 0.020$ & $0.752 \pm 0.019$ & $0.871 \pm 0.013$\\
Age & [Gyr] & $3.901 \pm 0.699$ & $12.153 \pm 1.128$ & $2.228 \pm 1.378$ & $4.189 \pm 0.155$ & $5.530 \pm 4.719$ & $12.071 \pm 0.371$ \\
FWHM & [km s$^{-1}$] & $8.615 \pm 0.010$ & $6.444 \pm 0.016$ & $9.407 \pm 0.018$ & $7.393 \pm 0.009$ & $7.539 \pm 0.018$ & $6.821 \pm 0.011$\\
$\log\mathrm{R'_{HK}}$& & $-5.02 \pm 0.06$ & $-5.03 \pm 0.18$ & $-4.87 \pm 0.65$ & $-5.12 \pm 0.10$ & $-4.55 \pm 0.03$ & $-5.05 \pm 0.10$\\
$v\sin i$ & [km s$^{-1}$] & $3.8 \pm 0.5$ & $<1$ & $4.7 \pm 0.5$ & $2.6 \pm 0.5$ & $<1$ & $<1$\\
$P_\mathrm{rot}$ & [d] &$14 \pm 1$ & $38 \pm 6$ & $10 \pm 2$ & $32 \pm 5$ & $22 \pm 2$ & $22 \pm 4$ \\
\hline
\end{tabular}
\begin{flushleft}{\footnotesize
\textbf{Notes:}
 $\,^{(a)}$ Original (not dereddened) value from Hipparcos \citep{Hipparcos}. $^{(b)}$ Originally within the defined volume limit (57.5 pc), the new Gaia parallax places it slightly outside of the sample.} $^{(c)}$ Calculated via theoretical isochrones \citep{Bressan2012}.\end{flushleft}
\end{table*}

\section{Radial velocity data and orbital solutions}
\label{sec:RV_obsol}
The RV variations are examined using tools provided by the Data and Analysis Center for Exoplanets (DACE)\footnote{\url{https://dace.unige.ch}}. We use the data obtained before the fibre upgrade in 2015 as the reference systemic velocity, which we denote as $\gamma_{03}$ and introduce an offset constant for the data obtained after the fibre upgrade, which we denote as $\gamma_{15}$. To ensure the observed RV variation is caused by a companion, its correlations with the stellar activity indicators and the bisector span are inspected. \cite{Delisle2018} describe the computation of activity cycles as part of a detrending process to remove systematic effects from RV data. A model is fitted to the systematic effects ---including activity cycles--- and is subtracted from the RV data to improve the sensitivity with which planets are detected. As our main focus is on long-period companions, inspection of the correlations is sufficient. For periods shorter than a few months, the RV variation is also compared to the rotational period of the star and the bisector span to infer whether the signal might originate from starspots or plages. After fitting away the activity signals detected in the periodograms, further periodic RV variations in the data are searched for using the false alarm probability (FAP) of the periodogram as an indicator of the statistical significance of the candidate signal. The Keplerian model is computed as described in \cite{Delisle2016}. The values of the FAP are calculated analytically following \citep{Baluev2008}. When fitting multiple Keplerian solutions, the stellar activity indicators are again examined before each new fit. An MCMC is used to specify the orbital solution and its uncertainties, the algorithm for which is defined in \cite{Diaz2014, Diaz2016}.

A magnetic cycle is discussed in subsection \ref{sec:Mag_Cyc}, as an example of how to distinguish stellar activity from long-period exoplanets. In subsection \ref{sec:Uncertain} we present the uncertain solution of HD3964, where the magnetic cycle hinders the accurate determination of the RV amplitude. The orbital solutions of the exoplanets without strong entanglement with the star's magnetic cycle are presented in subsection \ref{sec:planets}, while those of the brown dwarf candidates and binaries are presented in subsections \ref{sec:browndwarfs} and \ref{sec:binaries}. For all solutions presented, we fixed the HARPS instrumental error to 0.75 m s$^{-1}$. This value is close to some of the lowest residual RMSs we obtain from the literature; see e.g. \cite{Lovis2006n}. The true mass and inclination derived from astrometry in section \ref{sec:Orvara} classify some of the brown dwarf companions as stellar binaries; we therefore introduce these as candidates.

\subsection{Magnetic cycle of HD11608}
\label{sec:Mag_Cyc}
HD11608 was part of our analysis because the RVs show a prominent peak above 0.1\% FAP at $\sim3950$ days in the periodogram. However, after inspecting the stellar activity correlation values, we find that it is more likely activity induced.

\begin{figure}
    \centering
    \includegraphics[width=\linewidth]{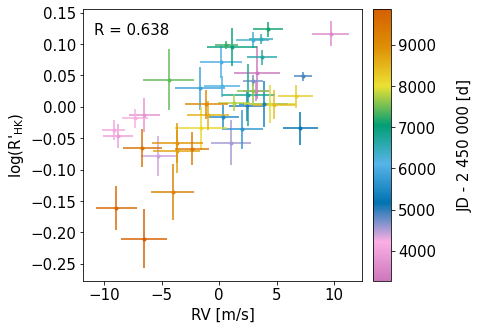}
    \caption{Correlation between $\log\mathrm{R'_{HK}}$ and the RV values of HD11608.}
    \label{fig:HD11608_CORR}
\end{figure}

\begin{figure}
    \centering
    \includegraphics[width=\linewidth]{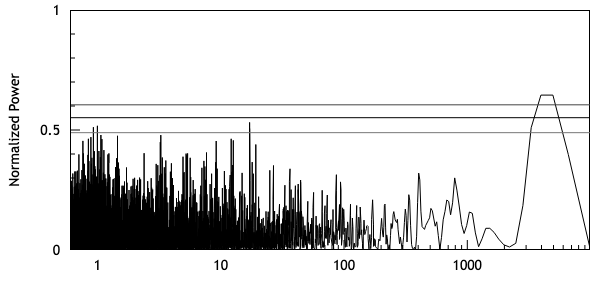}
    \includegraphics[width=\linewidth]{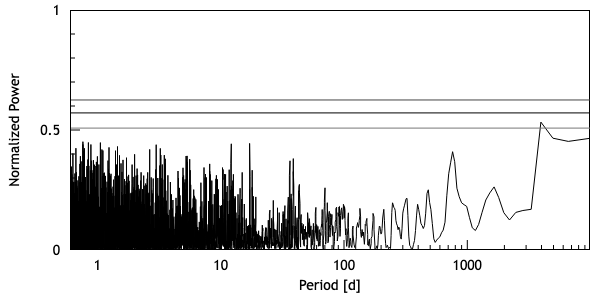}
    \caption{Periodograms of the RV data of HD11608 before (top) and after (bottom) detrending for $\log\mathrm{R'_{HK}}$ and the CCF-Bissector with a Gaussian low-pass filter (timescale 1 yr). The horizontal lines are the 10\%, 1\%, and 0.1\% FAP levels.}
    \label{fig:HD11608_Perio}
\end{figure}

The observations show high ($\gtrsim 0.5$) correlations with many activity indicators, such as the S-index, Ca-index, and $\log \mathrm{R'_{HK}}$ (see Fig. \ref{fig:HD11608_CORR}), which all have peaks in the periodogram above 0.1\% FAP at $\sim 3300$ days. After detrending for $\log \mathrm{R'_{HK}}$ with a Gaussian low-pass filter (timescale 0.5 yr) the peak reduces to below 10\% FAP. When using a longer timescale of 1 yr, the peak remains above 1\% FAP, but the residuals correlate with the CCF-Contrast, with a coefficient of -0.21. When subsequently detrending the CCF-Bissector, all correlation coefficients fall below 0.1 and the RV peak stays above 10\% FAP, as visible in Figure \ref{fig:HD11608_Perio}. The corresponding Keplerian solution has either an unlikely offset between the two datasets (21 m s$^{-1}$) or, when fixing the offset to 11 m s$^{-1}$ ---which is a more plausible offset for K1/K2V \citep{LoCurto2015}--- does not converge and finds a highly eccentric orbit with a period of decades. As neither solution is convincing, we conclude that the RV variations of HD11608 are most likely caused by stellar activity. If there is a companion present, its RV signal is heavily entangled with its magnetic cycle.


\subsection{Uncertain orbital solution of HD3964}
\label{sec:Uncertain}
\begin{figure}
    \centering
    \includegraphics[width=.93\linewidth]{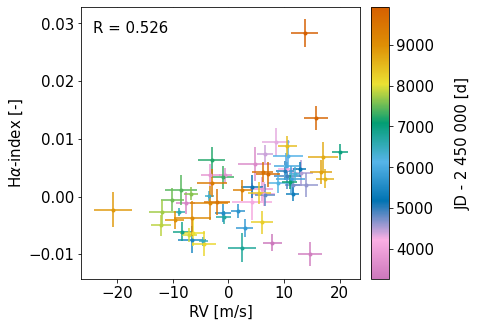}
    \caption{Correlation between H$\alpha$ and the RV values of HD3964.}
    \label{fig:HD3964_CorrRVHA}
\end{figure}

\begin{figure}
    \centering
    \begin{subfigure}[b]{1.0\linewidth}
        \includegraphics[width=\linewidth]{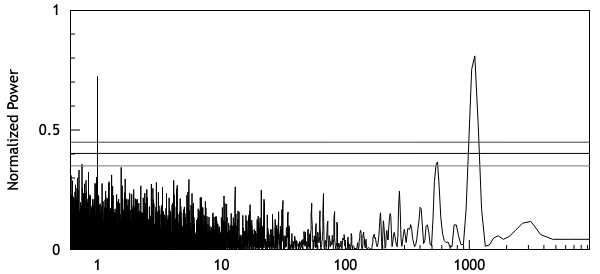}
    \end{subfigure}
    \begin{subfigure}[b]{1.0\linewidth}
        \includegraphics[width=\linewidth]{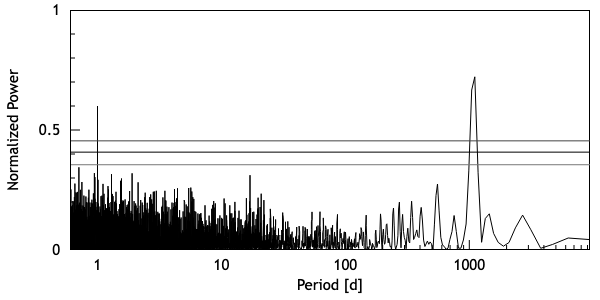}
    \end{subfigure}
    \caption{Periodograms of HD3964 for the initial RV data (up) and the RV data detrended for H$\alpha$ (down). The horizontal lines are the 10\%, 1\%, and 0.1\% FAP levels.}
    \label{fig:HD3964_Perio}
\end{figure}

HD3964 is a relatively quiet star ($\log\mathrm{R'_{HK}} = -4.87$). The $\log\mathrm{R'_{HK}}$ periodogram has a peak at 23.5 days, which corresponds to the stellar rotation period (see Table \ref{table:stellar_parameters}). The RV variation shows a significant correlation with H$\alpha$, as is visible in the correlation values in Figure \ref{fig:HD3964_CorrRVHA}. The H$\alpha$-index has a peak in the periodogram around 1150 days, which is similar to the detected long-period RV variation at 1086 days. 
There is a clear magnetic cycle visible in the RV variation of HD3964, but we cannot exclude a distant companion as well. After detrending for the H$\alpha$-index, the RV variation peak in the periodogram remains above 0.1\% FAP level at a period of 1086 days, as is visible in Figure \ref{fig:HD3964_Perio}, where the first periodogram is before and the second is after detrending the H$\alpha$-index. The two other strong peaks (both above 10\% FAP before detrending) are the aliases of the 1086 day period. As the periods of the magnetic cycle and possible companion are similar, there is no way to accurately determine the RV amplitude. Therefore, we present the orbital solution of HD3964 b with caution.

\begin{figure}
    \centering
    \includegraphics[width=\linewidth]{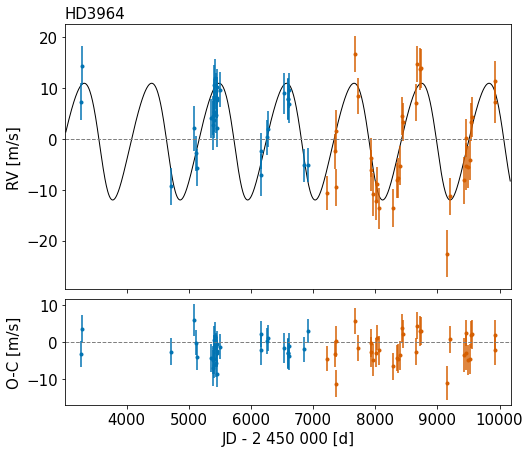}
    \caption{Radial velocity observations of HD3964 versus time and the residuals to the best-fit model after detrending. HARPS data from before
the fibre upgrade in 2015 are shown in blue  and those from after the upgrade are shown in orange.}
    \label{fig:HD3964_time}
\end{figure}

If the magnetic cycle is not intertwined with the companion signal, the detected long-period variation at 1086 days can be attributed to the presence of a distant Jupiter-like planet of 0.58 $M_J$ minimum mass. The period of the RV variation does not correspond to the rotational period of the star. Figure \ref{fig:HD3964_time} shows the RV variation of HD3964 with the proposed orbital solution versus time; the residuals (O-C) are included. 

The residuals show a correlation with the CCF-Bissector with a correlation coefficient of $\sim$0.3. However, the CCF-Bissector does not show any peaks above 10\% FAP in its periodogram. The average variation in RV signal changes after the 2015 fibre upgrade (6.2 m s$^{-1}$ before, 9.8 m s$^{-1}$ after), which is also visible in Figure \ref{fig:HD3964_time}. The stellar jitter of the model is equal to 3.1 m s$^{-1}$ and the $\sigma$(O-C) is 3.49 m s$^{-1}$. 

A drift does not improve the model; therefore, if there is a second companion, it should be at a short period. More extensive RV follow-up observations could provide more insight into the origin of the remaining fluctuations. Complementary RV measurements in the near-infrared (NIR) may help to disentangle stellar activity using the chromaticity of stellar active regions.

\subsection{Planetary systems}
\label{sec:planets}

\subsubsection{HD94771}
\begin{figure}
    \centering
    \includegraphics[width=\linewidth]{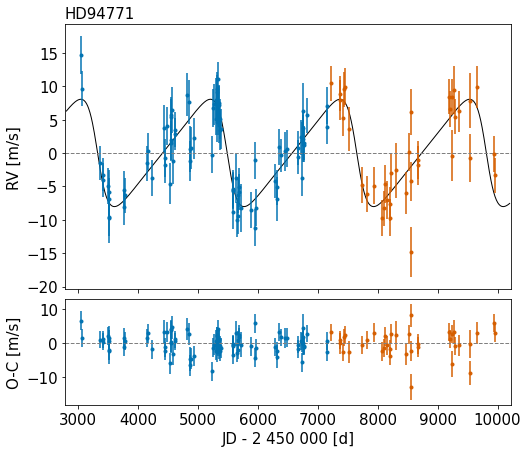}
    \caption{Radial velocity observations of HD94771 versus time and the residuals to the best-fit model. Line colours are as in Fig.5.}
    \label{fig:HD94771}
\end{figure}

The observations of the quiet relatively evolved (1.9 $R_{\odot}$) star HD94771 reveal no strong correlation with stellar activity indicators. However, there is a strong periodic RV variation above FAP 0.1\% for 2164 days plus its 1 day and 1 year alias. The Keplerian solution corresponds to a  companion with a minimum mass of 0.53 $M_J$. There is no strong suggestion of a second companion in the current data: adding a drift does not improve the model and the 2.2 m s$^{-1}$ stellar jitter is towards the lower limit for a 1.2 $M_{\odot}$ star with log$(g) \sim $4 cm s$^{-2}$ \citep{Luhn2020}. We conclude that neither stellar activity nor the rotational period of HD94771 induces the observed RV variation. The signal is caused by a giant exoplanet. The low stellar jitter might imply the rotation of the star is seen pole on; if the whole system is misaligned, the mass is much higher. In combination with the eccentric orbit, this suggests HD94771 b is potentially a brown dwarf. However, it is also possible that the star is in a quiet part of its cycle. As mentioned in Sect. \ref{sec:Orvara}, there is no strong astrometric signal visible for HD94771 b, and so we favour the latter explanation. The observations cover three periods of the orbit of HD94471b, as visible in Figure \ref{fig:HD94771}. 

\subsubsection{HIP54597}
HIP54597 is a relatively quiet star. Only one observation shows excessive activity in comparison to the others. Therefore, after the pre-selection of the S/N $>$ 25 and the DRS QC, this observation was excluded from the analysis. Stellar activity indicators S-index, H$\alpha$-index, log R$'_{\mathrm{HK}}$ , and $P_{\mathrm{rot}}$ suggest a modulation above 10\% FAP level at a slightly shorter period ($\sim$2825 days) than the RV variation (3250 days), and have correlation coefficients of approximately -0.25. 

\begin{figure}
    \centering
    \includegraphics[width=\linewidth]{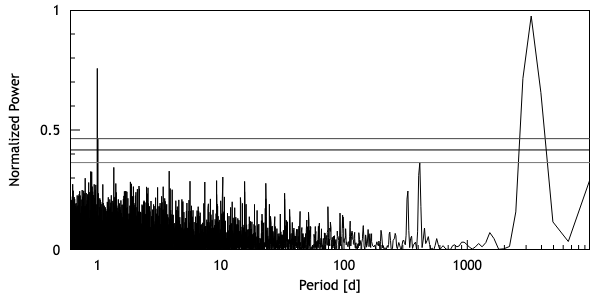}
    \caption{Radial velocity periodogram of HIP54597 after detrending for the S-index and the CCF-FWHM. The horizontal lines are the 10\%, 1\%, and 0.1\% FAP levels.}
    \label{fig:HIP54597_Perio}
\end{figure}

\begin{figure}
    \centering
\includegraphics[width=\linewidth]{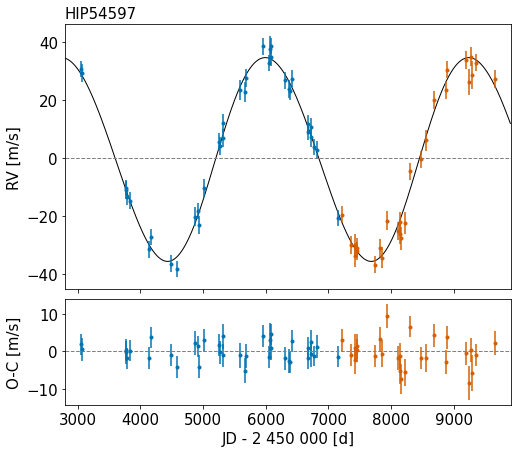}
    \caption{Radial velocity observations of HIP54597. Line colours are as in Fig.5.}
    \label{fig:HIP54597}
\end{figure}

However, after detrending for the S-index and the CCF-FWHM, the RV signal stays above FAP 0.1\% (see Fig. \ref{fig:HIP54597_Perio}); moreover, the detrending does not significantly affect the periodogram, and there are only some minor changes in the model ($\Delta \chi^2_{\mathrm{red}} = 0.01$). The correlation is small enough and the difference in periods large enough to conclude that the observed RV variation is not caused by stellar activity or the rotational period of the star, but by a companion with a minimum mass of 2.01 $M_J$  . The Keplerian solution is shown in Figure \ref{fig:HIP54597}. There are no strong indications of an additional companion, a drift does not improve the model, and 2.4 m s$^{-1}$ stellar jitter is relatively low for a K5V star \citep{Luhn2020}. The correlations suggest that there may be a magnetic cycle at play. 


\begin{table*}
\caption{Orbital and physical parameters for the exoplanet companions presented in this paper.}\label{table:planet_parameters}
\centering
\setlength\extrarowheight{3pt}
\begin{tabular}{l l c c c : c  c c c c c c c c} 
\hline\hline   
\multicolumn{2}{l}{Parameter} & HD3964 b$^{(a)}$ & HD94771 b & HIP54597 b & BD-210397 b & BD-210397 c \\ \hline
$P$ & [days] & $1086^{+9}_{-10}$ & $2164^{+21}_{-20}$ & $3250 \pm 17$ & $1891^{+56}_{-48}$ & $6360^{+6260}_{-711}$\\
$T_P$ & [rBJD] & $55785^{+134}_{-108}$ & $55485^{+40}_{-42}$ & $54994^{+264}_{-248}$ & $56490^{+255}_{-372}$ & $57479^{+2919}_{-1137}$ \\
$e$ & & $0.13^{+0.08}_{-0.07}$ & $0.39^{+0.05}_{-0.04}$ & $0.04 \pm 0.02$ & $0.1 \pm 0.1$ & $0.1^{+0.2}_{-0.1}$\\
$\omega$ & [$^{\circ}$] & $110^{+46}_{-37}$ & $90 \pm 8$ & $245^{+28}_{-29}$ & $-55^{+39}_{-61}$ & $6^{+103}_{-147}$\\
$K$ & [m s$^{-1}$] & $11 \pm 1$ & $8.0 \pm 0.4$ & $35.1 \pm 0.6$ & $14 \pm 2$ & $34^{+12}_{-3}$ \\
$m_2 \sin i$ & [$M_{J}$] & $0.58^{+0.06}_{-0.05}$ & $0.53 \pm 0.03$ & $2.01\pm 0.03$ & $0.7 \pm 0.1$ & $2.4^{+1.5}_{-0.2}$\\
$m_2 \sin i$ & [$M_{\oplus}$] & $183^{+19}_{-17}$ & $169 \pm 9$ & $639\pm 11$ & $214^{+31}_{-32}$ & $759^{+485}_{-76}$\\
$a_2$ & [AU] & $2.07 \pm 0.02$ & $3.48 \pm 0.03$ & $3.81\pm 0.02$ & $2.63^{+0.06}_{-0.05}$ & $5.9^{+3.4}_{-0.5}$\\
$S$ & [$S_{\oplus}$] & $0.207 \pm 0.005 $ & $0.271 \pm 0.006$ & $0.0143 \pm 0.0002$ & $0.020 \pm 0.001$ & $0.0039^{+0.0045}_{-0.0007}$\\
\hline
$\gamma_{03}$ & [m s$^{-1}$] & $-11825 \pm 1$ & $21117.2 \pm 0.3$ & $51751.8 \pm 0.5$ & \multicolumn{2}{c}{$3191^{+3}_{-23}$}\\
$\gamma_{15}$ & [m s$^{-1}$] & $-11807 \pm 1$ & $21130.9 \pm 0.5$ & $51763.1 \pm 0.7$ & \multicolumn{2}{c}{$3210^{+5}_{-9}$} \\
$\gamma_\mathrm{HIR}$ & [m s$^{-1}$] & - & - & - & \multicolumn{2}{c}{$-3 \pm 3$}\\
Jitter & [m s$^{-1}$] & $3.1^{+0.5}_{-0.4}$ & $2.2^{+0.3}_{-0.2}$ & $2.4 \pm 0.4$ & \multicolumn{2}{c}{$8.9^{+0.8}_{-0.7}$} \\
\hline
$\sigma (\mathrm{O-C})$ & [m s$^{-1}$] & 3.49 & 2.84 & 2.97 & \multicolumn{2}{c}{8.29}\\ 
$\chi^2_{\mathrm{red}}$ & & 1.16 & 1.17 & 1.21 & \multicolumn{2}{c}{1.21}\\ \hline 
\end{tabular}
\begin{flushleft}{\footnotesize \textbf{Notes:} Errors are 1$\sigma$ Monte Carlo uncertainties. $\sigma \mathrm{(O-C)}$ is the residual noise after orbital fitting the observed drift with time.\\
$^{(a)}$ The orbital solution is presented with caution, as it comes with a magnetic cycle with a similar period.}\end{flushleft}
\end{table*}

\subsubsection{BD-210397}
\begin{figure}
    \centering
    \includegraphics[width=\linewidth]{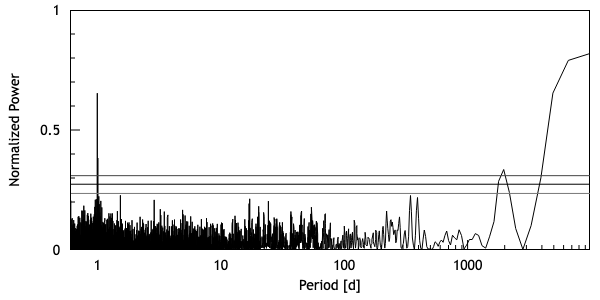}
    \caption{Radial velocity data periodogram of BD-210397 after detrending for the S-index. The horizontal lines are the 10\%, 1\%, and 0.1\% FAP levels.}
    \label{fig:BD-210397-Sindex}
\end{figure}

There are 83 HARPS observations of the late K star BD-210397 that pass the DRS QC. One data point was re-reduced with a K5 mask to match the rest of the observations. The HARPS observations show a negative correlation with the H$\alpha$-, Ca-, Na-, and S-index, but their periodicity (respectively $\sim$5800, $\sim$5500, $\sim$5400, and $\sim$5250 days) is not similar to the RV variation (1891 and 6360 days). After detrending for the S-index, the correlation coefficients are reduced to below 0.1 and the two RV variations remain visible in the periodogram (see Fig. \ref{fig:BD-210397-Sindex}). We conclude that the mentioned correlations are not the origin of the RV variation. We include 18 public HIRES \citep{HIRES} observations to the analysis for better convergence of the fit.

There are two long-period RV variations visible in the observations, which are attributed to a  companion with a minimum mass of 0.7 $M_J$ and a period of  1891 days (BD-210397 b) and another companion
(BD-210397 c) of $2.4 M_J$ minimum mass and a period of  6360 days. When fitting BD-210397 c first, the peak of BD-210397 b reduces from above 0.1\% to below 10\% FAP. However, the model including only BD-210397 c comes with an unrealistic offset (-14 m s$^{-1}$) between the HARPS data before and after the fibre upgrade \cite{LoCurto2015} and a stellar jitter of 10 m s$^{-1}$. When including BD-210397 b, the stellar jitter reduces to 8.9 m s$^{-1}$, the offset is more plausible at 19 m s$^{-1}$, and the $\chi^2_{\mathrm{red}}$ improves from 25 to 17 (without including stellar jitter). The $\ell1$-periodogram \citep{l1periodogram}, also finds the presence of the 1891 day signal after fitting the 6360 day signal. The $\ell$1-periodogram is a variant of the Lomb-Scargle periodogram, but instead of minimising the $\ell$2 norm (sum of squares), it minimises the $\ell$1 norm (sum of absolute values). This allows the $\ell$1 periodogram to be less sensitive to outliers and non-Gaussian noise.

\begin{figure}
    \begin{center}
    \includegraphics[width=\linewidth]{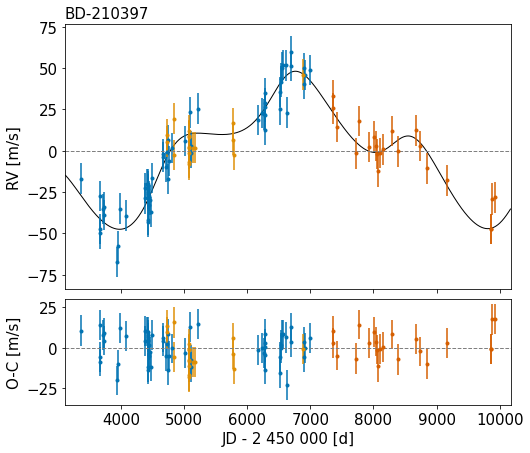}
    \end{center}
    \caption{Radial velocity curves for BD-210397, with blue and orange indicating HARPS data before and after the fibre upgrade in 2015, respectively (right). The public HIRES data are labelled in yellow in the left panel.}
    \label{fig:BD-210397}
\end{figure}

The model is presented in Table \ref{table:planet_parameters}, and the solution is visible in Figure \ref{fig:BD-210397}. The period of BD-210397 c is not well constrained; more observations are required. As the $\log \mathrm{R'_{HK}}$ value of BD-210397 is undefined, we cannot directly conclude as to whether or not the high stellar jitter (8.9 m s$^{-1}$) is caused by activity; although heavily uncertain, the age (6.2 $\pm$ 4.7 Gyr) suggests it is not. However, the standard deviation of the FWHM ($\pm$ 0.022 km s$^{-1}$) is large in comparison to the other targets presented in this paper, which indicates that BD-210397 might be on the more active side. Combining the $S$-index amplitude with the stellar mass 0.679 $M_{\odot}$ and log(g) = 4.67 cm s$^{-1}$, the stellar jitter is expected to be within the range of 3-9 m s$^{-1}$ \citep{Luhn2020}. The stellar jitter found is high but within the expected range. 

Another explanation for the stellar jitter could be a very short-period companion. The periodogram indeed shows a $\sim 0.5$ day signal below 10\% FAP level, with its aliases also present. When adding this signal, the jitter is reduced from $\sim$8.9 m s$^{-1}$ to $\sim$7 m s$^{-1}$. A short-period companion helps to reduce the jitter, but there are not enough data points to be sure. To determine whether the high stellar jitter is caused by another companion, by sampling or by aliasing effects, follow-up measurements with short time intervals are required. 

The HIRES and HARPS O-C values are visible in Figure \ref{fig:BD-210397}. Where for HARPS the average (absolute) O-C is 8 m s$^{-1}$, HIRES on average varies from the model by 10 m s$^{-1}$. The HIRES data do agree with the model
found but cover a very limited time span. 


\subsubsection{HD74698}
The RVs of the quiet star HD74698 do not show a high correlation with stellar activity indicators apart from the CCF-FWHM (0.26), which has a period $\sim$8500 days, and disappears when changing the offset between the two datasets. This is even more evident when binning the data every 60 days. As it is dependent on offset, the CCF-FWHM is not detrended and the offset is fixed to 15 m s$^{-1}$. This value is expected for a G5V star \citep{LoCurto2015}, reduces the correlation between the RV and CCF-FWHM to almost zero, and is the best-fit offset found by the binned data. There are two signals above 0.1\% FAP level visible in the RV periodogram, $P_b = 15.017$ days ($K_b = 5.8$ m s$^{-1}$) and $P_c = 3449$ days ($K_c = 5.3 $ m s$^{-1}$). After adding the first Keplerian model, the correlations with the stellar activity indicators remain low, indicating that both periods do not originate from activity. There is a third signal visible in the periodogram at $\sim$1000 days ($K \sim$6.5 m s$^{-1}$); it does not correspond to any activity indicator. Apart from DACE, we used two independent programs: KIMA \citep{KIMA} and the $\ell_1$-periodogram \citep{l1periodogram}, to verify the model. Both programs also suggest the presence of the 1000 day signal. KIMA finds the three-planet solution to be the most likely, and $\ell_1$-periodogram finds all three periods but does not suggest the 1000 day signal as a prominent one. As the 1000 day signal is below 10\% FAP level in the DACE periodogram and strongly depends on the offset between the two datasets, it does not appear sufficiently robust to be included in the analysis.

\begin{figure}
    \centering
    \includegraphics[width=\linewidth]{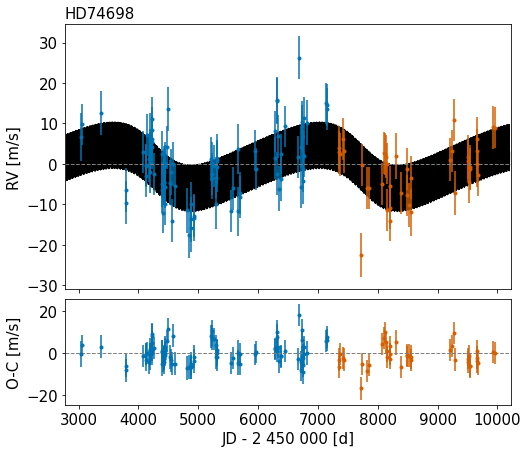}
    \caption{Radial velocity observations of HD74698, including two signals: the 3530 day period planet with a minimum mass of 0.38 $M_J$ and a 15.017 day period planet with a minimum mass of 0.08 $M_J$. HARPS data from before and after the fibre upgrade in 2015 are marked in blue and orange, respectively.}
    \label{fig:HD74698}
\end{figure}

\begin{figure}
    \centering
    \includegraphics[width=\linewidth]{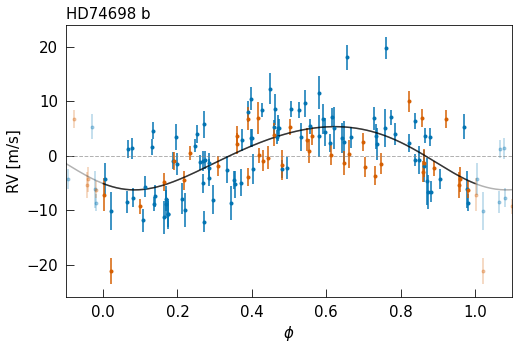}
    \caption{Phase-folded RV curves for HD74698 b after subtracting HD74698 c. HARPS data before and after the fibre upgrade in 2015  are marked  in blue and orange.}
    \label{fig:HD74698phs}
\end{figure}

$\ell_1$-periodogram is also used to calculate the periodograms of the stellar activity indicators, none of the stellar activities correspond to $P_b$, $P_c$, or the possible 1000 day signal. Figure \ref{fig:HD74698} shows the best-fit model, Figure \ref{fig:HD74698phs} the phase-folded solution for HD74698 b, and Table \ref{table:planet_parameters} the corresponding orbital parameters. Neither signal originates from stellar activity or the rotational period. We conclude that HD74698 b and c are exoplanetary companions. There are large variations still visible in the residuals, which are potentially caused by a 1000 day period companion. Additional observations can provide more insight into this compound system, for which there is substantial evidence of a third companion. As the minimum masses of the two found signals are in the range from Neptune to Saturn, this star is a very interesting target for follow-up observations. 


\begin{table}
\caption{Orbital and physical parameters for the exoplanet companions of HD74698.}\label{table:HD74698}
\centering
\setlength\extrarowheight{3pt}
\begin{tabular}{l l c c c c c : c c c c c c c} 
\hline\hline 
\multicolumn{2}{l}{Parameter} & HD74698 b & HD74698 c \\ \hline
$P$ & [days] & $15.017 \pm 0.002$ & $3449^{+189}_{-210}$\\ 
$T_P$ & [rBJD] & $55499^{+3}_{-2}$ & $57457^{+603}_{-659}$\\ 
$e$ & & $0.1 \pm 0.1$ & $0.2^{+0.2}_{-0.1}$ \\
$\omega$ & [$^{\circ}$] & $143^{+80}_{-46}$ & $103^{+126}_{-62}$\\
$K$ & [m s$^{-1}$] & $5.8^{+0.9}_{-0.8}$ & $5.3 \pm 0.8$\\
$m_2 \sin i$ & [$M_{J}$] & $0.07 \pm 0.01$ & $0.40 \pm 0.06$ \\
$m_2 \sin i$ & [$M_{\oplus}$] & $23 \pm 3$ & $126 \pm 19$\\
$a_2$ & [AU] & $0.121 \pm 0.001$ & $4.5 \pm 0.2$ \\
$S$ & [$S_{\oplus}$] & $123 \pm 2$ & $0.088 \pm 0.008$\\
\hline
$\gamma_{03}$ & [m s$^{-1}$] & \multicolumn{2}{c}{$38819.9 \pm 0.6$}\\
$\gamma_{15}\mathrm{^{(a)}}$ & [m s$^{-1}$] &  \multicolumn{2}{c}{$38834.9$}\\
Jitter & [m s$^{-1}$] &  \multicolumn{2}{c}{$4.9 \pm 0.4$}\\
\hline
$\sigma (\mathrm{O-C})$ & [m s$^{-1}$] & \multicolumn{2}{c}{5.11} \\
$\chi^2_{\mathrm{red}}$ & &  \multicolumn{2}{c}{1.11}\\ \hline 
\end{tabular}
\begin{flushleft}{\footnotesize \textbf{Notes:} Errors are 1$\sigma$ Monte Carlo uncertainties. $\sigma \mathrm{(O-C)}$ is the residual noise after orbital fitting the observed drift with time.\\
$^{(a)}$ Fixed difference between $\gamma_{15}$ and $\gamma_{03}$ of 15 m s$^{-1}$.}\end{flushleft}
\end{table}

\subsection{Brown dwarf candidates}
\label{sec:browndwarfs}
The mass limits of brown dwarfs, are subject to debate; between $\sim 13 M_J$ (limit of deuterium fusion) and 80 $M_J$ (limit of hydrogen fusion) is the generally applied range \citep{Spiegel2011, Chabrier2000}. However, \cite{Sahlmann2010, Sahlmann2011b} suggest that planet masses can go up to 25 $M_J$. Here we apply the commonly used limit of 13 $M_J$ in order to differentiate brown dwarfs from massive planets.
The eccentricities of brown dwarfs are usually larger than those of exoplanets. As they are faint, they are difficult to detect by direct imaging; here, long-term RV surveys like ours provide a means for their detection.

\subsubsection{HD62364}
\begin{figure}
    \centering
    \includegraphics[width=\linewidth]{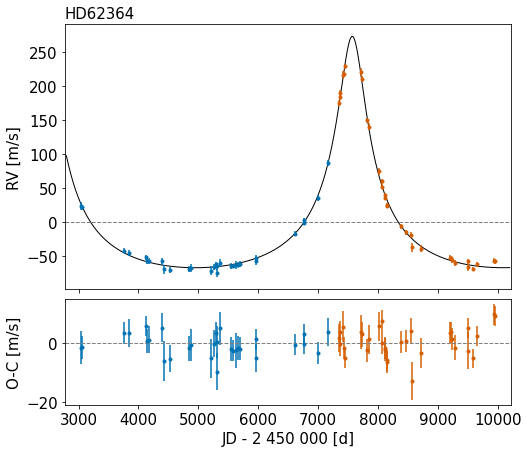}
    \caption{Radial velocity curves for HD62364, with blue and orange indicating
HARPS data before and after the fibre upgrade in 2015, respectively.}
    \label{fig:HD62364}
\end{figure}

The observations of the low-metallicity star HD62364 reveal a strong high-eccentric RV variation. With a 5138 day period, this signal corresponds to a companion with a minimum mass of 12.7 $M_J$  . The minimum mass of HD62364 b is at the edge of the range of brown dwarfs, but since its mass is higher with the inclination found (see section \ref{sec:Orvara}) and high eccentricity is more common among brown dwarfs, we conclude that this companion is probably a brown dwarf. The HARPS spectra cover the entire phase of HD62364 b, as is visible in Figure \ref{fig:HD62364}. There is no strong correlation with any stellar activity indicator and the rotational period is too short to create the observed fluctuation. There is no drift present and the remaining 3.0 m s$^{-1}$ stellar jitter is within the expected range for a 1.2 $M_{\star}$ star with log(g) $\sim$4 \citep{Luhn2020}.

\subsubsection{HD56380}
\begin{figure}
    \centering
    \includegraphics[width=\linewidth]{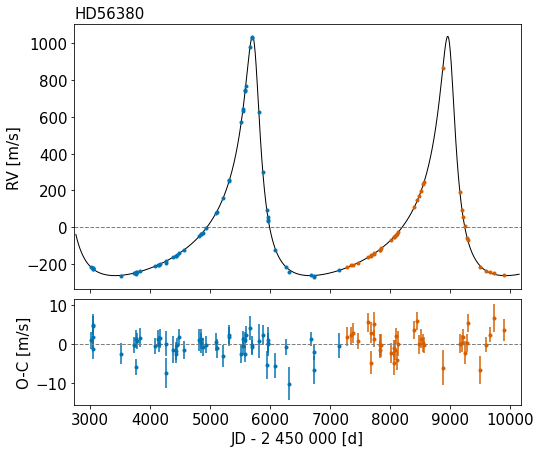}
    \caption{Radial velocity curves for HD56380, with blue and orange indicating HARPS data before and after the fibre upgrade in 2015, respectively.}
    \label{fig:HD56380}
\end{figure}

The inactive star HD56380 exhibits a strong RV variation above 0.1\% FAP level at 3254 days, corresponding to a 33.2 $M_J$ minimum mass companion, without strong stellar activity indicator correlations. The phase is well covered, as is visible in Figure \ref{fig:HD56380}. We conclude that the RV variation is not caused by activity or the star's rotational period but by a companion. The 1.2 m s$^{-1}$ stellar jitter of the Keplerian model is low for a $0.8 M_{\odot}$ star with log(g) $\sim$4.5 and a drift does not enhance the fit. 

\subsubsection{HD221638}
\begin{figure}
    \centering
    \includegraphics[width=\linewidth]{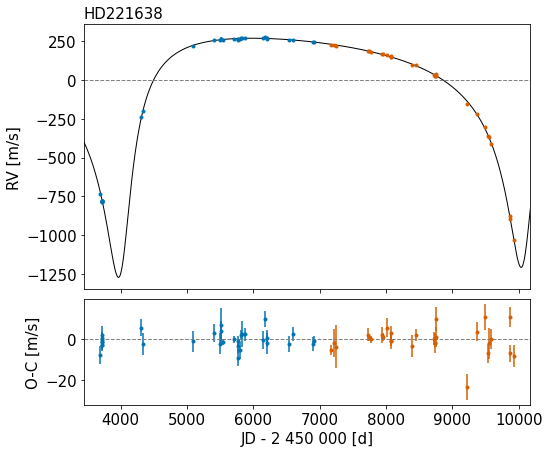}
    \caption{ Radial velocity curves for HD221638, with blue and orange indicating HARPS data before and after the fibre upgrade in 2015, respectively.}
    \label{fig:HD221638}
\end{figure}

The relatively quiet star HD221638 shows an RV variation with a period of $\sim$2 days that correlates  with the CCF-FWHM with a coefficient of 0.43 and with the S-index with a coefficient of 0.20. None of the periods come close to the RV variation at 6064 days, which stays above 0.1\% FAP level even after detrending the CCF-FWHM. The long-period RV variation corresponds to a companion with a minimum mass of  53 $M_J$. Adding a linear drift to the orbital parameters improves the fit by $\Delta \chi^2_{\mathrm{red}} = 0.08$, implying the presence of a  companion with an even longer period. After detrending the CCF-FWHM and adding a drift, there is no remaining stellar jitter. The phase is fully covered and shown in Figure \ref{fig:HD221638}.

\subsubsection{HD33473A}
The companion of the evolved star\footnote{Though given as a dwarf in Table \ref{table:stellar_parameters}, the $\log(g)$ and $R$ suggest it is evolved, and the spectral type is a Hipparcos value and does not take into account the more recent parameters from Gaia and/or others.} ($2.3 R_{\odot}$) HD33473A was discovered by \citep{Moutou2011}. However, the Keplerian solution was incomplete as the observations covered a small fraction of the period and assumed an additional long-term drift.

The increased number of observations allows us to present a significantly improved orbital solution ($\Delta \chi^2_{\mathrm{red}} = 2.25$). As the phase is not fully covered and the model has difficulties converging when the offset is allowed to vary, we fix the offset between the two datasets to 14 m s$^{-1}$. This offset corresponds to an approximation for its spectral type G3V, derived from the values mentioned for G2V and G4V in \cite{LoCurto2015}. When excluding one observation with excessive activity ($\Delta\log\mathrm{R'_{HK}} \sim -0.4$), the RV variation correlates with the CCF-Contrast with a coefficient of 0.44 (no period above 10\% FAP) and with the CCF-FWHM with a coefficient of -0.49. The CCF-FWHM shows a long period below 10\% FAP, which appears to be increasing outside of the range of the  span of the observations. 

\begin{figure}
    \centering
    \includegraphics[width=\linewidth]{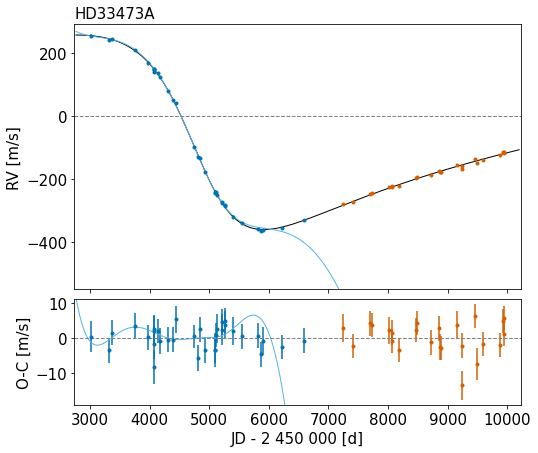}
    \caption{Radial velocity curves for HD33473A, which was originally discovered by \cite{Moutou2011}. Blue and orange colours indicate HARPS data before and after the fibre upgrade in 2015, the light blue line is the orbital solution as found by \cite{Moutou2011}, and the black line is the orbital solution as proposed in this paper.}
    \label{fig:HD33473A}
\end{figure}

After detrending the CCF-FWHM, the strong signal of 50 years remains visible in the periodogram. This period corresponds to a 
companion with a minimum mass of 38.3 $M_J$, which is significantly heavier than initially thought (7.2 $\pm$ 0.3 $M_J$). The rotational period of the star (32 days) cannot account for the signal, nor can HD33473B, the stellar companion at a separation of 10 arcsecs \citep{Chaname2012}. Though a stellar companion justifies the inclusion of a long-term drift in the Keplerian model, we decided not to include one, as a linear drift does not improve the model. The phase is still not entirely covered, as shown in Figure \ref{fig:HD33473A}; more observations could further improve the orbital solution. We refer to HD33473C as the companion to HD33473A, which was previously designated as HD33473A b by \citep{Moutou2011}.

\begin{table*}
\caption{Orbital and physical parameters for the brown dwarf candidates presented in this paper.}\label{table:brown_dwarfs}
\centering
\setlength\extrarowheight{3pt}
\begin{tabular}{l l c c c c c c c c c c c} 
\hline\hline   
\multicolumn{2}{l}{Parameter} & HD62364 b & HD56380B & HD221638B & HD33473C\\ \hline
$P$ & [days] & $5138^{+24}_{-22}$ & $3254.0\pm 0.5$ & $6064^{+17}_{-12}$ & $18236^{+1159}_{-651}$\\
$T_P$ & [rBJD] & $57575\pm5$ & $55738.1\pm0.4$ & $54013^{+8}_{-12}$ & $54859^{+19}_{-14}$\\
$e$ & & $0.607\pm0.005$ & $0.6507 \pm 0.0005$ & $0.694^{+0.004}_{-0.003}$ & $0.56^{+0.02}_{-0.01}$\\
$\omega$ & [$^{\circ}$] & $1.2\pm0.8$ & $23.83^{+0.08}_{-0.07}$ & $200^{+1}_{-2}$ & $108^{+2}_{-1}$\\
$K$ & [m s$^{-1}$] & $170\pm2$ & $650.5\pm0.8$ & $758^{+11}_{-9}$ & $308^{+3}_{-2}$\\
$m_2\sin i$ & [$M_J$] & $12.7\pm0.2$ &$33.2\pm 0.2$ & $53 \pm 1$ & $38.3^{+0.7}_{-0.6}$\\
$m_2$$^{(a)}$ & [$M_J$] & ${18.77}_{-0.63}^{+0.66}$ & ${375.3}_{-8.4}^{+8.6}$ & ${110.0}_{-3.7}^{+3.9}$ & ${271.0}_{-3.8}^{+3.9}$ \\
$i$$^{(a)}$ & [$^\circ$] & ${137.3}_{-1.6}^{+1.5}$ & ${5.96}\pm 0.10$ & ${30.68}_{-0.76}^{+0.82}$ & ${170.79}_{-0.16}^{+0.15}$ \\
$a_2$ & [AU] & $6.15^{+0.04}_{-0.03}$ & $4.17\pm 0.01$ & $6.87 \pm 0.08$ & $14.7^{+0.6}_{-0.4}$\\
$S$ & [$S_{\odot}$] & $0.072 \pm 0.001$ & $0.0322 \pm 0.0006$ & $0.044 \pm 0.001$ & $0.024 \pm 0.002$ \\
\hline
$\gamma_{03}$ & [m s$^{-1}$] & $35509.2^{+0.9}_{-1.0}$ & $90952.7\pm 0.3$ & $1119^{+5}_{-3}$ & $44265^{+8}_{-5}$\\
$\gamma_{15}$ & [m s$^{-1}$] & $35526\pm 1$ & $90964.8\pm 0.4$ & $1130^{+5}_{-4}$ & $44279^{(b)}$\\
Jitter & [m s$^{-1}$] & $3.0\pm0.5$ & $1.2^{+0.4}_{-0.5}$ & - & $3.2^{+0.5}_{-0.4}$\\
Drift & [m s$^{-1}$ yr$^{-1}$] & - & - & $4^{+1}_{-2}$ & -\\ 
\hline
$\sigma (\mathrm{O-C})$ & [m s$^{-1}$] & 3.70 & 2.53 & 3.33 & 3.47\\
$\chi^2_{\mathrm{red}}$ & & 1.06 & 1.30 & 1.28 & 1.25\\\hline 
\end{tabular}
\begin{flushleft}{\footnotesize \textbf{Notes:} Errors are 1$\sigma$ Monte Carlo uncertainties. $\sigma \mathrm{(O-C)}$ is the residual noise after orbital fitting the observed drift with time.\\
$^{(a)}$: Results from astrometric solution}; see section \ref{sec:Orvara}. 
$^{(b)}$: Fixed difference between $\gamma_{15}$ and $\gamma_{03}$ of 14 m s$^{-1}$. \end{flushleft}
\end{table*}

\subsection{Binaries}
\label{sec:binaries}
Long-term RV monitoring allows the detection of binaries not yet resolved by direct imaging because they are too faint and their separation is too small. However, as their periods are on the order of decades, the phases of the binaries are not always fully covered. To confirm the origin of the RV variation, we searched for a second component in the CCFs, and resolve binaries as SB2; see \cite{Bouchy2016} for an in-depth explanation of the applied method. Briefly, the CCFs are recomputed with an M mask to optimise the detectability of the low-mass companion. The CCFs are shifted to the systemic velocity $V_0$ and their average is subtracted to remove the first component. The expected radial velocity of the second component is defined by the radial velocity of the main component $V_1$, the systemic velocity $V_0$, and the mass ratio $q = M_{\mathrm{pl}} / M_{\star}$. The residual CCFs are shifted to this expected value and again averaged. Consequently, the secondary is expected to be at $V_0$. We inspected various combinations of $V_0$ and $q;$  by searching for the deepest peak at the varying location of $V_0$, we found which combination  best matches the CCFs. The range of $V_0$ is defined by what is feasible according to the RV model. In this section, we present the finding of one stellar binary (HD11938)  confirmed using this method, and constrain its inclination $i$. The other presented stellar binary (HD61383) cannot be confirmed this way due to the blending of the primary and secondary peaks in the CCF.

\subsubsection{HD11938}
\begin{figure}
    \centering
    \includegraphics[width=\linewidth]{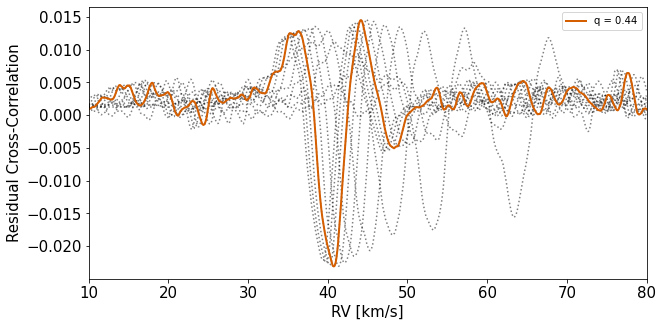}
    \caption{Averaged CCF residuals of HD11938 for different radial velocity shifts. The deepest peak (orange curve) occurs for a shift corresponding to a mass ratio of 0.44.}
    \label{fig:HD11938_ccf}
\end{figure}

There are 25 observations of the active star HD11938 that pass the DRS quality control and have S/N $>$ 25. One measurement was reduced with a G2 mask instead of a K5 mask and was therefore reprocessed. As there are only 25 observations that do not fully cover the phase of HD11938B we fixed the offset between the HARPS data before and after the fibre upgrade to 15 m s$^{-1}$ as suggested by \cite{LoCurto2015} for K4/5V spectral types. There is a clear signal in the periodogram of 35 years. The stellar binary is also visible in the CCFs, where the best solution is at a $V_0$ of $40.7 \pm 0.2$ km s$^{-1}$ ($= \gamma_{03}$). The mass ratio $q$ in that case is $0.44 \pm 0.03$, corresponding to a secondary mass of $0.33 \pm 0.03$ $M_{\odot}$. Figure \ref{fig:HD11938_ccf} shows the average CCF residual for this $V_0$ and $q$ combination in red.

\begin{figure}
    \centering
    \includegraphics[width=\linewidth]{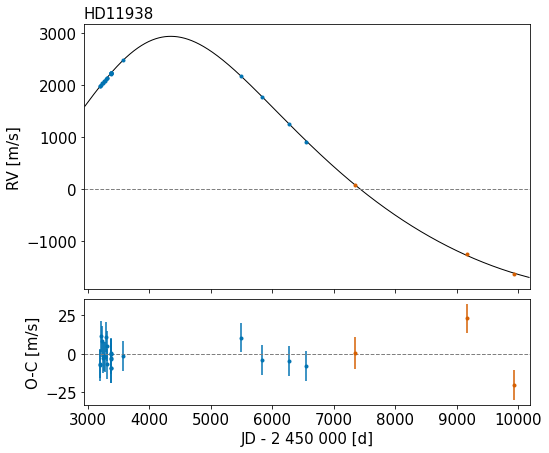}
    \caption{Radial velocity curves for HD11938, with blue and orange indicating HARPS data before and after the fibre upgrade in 2015.}
    \label{fig:HD11938}
\end{figure}

HD11938 is an active star (log$\mathrm{R'_{HK}} \sim -4.5$) with correlations with all stellar activity indicators, but none have periodical signals above 10\% FAP level in the periodogram. This strong RV variation signal, with an amplitude of 2.41 km s$^{-1}$, is therefore not caused by activity. We fix $\gamma_{03}$ to the found systemic velocity and its error margins $V_0 = 40.7 \pm 0.2$ km s$^{-1}$ in three iterations (i.e. 40.5, 40.7, and 40.9 km s$^{-1}$). The model presented in Table \ref{table:binaries} corresponds to 40.7 km s$^{-1}$, and the presented errors correspond to the models found for 40.5 and 40.9 km s$^{-1}$. This is done because the MCMC model drifts away from the found systemic velocity when using a prior. We note that the period is heavily dependent on the systemic velocity. The solution is visible in Figure \ref{fig:HD11938}. The found minimum mass $m \sin i = 0.21 M_{\odot}$ corresponds to an inclination of $39.33^{\circ} \pm 0.07 ^{\circ}$. The model does not include detrending or a drift, as both do not improve the model. The relatively large 9 m s$^{-1}$ stellar jitter agrees with an active star (without detrending). We conclude that the rotation period cannot account for the signal nor can the stellar activity. The RV variation belongs to a stellar binary companion, separated by 262 mas from its host star. The Gaia archive does not show a secondary within this distance range. 

\subsubsection{HD61383}
\begin{figure}
    \centering
    \includegraphics[width=\linewidth]{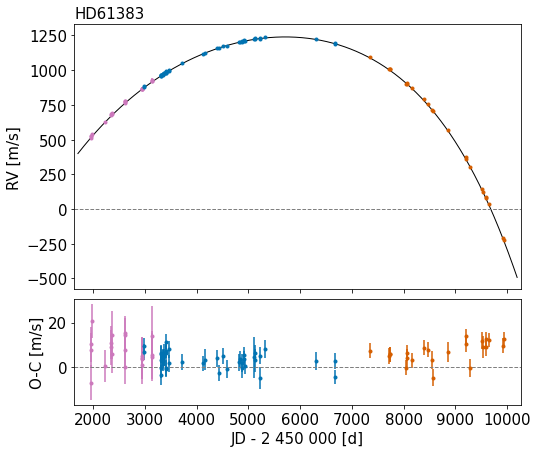}
    \caption{Radial velocity curves for HD61383, with purple indicating CORALIE data, and blue and orange showing HARPS data before and after the fibre upgrade in 2015.}
    \label{fig:HD61383}
\end{figure}

There are 66 HARPS observations of the metal-poor quiet star HD61383 that pass the DRS QC. The average photon noise is 3.1 m s$^{-1}$. To increase phase coverage, the analysis also includes 20 observations from CORALIE \citep{CORALIE} with an average photon noise of 6.3 m s$^{-1}$. The CORALIE instrumental error is fixed at 5.0 m s$^{-1}$. The phase of the data is not fully covered; this would require  at least 20 more years of observations, as is visible in Figure \ref{fig:HD61383}. Due to the incomplete coverage of the phase, the model has difficulties converging. We fix the offset between the HARPS data to 14 m s$^{-1}$, as HD61383 is a G3V star. There is a strong peak visible in the periodogram at 39.1 years corresponding to a minimum mass of 0.18 $M_{\odot}$. 

The HARPS observations weakly correlate with almost all stellar activity indicators and though there are some with long-term trends, the amplitude of the RV variation is too large to be caused by stellar activity. The binary is barely visible in the CCFs as a secondary peak, but the amplitude is small in comparison to the noise, as it is in most CCFs blended with the primary peak. With the currently available data, we present the orbit, as shown in Table \ref{table:binaries} and Figure \ref{fig:HD61383}.

\begin{table}
\caption{Orbital and physical parameters for the binaries presented in this paper.}\label{table:binaries}
\centering
\setlength\extrarowheight{3pt}
\begin{tabular}{l l c c c c c c c c c c c} 
\hline\hline   
\multicolumn{2}{l}{Parameter} & HD11938B & HD61383B\\ \hline
$P$ & [years] & $35.0^{+3.7}_{-3.6}$ & $39.1_{-0.9}^{+0.5}$\\
$T_P$ & [rBJD] & $53864_{-176}^{+94}$ & $62023^{+46}_{-155}$ \\
$e$ & & $0.23^{+0.03}_{-0.04}$ & $0.35 \pm 0.01$\\
$\omega$ & [$^{\circ}$] & $-22^{+4}_{-8}$ & $169.0 \pm 0.6$\\
$K$ & [m s$^{-1}$] & $2413^{+93}_{-91}$ & $1887^{+70}_{-164}$\\
$m_2 \sin i$ & [$M_{\odot}$] & $0.21 \pm 0.01$ & $0.18^{+0.01}_{-0.02}$\\
$m_{2,\mathrm{CCF}}$ & [$M_{\odot}$] & $0.33 \pm 0.03$ & - \\
$i_{\mathrm{CCF}}$ & [$^{\circ}$] & $39.33 \pm 0.07$& -\\
$m_2$$^{(a)}$ & [$M_{\odot}$] & $0.386 \pm 0.007$ & $0.190 \pm 0.003$\\
$i$$^{(a)}$ & [$^\circ$] & ${64.0}_{-2.4}^{+2.6}$ & ${86.2}\pm 2.9$ \\
$a_2$ & [AU] & $10.6 \pm 0.8$ & $11.7^{+0.1}_{-0.2}$\\
\hline
$\gamma_{03}$ & [m s$^{-1}$] & $40729 \pm 200^{(b)}$ & $51775^{+94}_{-25}$\\
$\gamma_{15}$ & [m s$^{-1}$] & $40744^{(c)}$ & $51789^{(d)}$ \\
$\gamma_{\mathrm{COR}}$ & [m s$^{-1}$] & - & $51734 \pm 3$\\
Jitter & [m s$^{-1}$] & $9^{+5}_{-2}$ & $2.9 \pm 0.4$\\
\hline
$\sigma (\mathrm{O-C})$ & [m s$^{-1}$] & 6.04 & 3.85 \\
$\chi^2_{\mathrm{red}}$ & & 1.33 & 1.01 \\ \hline 
\end{tabular}
\begin{flushleft}{\footnotesize \textbf{Notes:} Errors are 1$\sigma$ Monte Carlo uncertainties. $\sigma \mathrm{(O-C)}$ is the residual noise after orbital fitting the observed drift with time.\\
$^{(a)}$ Results from astrometric solution}; see section \ref{sec:Orvara}.\\
$^{(b)}$ Offset as found by second component in CCF.\\
$^{(c)}$ Fixed difference between $\gamma_{15}$ and $\gamma_{03}$ of 15 m s$^{-1}$.\\
$^{(d)}$ Fixed difference between $\gamma_{15}$ and $\gamma_{03}$ of 14 m s$^{-1}$.
\end{flushleft}
\end{table}

\section{True mass and inclination from astrometry}
\label{sec:Orvara}
Astrometry is a promising technique for measuring the inclinations of our systems, and the single measurement precision of Gaia in particular should allow such measurements ($10 \upmu$as, based on Gaia EDR3 \citeauthor{GaiaDR3e} \citeyear{GaiaDR3e}). The S/N reached by Gaia can be estimated by combining the astrometric signal (approximately $M_{\mathrm{pl}} M_{\star}^{-1} a_{\mathrm{pl}} d^{-1}$ for $M_{\mathrm{pl}} \ll M_{\star}$) with the single measurement precision and the number of Gaia measurements from the Gaia Observation Forecast Tool\footnote{\url{http://gaia.esac.esa.int/gost/}}. All presented targets have an expected S/N of above 6.2, the minimum value needed to retrieve an astrometric orbit from combined astrometry and radial velocity data \citep{Sahlmann2011}. However, it is not certain that Gaia covers the full orbit, and as the presented periods are on the order of several years, depending on the companion mass, the signal might get included in the proper motion. The Gaia excess noise values and their significance are an indication of the astrometric signal. For all presented targets, the excess noise significance is larger than 2, meaning there is a significant astrometric signal seen by Gaia \citep{Lindegren2012}. 

As the individual Gaia observations have not yet been released, we are limited to the currently available proper motion and RA/Dec positions. To derive the true mass and inclination, we use the Python package \texttt{orvara} \citep{Orvara}, which can fit Keplerian orbits by utilising a combination of radial velocity, relative astrometry, and available absolute astrometry data. Here we combine the radial velocity observations presented in this paper and the absolute astrometry data that comes from the Hipparcos-Gaia Catalog of Accelerations (HGCA, \citeauthor{HGCAeDR3} \citeyear{HGCAeDR3}). In addition, we use an extension of \texttt{orvara} that allows priors on orbital periods and semi-major axes,\footnote{\url{https://github.com/nicochunger/orvara/tree/period-prior}} and fix the offset between the radial velocity data to the offsets presented in section \ref{sec:RV_obsol}, as \texttt{orvara} otherwise finds values outside of the expected offset range. See Appendix \ref{app:orvara-res} for the resulting corner plots, and Appendix \ref{app:orvara-res2} for the proper motion plots.

For most of the exoplanets, the amplitude and period are too low to properly fit the Hipparcos-Gaia proper motions. Apart from HIP54597 b, HD74698 c, and BD-210397 c, the mass and inclination of the exoplanets cannot be constrained. For HIP54597 b, the constraint is weak and varies between an inclination of 120$^{\circ}$ and 60$^{\circ}$. This might increase the mass of the planet by 10\%-30\% but not significantly more. The constraint of BD-210397 c is slightly better, with peaks at 50$^{\circ}$ and 130$^{\circ}$, favouring the former, and a mass of 2.7 $M_J$. The inclination of HD74698 c is $90^{\circ} \pm 33^{\circ}$; though this is not well-constrained, it shows the minimum mass is likely equal to its true mass. For HIP54597 b, HD74698 c, and BD-210397 c, the inclination is not well-constrained. Their \texttt{orvara} results are therefore not included in Table \ref{table:planet_parameters}. However, we do conclude that their companions have masses in the planetary range. The convergence is better for our more massive companions, and with the found models, three of the four brown dwarf candidates are found to be stellar binaries (HD56380B, HD221638B, and HD33473C). HD61383 has an edge-on orbit ($i=86.2^{\circ}$), and its true mass is very close to the minimum mass. For HD11938, \texttt{orvara} finds an inclination of 64.0$^{\circ}$, corresponding to a mass of 0.386 $M_{\odot}$, which is slightly higher than the mass found by the secondary peak in the CCF. \texttt{Orvara} does not find the same RV orbital parameters as the simple RV analysis, which causes discrepancies between all presented minimum masses and true masses. For the vast majority of our stars, the difference is negligible. For HD11938, on the other hand, \texttt{orvara} finds a different systemic velocity. This stems from the fact that the orbit is incomplete and in the RV model the systemic velocity is fixed to the value found by the second component in the CCF. When the systemic velocity is let free, the period can increase up to 41 years and the minimum mass increases to 0.30 $M_{\odot}$, which corresponds better to the model found by \texttt{orvara}.

The true masses found are in agreement with the Gaia  renormalised unit weight error (RUWE) values, which are expected to be 1 for well-behaved single-star solutions. Values significantly greater than 1, with a limit generally placed at 1.4, either indicate the star is not a single star or is otherwise problematic. All presented stars have RUWE values below 1.4, except HD56380 (1.8348), which corresponds to one of our most massive companions (0.36 $M_{\odot}$). The other massive companion HD11938B does not present a large RUWE, probably due to its very long orbital period (35 years).

Considering that HD56380B is our second most massive companion, we decided to reinspect the CCF following the method described in section \ref{sec:binaries} and identified a second component compatible with a massive companion. However, considering that both components are always blended (within 2 km/s), the method does not allow us to derive a precise mass ratio.

\section{Discussion and conclusions}
\label{sec:Conclusions}
\begin{figure*}
    \centering
    \includegraphics[width=\linewidth]{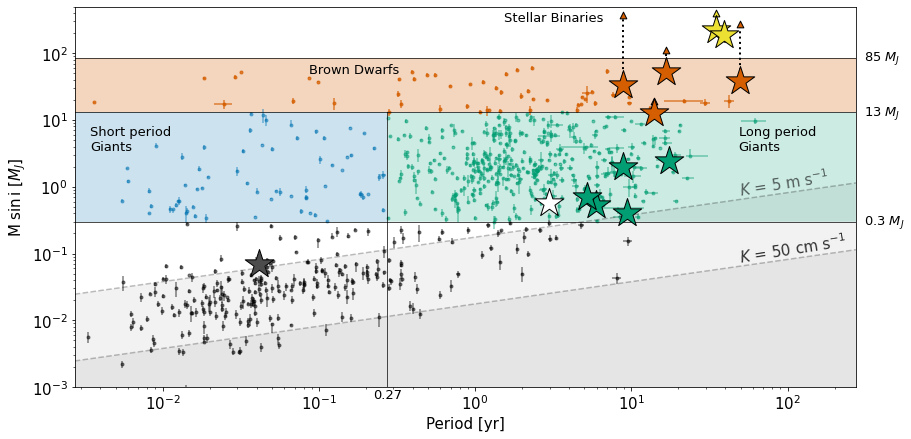}
    \caption{Overview of the discovered companions, including the uncertain planetary signal HD3964 b (white star). Star markers indicate companions discussed in this paper, and colours differentiate between short-period (black) and long-period giants (green), brown dwarfs (orange), and stellar binaries (yellow). Arrows indicate the true masses as found by \texttt{orvara}. Round markers are previously detected (using the RV method) exoplanets and brown dwarfs with well-constrained ($> 3\sigma$) masses and periods from the exoplanet.eu archive. The grey shaded regions indicate the RV precision detection limits 50 cm s$^{-1}$ and 5 m s $^{-1}$ for $1 M_{\odot}$. The giant planets ($0.3 M_J < M < 13 M_J$) are highlighted where the difference between the short-period (blue-shaded) and long-period (green-shaded) objects is defined at 100 days. The brown dwarf region is shaded orange.}
    \label{fig:Overview}
\end{figure*}

In this paper, we discuss the RV variations of 12 stars. We present the discovery of six exoplanets, one uncertain exoplanet candidate, one brown dwarf, four stellar binaries and the improved orbital solution of the stellar binary HD33473C discovered  earlier by \cite{Moutou2011}. An overview of the detections is presented in Figure \ref{fig:Overview}. 

The biggest difficulty for long-period companions arises from the RV signal entanglement with the star's magnetic cycle. Therefore, for all proposed companions, we discuss the possibility of the RV variation originating from activity. For comparison, we also present an activity-induced signal (HD11608) and a possible exoplanet signal with strong entanglement with its magnetic cycle (HD3964). These two stars show the strength of investigating the correlation with stellar activity indicators. Detrending reduces the activity-induced component but makes it difficult to constrain the orbital period. As HD3964 has an RV period (1086 days) similar to the H$\alpha$-index period (1150), there is no way to accurately determine the RV signature of the possible companion.

HD56380 and HD61383 are relatively old ($>12$ Gyr). In principle, this is not inconsistent with their average lifetime, but it is uncommon to get so close to the upper age limit for these spectral types. A possible explanation could be that the light of the secondary was blended in the observations, causing the age of the primary star to be overestimated. The ages originate from \cite{DelgadoMena2019}, who use isochrones based on $T_\mathrm{eff}$, $[$Fe/H$],$ and $V$ magnitudes. The latter is sensitive to blending. As HD56380 and HD61383 are binaries, and given that we find that the secondary peak in the RV CCF is blended with the primary peak  for both sources, the age could indeed be influenced by the secondary.

After determining the orbital solutions, overlap with transit data was examined for all companions as expected by the combination of the period and time of periastron. There are no CHEOPS light curves for the presented stars, and the TESS light curves have no superposition with the potential transits. The only companion to have superposition with the TESS sectors is HD74698 b, which does not show a transit or periodicity of 15 days: the orbit is not well-aligned with our line of sight. As the potential transits have uncertainties on the order of months, and for some of our targets even on the order of years, we also used a more general approach and looked at the Lomb Scargle periodograms of the TESS light curves. Again, no periodicities were found, meaning there are no observed transits for our targets, but also that there are no activity-induced signals with the same periods as our companions. 

Long-period companions provide interesting systems for follow-up observations. HD3964 has a remaining RV signal of 4 m s$^{-1}$, which is likely caused by activity given that the CCF-Bissector is correlated to the residuals, but this could be better constrained by follow-up observations. HD94771 does not show any strong signs of multiple companions and has a relatively high eccentricity, making the stability of a potential multi-planet system less likely. For HIP54597, there is no significant evidence of additional planets, but there is for both BD-210397 and HD74698. As BD-210397 includes a high jitter and HD74698 shows a 1000 day period in the periodogram, both targets are very interesting targets for follow-up observations. Apart from BD-210397, none of the exoplanet's Keplerian models show a high jitter. For BD-210397, the activity is unknown, but according to the $S$-index, stellar mass, and surface gravity, the jitter falls within the expected range \citep{Luhn2020}. If the jitter is caused by a companion, it is a very short period signal ($\sim $0.5 days); this can be verified by observations at short time intervals.

It is common for brown dwarfs to have higher eccentricities than exoplanets. According to the NASA exoplanet archive, 94\% of the confirmed exoplanets\footnote{Defining `confirmed' as companions with well-constrained masses and periods $(> 3 \sigma)$.} have eccentricities $e \leq 0.5$, versus 69\% of the brown dwarfs. For $e \leq 0.3,$ this value decreases to 84\% for the exoplanets and to 56\% for the brown dwarfs. Brown dwarfs are less likely to have eccentricities below 0.3 than exoplanets. Our results are in agreement. Only one of the proposed exoplanets (HD94771
b) has an eccentricity of larger than 0.3, while HD62364 has a high eccentricity (0.607). When comparing long-period giant planets to short-period giant planets, placing the limit at 100 days (see also Figure \ref{fig:Overview}), a similar trend occurs: 90\% of the short-period giant planets have eccentricities $e \leq 0.3$, versus 72\% of the long-period giant planets.

According to the planet-metallicity correlation, host stars with higher metallicities are more likely to host a giant planet \citep{Fischer2005}, which is particularly true for hot Jupiters \citep{Osborn2020}. The short-period giant planets ($< 100$ days) have parent stars with an average metallicity of 0.11 dex. This value is 0.04
dex on average for stars hosting  long-period giant planets, a similar value to our small sample of presented stars hosting (long-period) giant planets (0.05 dex). This is in agreement with the findings of \cite{Adibekyan2013}: giant planets orbiting metal-poor stars have longer periods than those in metal-rich systems. 
The findings of \citep{Santos2017} highlight the possibility that stars with massive giant planets ($M \gtrsim 4 M_J$) do not follow the same metallicity trend as stars with lower mass giant planets. The stars hosting massive giant planets are on average more metal-poor. Though all planets discussed in the present paper have minimal masses of below 4 $M_J$, this may indicate that HIP54597 b ([Fe/H] $= -0.22$, $m_2\sin i = 2.01 M_J$) falls in the massive giant category; however, the results from \texttt{orvara} show this is unlikely. 

The discovery of very long-period companions requires years of observations. With the present research, we probe a region of a largely unknown population. The current baseline of 19 years allows us to detect Saturn-mass companions with periods of 10 years and brown dwarfs and stellar binaries with periods of up to 50 years. Though the RV observations do not cover the full phase of the found stellar binaries, we are able to constrain the orbital parameters and confirm the origin of HD11938B using the secondary peak in the combined CCFs, and of HD62364 b, HD5680B, HD221638B, HD33473C, HD11938B, and HD61383B via absolute astrometry. As by-products of this survey, the brown dwarfs and binaries demonstrate the effectiveness of a long baseline. The impact of continuing with observations is made clear by the difference between the orbital solution of  HD33473C found by \citet{Moutou2011} and the solution for this object presented
 here. Incomplete orbits are prone to errors. Following indications by \cite{Rosenthal2022} and \cite{ZhuWu2018}, the long-period Jupiters detected in this work are prime targets for the detection of low-mass inner planets.

\begin{acknowledgements}
      The authors thank the ESO staff at La Silla for their diligent and competent help during the observations and for the effort to maintain the instrument operating and stable for so many years. We thank Emanuela Pompei for providing helpful comments. The HARPS spectrograph was built by the contributions of the Swiss FNRS, the Geneva University, the French Institut National des Sciences de l'Univers (INSU) and ESO. This research made use of the Simbad database, operated at the CDS, Strasbourg, France.
      This work has been carried out within the framework of the NCCR PlanetS supported by the Swiss National Science Foundation. NCS acknowledges the support from the European Research Council through grant agreement 101052347 (FIERCE). This work was supported by FCT - Funda\c{c}\~ao para a Ci\^encia e a Tecnologia through national funds and by FEDER through COMPETE2020 - Programa Operacional Competitividade e Internacionaliza\c{c}\~ao by these grants: UIDB/04434/2020; UIDP/04434/2020. This publication makes use of The Data \& Analysis Center for Exoplanets (DACE), which is a facility based at the University of Geneva (CH) dedicated to extrasolar planets data visualisation, exchange and analysis. DACE is a platform of the Swiss National Centre of Competence in Research (NCCR) PlanetS, federating the Swiss expertise in Exoplanet research. The DACE platform is available at https://dace.unige.ch.
\end{acknowledgements}

\bibliographystyle{aa} 
\bibliography{46203} 

\begin{thebibliography}{62}
\expandafter\ifx\csname natexlab\endcsname\relax\def\natexlab#1{#1}\fi

\bibitem[{Adibekyan {et~al.}(2013)Adibekyan, Figueira, Santos, Mortier,
  Mordasini, Delgado~Mena, Sousa, Correia, Israelian, \&
  Oshagh}]{Adibekyan2013}
Adibekyan, V.~Z., Figueira, P., Santos, N.~C., {et~al.} 2013, Astronomy {\&}
  Astrophysics, 560, A51

\bibitem[{Baluev(2008)}]{Baluev2008}
Baluev, R.~V. 2008, Monthly Notices of the Royal Astronomical Society, 385,
  1279

\bibitem[{Boisse {et~al.}(2012)Boisse, Pepe, Perrier, Queloz, Bonfils, Bouchy,
  Santos, Arnold, Beuzit, D{\'{i}}az, Delfosse, Eggenberger, Ehrenreich,
  Forveille, H{\'{e}}brard, Lagrange, Lovis, Mayor, Moutou, Naef, Santerne,
  S{\'{e}}gransan, Sivan, \& Udry}]{JA_Boisse2012}
Boisse, I., Pepe, F., Perrier, C., {et~al.} 2012, Astronomy {\&} Astrophysics,
  545, A55

\bibitem[{Bouchy {et~al.}(2009)Bouchy, H{\'{e}}brard, Udry, Delfosse, Boisse,
  Desort, Bonfils, Eggenberger, Ehrenreich, Forveille, Lagrange, Le~Coroller,
  Lovis, Moutou, Pepe, Perrier, Pont, Queloz, Santos, S{\'{e}}gransan, \&
  Vidal-Madjar}]{Bouchy2009}
Bouchy, F., H{\'{e}}brard, G., Udry, S., {et~al.} 2009, Astronomy {\&}
  Astrophysics, 505, 853

\bibitem[{Bouchy {et~al.}(2016)Bouchy, S{\'{e}}gransan, D{\'{i}}az, Forveille,
  Boisse, Arnold, Astudillo-Defru, Beuzit, Bonfils, Borgniet, Bourrier,
  Courcol, Delfosse, Demangeon, Delorme, Ehrenreich, H{\'{e}}brard, Lagrange,
  Mayor, Montagnier, Moutou, Naef, Pepe, Perrier, Queloz, Rey, Sahlmann,
  Santerne, Santos, Sivan, Udry, \& Wilson}]{Bouchy2016}
Bouchy, F., S{\'{e}}gransan, D., D{\'{i}}az, R.~F., {et~al.} 2016, Astronomy
  {\&} Astrophysics, 585, A46

\bibitem[{Brandt(2021)}]{HGCAeDR3}
Brandt, T.~D. 2021, The Astrophysical Journal Supplement Series, 254, 42

\bibitem[{Brandt {et~al.}(2021)Brandt, Dupuy, Li, Brandt, Zeng, Michalik,
  Bardalez~Gagliuffi, \& Raposo-Pulido}]{Orvara}
Brandt, T.~D., Dupuy, T.~J., Li, Y., {et~al.} 2021, The Astronomical Journal,
  162, 186

\bibitem[{Bressan {et~al.}(2012)Bressan, Marigo, Girardi, Salasnich, Dal~Cero,
  Rubele, \& Nanni}]{Bressan2012}
Bressan, A., Marigo, P., Girardi, L., {et~al.} 2012, Monthly Notices of the
  Royal Astronomical Society, 427, 127

\bibitem[{Bryan {et~al.}(2019)Bryan, Knutson, Lee, Fulton, Batygin, Ngo, \&
  Meshkat}]{JA_Bryan2019}
Bryan, M.~L., Knutson, H.~A., Lee, E.~J., {et~al.} 2019, The Astronomical
  Journal, 157, 52

\bibitem[{Chabrier \& Baraffe(2000)}]{Chabrier2000}
Chabrier, G. \& Baraffe, I. 2000, Annual Review of Astronomy and Astrophysics,
  38, 337

\bibitem[{Chanam{\'{e}} \& Ram{\'{i}}rez(2012)}]{Chaname2012}
Chanam{\'{e}}, J. \& Ram{\'{i}}rez, I. 2012, The Astrophysical Journal, 746,
  102

\bibitem[{Dalal {et~al.}(2021)Dalal, Kiefer, H{\'{e}}brard, Sahlmann, Sousa,
  Forveille, Delfosse, Arnold, Astudillo-Defru, Bonfils, Boisse, Bouchy,
  Bourrier, Brugger, Cort{\'{e}}s-Zuleta, Deleuil, Demangeon, D{\'{i}}az, Hara,
  Heidari, Hobson, Lopez, Lovis, Martioli, Mignon, Mousis, Moutou, Rey,
  Santerne, Santos, S{\'{e}}gransan, Str{\o}m, \& Udry}]{LP_Dalal2021}
Dalal, S., Kiefer, F., H{\'{e}}brard, G., {et~al.} 2021, Astronomy {\&}
  Astrophysics, 651, A11

\bibitem[{Delgado~Mena {et~al.}(2019)Delgado~Mena, Moya, Adibekyan, Tsantaki,
  Gonz{\'{a}}lez~Hern{\'{a}}ndez, Israelian, Davies, Chaplin, Sousa, Ferreira,
  \& Santos}]{DelgadoMena2019}
Delgado~Mena, E., Moya, A., Adibekyan, V., {et~al.} 2019, Astronomy {\&}
  Astrophysics, 624, A78

\bibitem[{Delgado~Mena {et~al.}(2017)Delgado~Mena, Tsantaki, Adibekyan, Sousa,
  Santos, Gonz{\'{a}}lez~Hern{\'{a}}ndez, \& Israelian}]{DelgadoMena2017}
Delgado~Mena, E., Tsantaki, M., Adibekyan, V.~Z., {et~al.} 2017, Astronomy {\&}
  Astrophysics, 606, A94

\bibitem[{Delisle {et~al.}(2016)Delisle, S{\'{e}}gransan, Buchschacher, \&
  Alesina}]{Delisle2016}
Delisle, J.-B., S{\'{e}}gransan, D., Buchschacher, N., \& Alesina, F. 2016,
  Astronomy {\&} Astrophysics, 590, A134

\bibitem[{Delisle {et~al.}(2018)Delisle, S{\'{e}}gransan, Dumusque, Diaz,
  Bouchy, Lovis, Pepe, Udry, Alonso, Benz, Coffinet, Collier~Cameron, Deleuil,
  Figueira, Gillon, Lo~Curto, Mayor, Mordasini, Motalebi, Moutou, Pollacco,
  Pompei, Queloz, Santos, \& Wyttenbach}]{Delisle2018}
Delisle, J.-B., S{\'{e}}gransan, D., Dumusque, X., {et~al.} 2018, Astronomy
  {\&} Astrophysics, 614, A133

\bibitem[{Demangeon {et~al.}(2021)Demangeon, Dalal, H{\'{e}}brard, Nsamba,
  Kiefer, Camacho, Sahlmann, Arnold, Astudillo-Defru, Bonfils, Boisse, Bouchy,
  Bourrier, Campante, Delfosse, Deleuil, D{\'{i}}az, Faria, Forveille, Hara,
  Heidari, Hobson, Lopez, Moutou, Rey, Santerne, Sousa, Santos, Str{\o}m,
  Tsantaki, \& Udry}]{LP_Demangeon2021}
Demangeon, O. D.~S., Dalal, S., H{\'{e}}brard, G., {et~al.} 2021, Astronomy
  {\&} Astrophysics, 653, A78

\bibitem[{D{\'{i}}az {et~al.}(2014)D{\'{i}}az, Almenara, Santerne, Moutou,
  Lethuillier, \& Deleuil}]{Diaz2014}
D{\'{i}}az, R.~F., Almenara, J.~M., Santerne, A., {et~al.} 2014, Monthly
  Notices of the Royal Astronomical Society, 441, 983

\bibitem[{D{\'{i}}az {et~al.}(2016)D{\'{i}}az, S{\'{e}}gransan, Udry, Lovis,
  Pepe, Dumusque, Marmier, Alonso, Benz, Bouchy, Coffinet, Collier~Cameron,
  Deleuil, Figueira, Gillon, Lo~Curto, Mayor, Mordasini, Motalebi, Moutou,
  Pollacco, Pompei, Queloz, Santos, \& Wyttenbach}]{Diaz2016}
D{\'{i}}az, R.~F., S{\'{e}}gransan, D., Udry, S., {et~al.} 2016, Astronomy {\&}
  Astrophysics, 585, A134

\bibitem[{Dumusque {et~al.}(2011)Dumusque, Lovis, S{\'{e}}gransan, Mayor, Udry,
  Benz, Bouchy, Lo~Curto, Mordasini, Pepe, Queloz, Santos, \&
  Naef}]{Dumusque2011c}
Dumusque, X., Lovis, C., S{\'{e}}gransan, D., {et~al.} 2011, Astronomy {\&}
  Astrophysics, 535, A55

\bibitem[{{ESA}(1997)}]{Hipparcos}
{ESA}. 1997, in ESA Special Publication, Vol. 1200, ESA

\bibitem[{{ESO}(2019)}]{HARPS_DRS24}
{ESO}. 2019, {HARPS Data Reduction Software User Manual (Issue 2.4)}, Tech.
  rep., European Southern Observatory, La Silla

\bibitem[{Faria {et~al.}(2018)Faria, Santos, Figueira, \& Brewer}]{KIMA}
Faria, J.~P., Santos, N.~C., Figueira, P., \& Brewer, B.~J. 2018, Journal of
  Open Source Software, 3, 487

\bibitem[{Feng {et~al.}(2019)Feng, Anglada-Escud{\'{e}}, Tuomi, Jones,
  Chanam{\'{e}}, Butler, \& Janson}]{JA_Feng2019}
Feng, F., Anglada-Escud{\'{e}}, G., Tuomi, M., {et~al.} 2019, Monthly Notices
  of the Royal Astronomical Society, 490, 5002

\bibitem[{Fischer \& Valenti(2005)}]{Fischer2005}
Fischer, D.~A. \& Valenti, J. 2005, The Astrophysical Journal, 622, 1102

\bibitem[{Flower(1996)}]{Flower1996}
Flower, P.~J. 1996, The Astrophysical Journal, 469, 355

\bibitem[{{Gaia Collaboration} {et~al.}(2021){Gaia Collaboration}, Brown,
  Vallenari, Prusti, de~Bruijne, Babusiaux, Biermann, Creevey, Evans, Eyer,
  Hutton, Jansen, Jordi, Klioner, Lammers, Lindegren, Luri, Mignard, Panem,
  Pourbaix, Randich, Sartoretti, Soubiran, Walton, Arenou, Bailer-Jones,
  Bastian, Cropper, Drimmel, Katz, Lattanzi, van Leeuwen, Bakker, Cacciari,
  Casta{\~{n}}eda, De~Angeli, Ducourant, Fabricius, Fouesneau, Fr{\'{e}}mat,
  Guerra, Guerrier, Guiraud, Jean-Antoine~Piccolo, Masana, Messineo, Mowlavi,
  Nicolas, Nienartowicz, Pailler, Panuzzo, Riclet, Roux, Seabroke, Sordo,
  Tanga, Th{\'{e}}venin, Gracia-Abril, Portell, Teyssier, Altmann, Andrae,
  Bellas-Velidis, Benson, Berthier, Blomme, Brugaletta, Burgess, Busso, Carry,
  Cellino, Cheek, Clementini, Damerdji, Davidson, Delchambre, Dell’Oro,
  Fern{\'{a}}ndez-Hern{\'{a}}ndez, Galluccio, Garc{\'{i}}a-Lario,
  Garcia-Reinaldos, Gonz{\'{a}}lez-N{\'{u}}{\~{n}}ez, Gosset, Haigron,
  Halbwachs, Hambly, Harrison, Hatzidimitriou, Heiter, Hern{\'{a}}ndez,
  Hestroffer, Hodgkin, Holl, Jan{\ss}en, Jevardat~de Fombelle, Jordan,
  Krone-Martins, Lanzafame, L{\"{o}}ffler, Lorca, Manteiga, Marchal, Marrese,
  Moitinho, Mora, Muinonen, Osborne, Pancino, Pauwels, Petit, Recio-Blanco,
  Richards, Riello, Rimoldini, Robin, Roegiers, Rybizki, Sarro, Siopis, Smith,
  Sozzetti, Ulla, Utrilla, van Leeuwen, van Reeven, Abbas, Abreu~Aramburu,
  Accart, Aerts, Aguado, Ajaj, Altavilla, {\'{A}}lvarez, {\'{A}}lvarez
  Cid-Fuentes, Alves, Anderson, Anglada~Varela, Antoja, Audard, Baines, Baker,
  Balaguer-N{\'{u}}{\~{n}}ez, Balbinot, Balog, Barache, Barbato, Barros,
  Barstow, Bartolom{\'{e}}, Bassilana, Bauchet, Baudesson-Stella, Becciani,
  Bellazzini, Bernet, Bertone, Bianchi, Blanco-Cuaresma, Boch, Bombrun,
  Bossini, Bouquillon, Bragaglia, Bramante, Breedt, Bressan, Brouillet,
  Bucciarelli, Burlacu, Busonero, Butkevich, Buzzi, Caffau, Cancelliere,
  C{\'{a}}novas, Cantat-Gaudin, Carballo, Carlucci, Carnerero, Carrasco,
  Casamiquela, Castellani, Castro-Ginard, Castro~Sampol, Chaoul, Charlot,
  Chemin, Chiavassa, Cioni, Comoretto, Cooper, Cornez, Cowell, Crifo, Crosta,
  Crowley, Dafonte, Dapergolas, David, David, de~Laverny, De~Luise, De~March,
  De~Ridder, de~Souza, de~Teodoro, de~Torres, del Peloso, del Pozo, Delbo,
  Delgado, Delgado, Delisle, Di~Matteo, Diakite, Diener, Distefano, Dolding,
  Eappachen, Edvardsson, Enke, Esquej, Fabre, Fabrizio, Faigler, Fedorets,
  Fernique, Fienga, Figueras, Fouron, Fragkoudi, Fraile, Franke, Gai, Garabato,
  Garcia-Gutierrez, Garc{\'{i}}a-Torres, Garofalo, Gavras, Gerlach, Geyer,
  Giacobbe, Gilmore, Girona, Giuffrida, Gomel, Gomez, Gonzalez-Santamaria,
  Gonz{\'{a}}lez-Vidal, Granvik, Guti{\'{e}}rrez-S{\'{a}}nchez, Guy, Hauser,
  Haywood, Helmi, Hidalgo, Hilger, H{\l}adczuk, Hobbs, Holland, Huckle,
  Jasniewicz, Jonker, Juaristi~Campillo, Julbe, Karbevska, Kervella, Khanna,
  Kochoska, Kontizas, Kordopatis, Korn, Kostrzewa-Rutkowska, Kruszy{\'{n}}ska,
  Lambert, Lanza, Lasne, Le~Campion, Le~Fustec, Lebreton, Lebzelter, Leccia,
  Leclerc, Lecoeur-Taibi, Liao, Licata, Lindstr{\o}m, Lister, Livanou, Lobel,
  Madrero~Pardo, Managau, Mann, Marchant, Marconi, Marcos~Santos, Marinoni,
  Marocco, Marshall, Martin~Polo, Mart{\'{i}}n-Fleitas, Masip, Massari,
  Mastrobuono-Battisti, Mazeh, McMillan, Messina, Michalik, Millar, Mints,
  Molina, Molinaro, Moln{\'{a}}r, Montegriffo, Mor, Morbidelli, Morel, Morris,
  Mulone, Munoz, Muraveva, Murphy, Musella, Noval, Ord{\'{e}}novic, Orr{\`{u}},
  Osinde, Pagani, Pagano, Palaversa, Palicio, Panahi, Pawlak,
  Pe{\~{n}}alosa~Esteller, Penttil{\"{a}}, Piersimoni, Pineau, Plachy, Plum,
  Poggio, Poretti, Poujoulet, Pr{\v{s}}a, Pulone, Racero, Ragaini, Rainer,
  Raiteri, Rambaux, Ramos, Ramos-Lerate, Re~Fiorentin, Regibo, Reyl{\'{e}},
  Ripepi, Riva, Rixon, Robichon, Robin, Roelens, Rohrbasser,
  Romero-G{\'{o}}mez, Rowell, Royer, Rybicki, Sadowski,
  Sagrist{\`{a}}~Sell{\'{e}}s, Sahlmann, Salgado, Salguero, Samaras,
  Sanchez~Gimenez, Sanna, Santove{\~{n}}a, Sarasso, Schultheis, Sciacca, Segol,
  Segovia, S{\'{e}}gransan, Semeux, Shahaf, Siddiqui, Siebert, Siltala, Slezak,
  Smart, Solano, Solitro, Souami, Souchay, Spagna, Spoto, Steele,
  Steidelm{\"{u}}ller, Stephenson, S{\"{u}}veges, Szabados, Szegedi-Elek,
  Taris, Tauran, Taylor, Teixeira, Thuillot, Tonello, Torra, Torra, Turon,
  Unger, Vaillant, van Dillen, Vanel, Vecchiato, Viala, Vicente, Voutsinas,
  Weiler, Wevers, Wyrzykowski, Yoldas, Yvard, Zhao, Zorec, Zucker, Zurbach, \&
  Zwitter}]{GaiaDR3e}
{Gaia Collaboration}, Brown, A. G.~A., Vallenari, A., {et~al.} 2021, Astronomy
  {\&} Astrophysics, 649, A1

\bibitem[{{Gaia Collaboration} {et~al.}(2022){Gaia Collaboration}, Vallenari,
  Brown, Prusti, de~Bruijne, Arenou, Babusiaux, Biermann, Creevey, Ducourant,
  Evans, Eyer, Guerra, Hutton, Jordi, Klioner, Lammers, Lindegren, Luri, \&
  Mignard}]{GaiaDR3}
{Gaia Collaboration}, Vallenari, A., Brown, A., {et~al.} 2022, Astronomy {\&}
  Astrophysics

\bibitem[{Gomes~da Silva {et~al.}(2021)Gomes~da Silva, Santos, Adibekyan,
  Sousa, Campante, Figueira, Bossini, Delgado-Mena, Monteiro, de~Laverny,
  Recio-Blanco, \& Lovis}]{GomesdaSilva2021}
Gomes~da Silva, J., Santos, N.~C., Adibekyan, V., {et~al.} 2021, Astronomy {\&}
  Astrophysics, 646, A77

\bibitem[{Hara {et~al.}(2017)Hara, Bou{\'{e}}, Laskar, \&
  Correia}]{l1periodogram}
Hara, N.~C., Bou{\'{e}}, G., Laskar, J., \& Correia, A. C.~M. 2017, Monthly
  Notices of the Royal Astronomical Society, 464, 1220

\bibitem[{Kiefer {et~al.}(2019)Kiefer, H{\'{e}}brard, Sahlmann, Sousa,
  Forveille, Santos, Mayor, Deleuil, Wilson, Dalal, D{\'{i}}az, Henry,
  Hagelberg, Hobson, Demangeon, Bourrier, Delfosse, Arnold, Astudillo-Defru,
  Beuzit, Boisse, Bonfils, Borgniet, Bouchy, Courcol, Ehrenreich, Hara,
  Lagrange, Lovis, Montagnier, Moutou, Pepe, Perrier, Rey, Santerne,
  S{\'{e}}gransan, Udry, \& Vidal-Madjar}]{LP_Kiefer2019}
Kiefer, F., H{\'{e}}brard, G., Sahlmann, J., {et~al.} 2019, Astronomy {\&}
  Astrophysics, 631, A125

\bibitem[{Lindegren {et~al.}(2012)Lindegren, Lammers, Hobbs, O’Mullane,
  Bastian, \& Hern{\'{a}}ndez}]{Lindegren2012}
Lindegren, L., Lammers, U., Hobbs, D., {et~al.} 2012, Astronomy {\&}
  Astrophysics, 538, A78

\bibitem[{Lo~Curto {et~al.}(2010)Lo~Curto, Mayor, Benz, Bouchy, Lovis, Moutou,
  Naef, Pepe, Queloz, Santos, Segransan, \& Udry}]{LoCurto2010}
Lo~Curto, G., Mayor, M., Benz, W., {et~al.} 2010, Astronomy and Astrophysics,
  512, A48

\bibitem[{Lo~Curto {et~al.}(2015)Lo~Curto, Pepe, Avila, Boffin, Bovay,
  Chazelas, Coffinet, Fleury, Hughes, Lovis, Maire, Manescau, Pasquini, Rihs,
  Sinclaire, \& Udry}]{LoCurto2015}
Lo~Curto, G., Pepe, F., Avila, G., {et~al.} 2015, The Messenger, 162, 9

\bibitem[{Lovis {et~al.}(2006)Lovis, Mayor, Pepe, Alibert, Benz, Bouchy,
  Correia, Laskar, Mordasini, Queloz, Santos, Udry, Bertaux, \&
  Sivan}]{Lovis2006n}
Lovis, C., Mayor, M., Pepe, F., {et~al.} 2006, Nature, 441, 305

\bibitem[{Luhn {et~al.}(2020)Luhn, Wright, Howard, \& Isaacson}]{Luhn2020}
Luhn, J.~K., Wright, J.~T., Howard, A.~W., \& Isaacson, H. 2020, The
  Astronomical Journal, 159, 235

\bibitem[{Mamajek \& Hillenbrand(2008)}]{Mamajek2008}
Mamajek, E.~E. \& Hillenbrand, L.~A. 2008, The Astrophysical Journal, 687, 1264

\bibitem[{Mayor {et~al.}(2003)Mayor, Pepe, Queloz, Bouchy, Rupprecht, Lo~Curto,
  Avila, Benz, Bertaux, Bonfils, Dall, Dekker, Delabre, Eckert, Fleury,
  Gilliotte, Gojak, Guzman, Kohler, Lizon, Longinotti, Lovis, Megevand,
  Pasquini, Reyes, Sivan, Sosnowska, Soto, Udry, van Kesteren, Weber, \&
  Weilenmann}]{HARPS}
Mayor, M., Pepe, F., Queloz, D., {et~al.} 2003, The Messenger, 114, 20

\bibitem[{Mayor \& Queloz(1995)}]{Mayor1995}
Mayor, M. \& Queloz, D. 1995, Nature, 378, 355

\bibitem[{Morbidelli(2018)}]{Morbidelli2018}
Morbidelli, A. 2018, in Handbook of Exoplanets (Cham: Springer International
  Publishing), 2523--2541

\bibitem[{Moutou {et~al.}(2014)Moutou, H{\'{e}}brard, Bouchy, Arnold, Santos,
  Astudillo-Defru, Boisse, Bonfils, Borgniet, Delfosse, D{\'{i}}az, Ehrenreich,
  Forveille, Gregorio, Labrevoir, Lagrange, Montagnier, Montalto, Pepe,
  Sahlmann, Santerne, S{\'{e}}gransan, Udry, \& Vanhuysse}]{JA_Moutou2014}
Moutou, C., H{\'{e}}brard, G., Bouchy, F., {et~al.} 2014, Astronomy {\&}
  Astrophysics, 563, A22

\bibitem[{Moutou {et~al.}(2011)Moutou, Mayor, Lo~Curto, S{\'{e}}gransan, Udry,
  Bouchy, Benz, Lovis, Naef, Pepe, Queloz, Santos, \& Sousa}]{Moutou2011}
Moutou, C., Mayor, M., Lo~Curto, G., {et~al.} 2011, Astronomy {\&}
  Astrophysics, 527, A63

\bibitem[{Naef {et~al.}(2005)Naef, Mayor, Beuzit, Perrier, Queloz, Sivan, \&
  Udry}]{Naef2005}
Naef, D., Mayor, M., Beuzit, J.~L., {et~al.} 2005, in ESA Special Publication,
  Vol. 560, 13th Cambridge Workshop on Cool Stars, Stellar Systems and the Sun,
  ed. F.~Favata, G.~A.~J. Hussain, \& B.~Battrick, 833

\bibitem[{Noyes {et~al.}(1984)Noyes, Hartmann, Baliunas, Duncan, \&
  Vaughan}]{Noyes1984}
Noyes, R.~W., Hartmann, L.~W., Baliunas, S.~L., Duncan, D.~K., \& Vaughan,
  A.~H. 1984, The Astrophysical Journal, 279, 763

\bibitem[{Osborn \& Bayliss(2020)}]{Osborn2020}
Osborn, A. \& Bayliss, D. 2020, Monthly Notices of the Royal Astronomical
  Society, 491, 4481

\bibitem[{Pollack {et~al.}(1996)Pollack, Hubickyj, Bodenheimer, Lissauer,
  Podolak, \& Greenzweig}]{Pollack1996}
Pollack, J.~B., Hubickyj, O., Bodenheimer, P., {et~al.} 1996, Icarus, 124, 62

\bibitem[{Queloz {et~al.}(2001)Queloz, Mayor, Udry, Burnet, Carrier,
  Eggenberger, Naef, Santos, Pepe, Rupprecht, Avila, Baeza, Benz, Bertaux,
  Bouchy, Cavadore, Delabre, Eckert, Fischer, Fleury, Gilliotte, Goyak, Guzman,
  Kohler, Lacroix, Lizon, Megevand, Sivan, Sosnowska, \& Weilenmann}]{CORALIE}
Queloz, D., Mayor, M., Udry, S., {et~al.} 2001, The Messenger, 105, 1

\bibitem[{Rey {et~al.}(2017)Rey, H{\'{e}}brard, Bouchy, Bourrier, Boisse,
  Santos, Arnold, Astudillo-Defru, Bonfils, Borgniet, Courcol, Deleuil,
  Delfosse, Demangeon, D{\'{i}}az, Ehrenreich, Forveille, Marmier, Moutou,
  Pepe, Santerne, Sahlmann, S{\'{e}}gransan, Udry, \& Wilson}]{JA_Rey2017}
Rey, J., H{\'{e}}brard, G., Bouchy, F., {et~al.} 2017, Astronomy {\&}
  Astrophysics, 601, A9

\bibitem[{Rosenthal {et~al.}(2022)Rosenthal, Knutson, Chachan, Dai, Howard,
  Fulton, Chontos, Crepp, Dalba, Henry, Kane, Petigura, Weiss, \&
  Wright}]{Rosenthal2022}
Rosenthal, L.~J., Knutson, H.~A., Chachan, Y., {et~al.} 2022, The Astrophysical
  Journal Supplement Series, 262, 1

\bibitem[{Sahlmann {et~al.}(2011{\natexlab{a}})Sahlmann, Lovis, Queloz, \&
  S{\'{e}}gransan}]{Sahlmann2011}
Sahlmann, J., Lovis, C., Queloz, D., \& S{\'{e}}gransan, D. 2011{\natexlab{a}},
  Astronomy {\&} Astrophysics, 528, L8

\bibitem[{Sahlmann {et~al.}(2010)Sahlmann, S{\'{e}}gransan, Queloz, \&
  Udry}]{Sahlmann2010}
Sahlmann, J., S{\'{e}}gransan, D., Queloz, D., \& Udry, S. 2010, Proceedings of
  the International Astronomical Union, 6, 117

\bibitem[{Sahlmann {et~al.}(2011{\natexlab{b}})Sahlmann, S{\'{e}}gransan,
  Queloz, Udry, Santos, Marmier, Mayor, Naef, Pepe, \& Zucker}]{Sahlmann2011b}
Sahlmann, J., S{\'{e}}gransan, D., Queloz, D., {et~al.} 2011{\natexlab{b}},
  Astronomy {\&} Astrophysics, 525, A95

\bibitem[{Santos {et~al.}(2017)Santos, Adibekyan, Figueira, Andreasen, Barros,
  Delgado-Mena, Demangeon, Faria, Oshagh, Sousa, Viana, \&
  Ferreira}]{Santos2017}
Santos, N.~C., Adibekyan, V., Figueira, P., {et~al.} 2017, Astronomy {\&}
  Astrophysics, 603, A30

\bibitem[{Santos {et~al.}(2002)Santos, Mayor, Naef, Pepe, Queloz, Udry, Burnet,
  Clausen, Helt, Olsen, \& Pritchard}]{Santos2002}
Santos, N.~C., Mayor, M., Naef, D., {et~al.} 2002, Astronomy {\&} Astrophysics,
  392, 215

\bibitem[{Sousa {et~al.}(2011)Sousa, Santos, Israelian, Mayor, \&
  Udry}]{Sousa2011}
Sousa, S.~G., Santos, N.~C., Israelian, G., Mayor, M., \& Udry, S. 2011,
  Astronomy {\&} Astrophysics, 533, A141

\bibitem[{Spiegel {et~al.}(2011)Spiegel, Burrows, \& Milsom}]{Spiegel2011}
Spiegel, D.~S., Burrows, A., \& Milsom, J.~A. 2011, The Astrophysical Journal,
  727, 57

\bibitem[{Sreenivas {et~al.}(2022)Sreenivas, Perdelwitz, Tal-Or, Trifonov,
  Zucker, \& Mazeh}]{JA_Sreenivas2022}
Sreenivas, K.~R., Perdelwitz, V., Tal-Or, L., {et~al.} 2022, Astronomy {\&}
  Astrophysics, 660, A124

\bibitem[{Tinney {et~al.}(2001)Tinney, Butler, Marcy, Jones, Penny, Vogt, Apps,
  \& Henry}]{Tinney2001}
Tinney, C.~G., Butler, R.~P., Marcy, G.~W., {et~al.} 2001, The Astrophysical
  Journal, 551, 507

\bibitem[{Vogt {et~al.}(1994)Vogt, Allen, Bigelow, Bresee, Brown, Cantrall,
  Conrad, Couture, Delaney, Epps, Hilyard, Hilyard, Horn, Jern, Kanto, Keane,
  Kibrick, Lewis, Osborne, Pardeilhan, Pfister, Ricketts, Robinson, Stover,
  Tucker, Ward, \& Wei}]{HIRES}
Vogt, S.~S., Allen, S.~L., Bigelow, B.~C., {et~al.} 1994, in Proc. SPIE 2198,
  Instrumentation in Astronomy VIII, ed. D.~L. Crawford \& E.~R. Craine,
  362--375

\bibitem[{Wittenmyer {et~al.}(2016)Wittenmyer, Butler, Tinney, Horner, Carter,
  Wright, Jones, Bailey, \& O’Toole}]{JA_Wittenmyer2016}
Wittenmyer, R.~A., Butler, R.~P., Tinney, C.~G., {et~al.} 2016, The
  Astrophysical Journal, 819, 28

\bibitem[{Yee {et~al.}(2017)Yee, Petigura, \& von Braun}]{Yee2017}
Yee, S.~W., Petigura, E.~A., \& von Braun, K. 2017, The Astrophysical Journal,
  836, 77

\bibitem[{Zhu \& Wu(2018)}]{ZhuWu2018}
Zhu, W. \& Wu, Y. 2018, The Astronomical Journal, 156, 92

\end{thebibliography}

\appendix
\FloatBarrier
\onecolumn
\section{Orvara corner plots}
\label{app:orvara-res}
\begin{figure}[H]
    \centering
    \includegraphics[width=.55\textwidth]{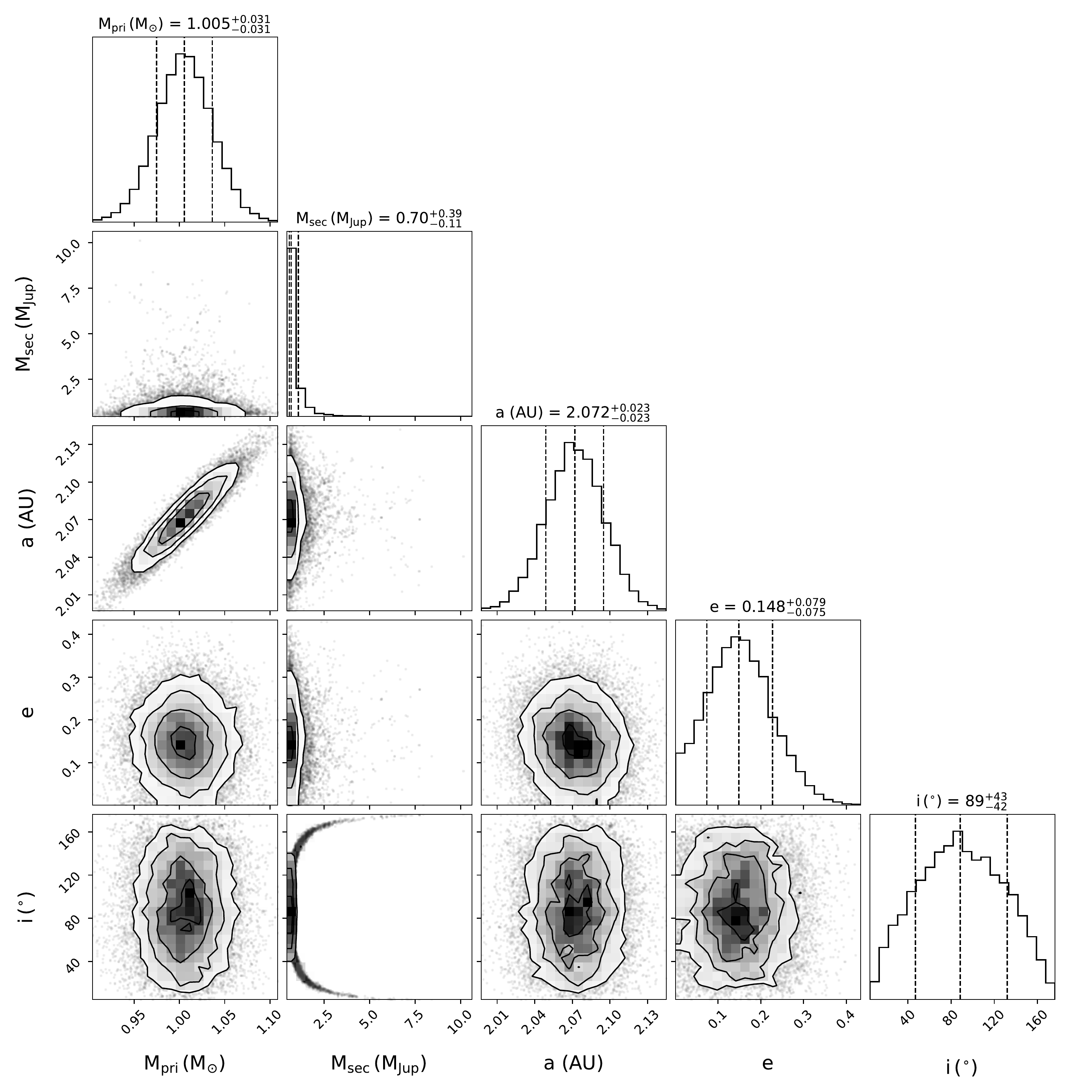}
    \caption{Corner plot with the \texttt{orvara}-derived primary star mass, companion mass, semimajor axis, eccentricity, and inclination of HD3964 b. There are no strong constraints on the secondary mass or inclination.}
\end{figure}
\begin{figure}[H]
    \centering
    \includegraphics[width=.55\textwidth]{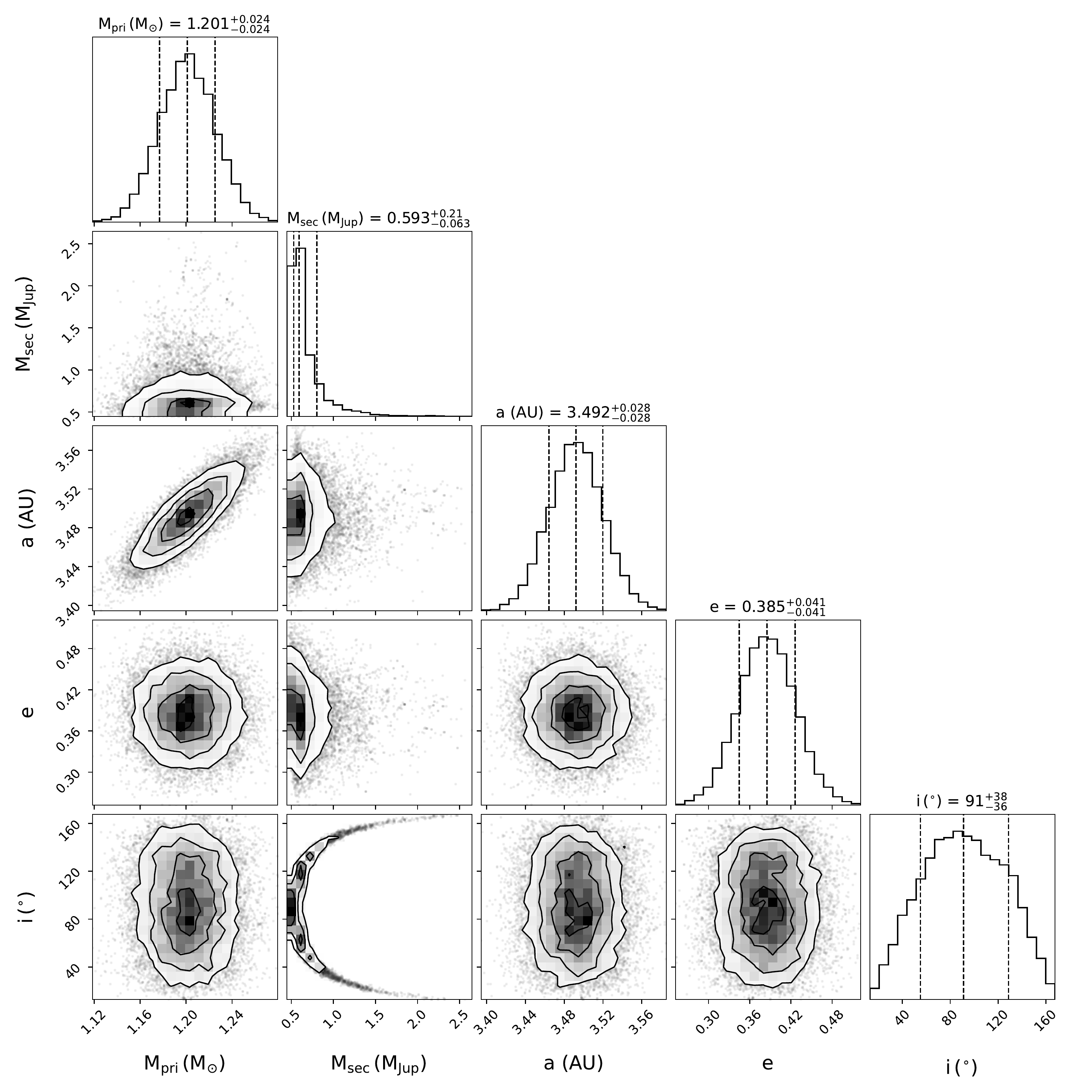}
    \caption{Corner plot with the \texttt{orvara}-derived primary star mass, companion mass, semimajor axis, eccentricity, and inclination of HD94771 b. There are again no strong constraints on the secondary mass or inclination.}
\end{figure}

\begin{figure}[H]
    \centering
    \includegraphics[width=.55\textwidth]{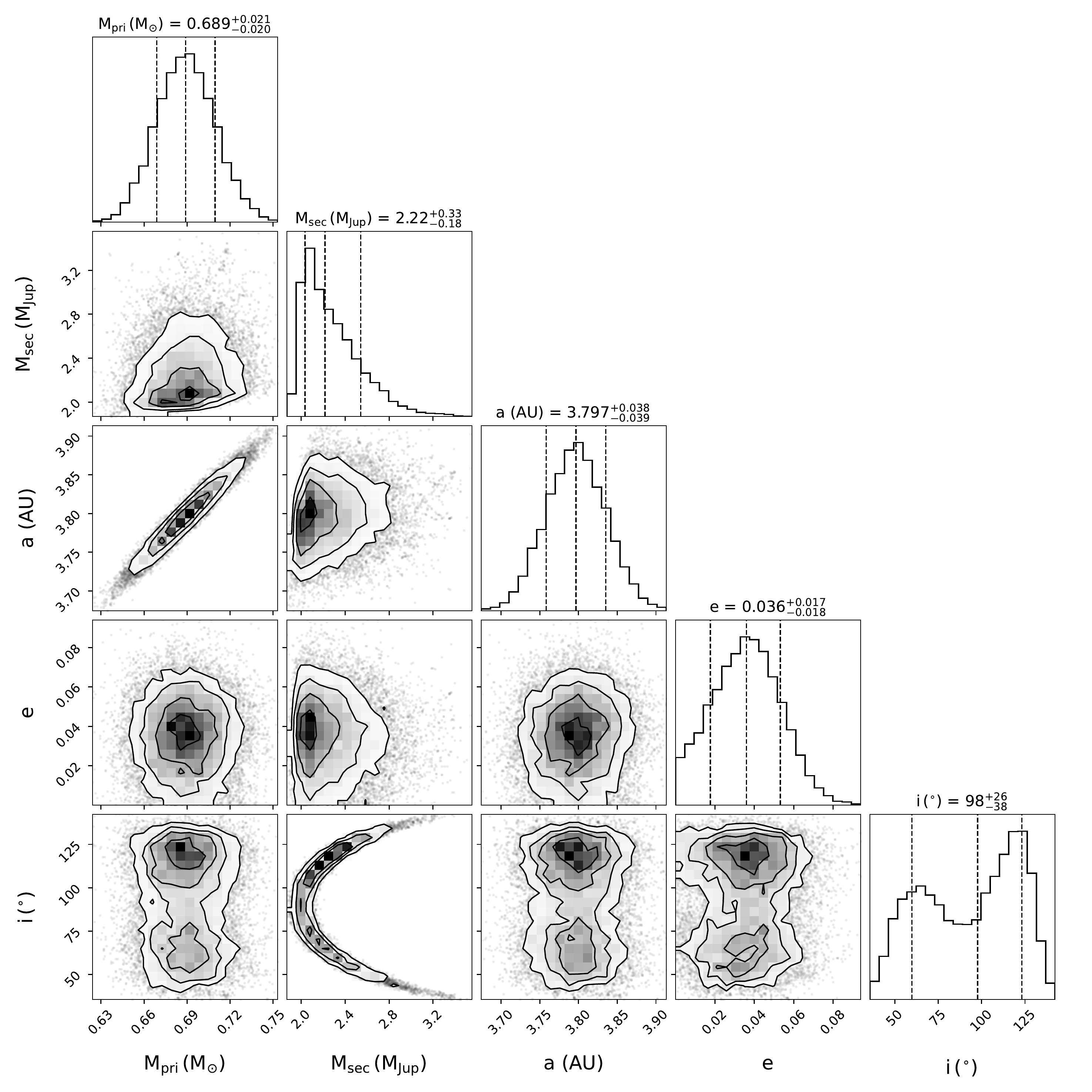}
    \caption{Corner plot with the \texttt{orvara}-derived primary star mass, companion mass, semimajor axis, eccentricity, and inclination of HIP54597 b. Though the mass and inclination are not well-constrained, the inclination varies between 120$^{\circ}$ and 60$^{\circ}$. The mass of the planet might increase by $\sim$10\%-30\% but not significantly more.}
\end{figure}

\begin{figure}[H]
    \centering
    \includegraphics[width=.55\textwidth]{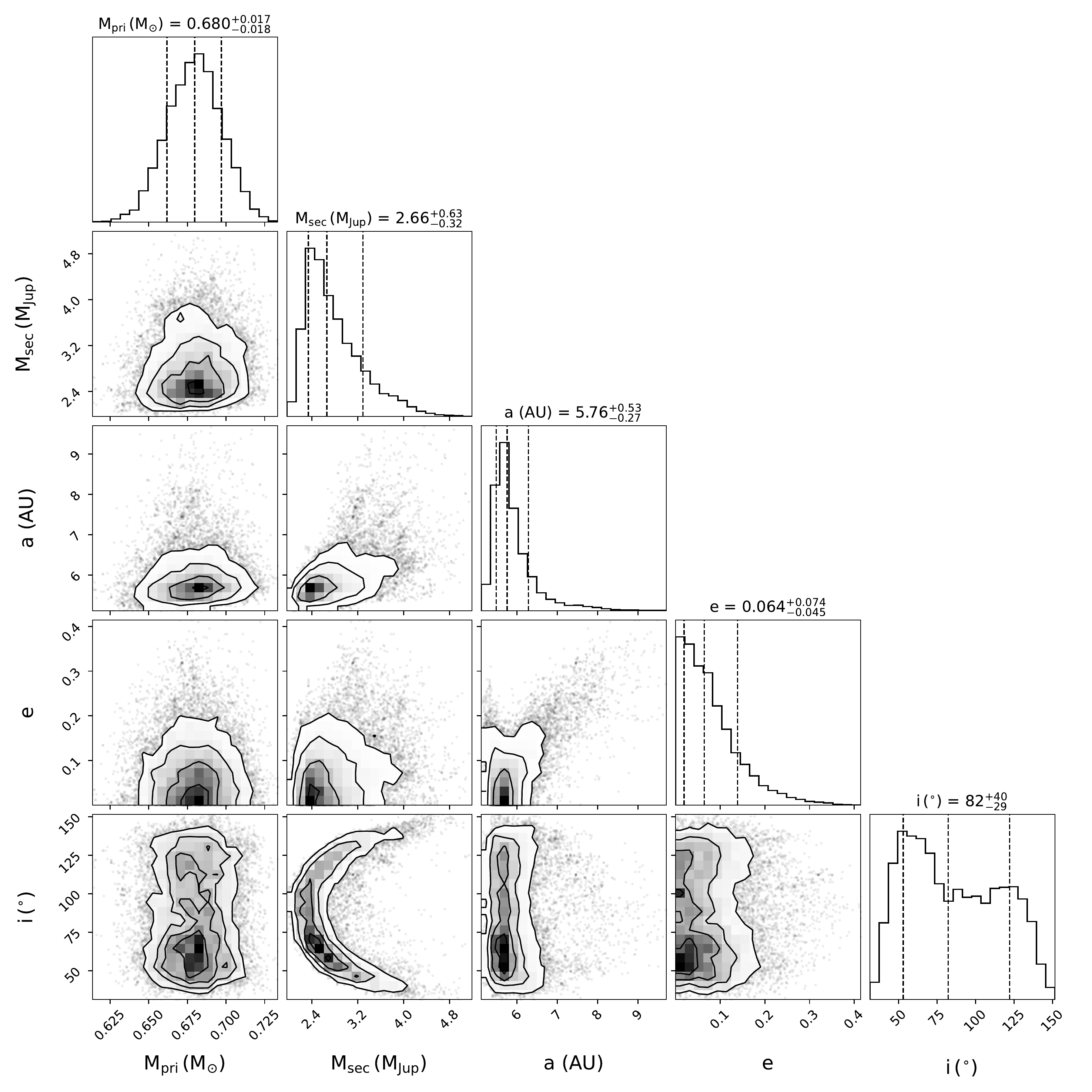}
    \caption{Corner plot with the \texttt{orvara}-derived primary star mass, companion mass, semimajor axis, eccentricity, and inclination of BD-210397 c. The inclination peaks at 50$^{\circ}$ and 130$^{\circ}$, favouring 50$^{\circ}$. The found mass is 2.7 $M_J$. This corner plot is for the long-period (7063 d) planet; the fit also includes BD-210397 b.}
\end{figure}

\begin{figure}[H]
    \centering
    \includegraphics[width=.55\textwidth]{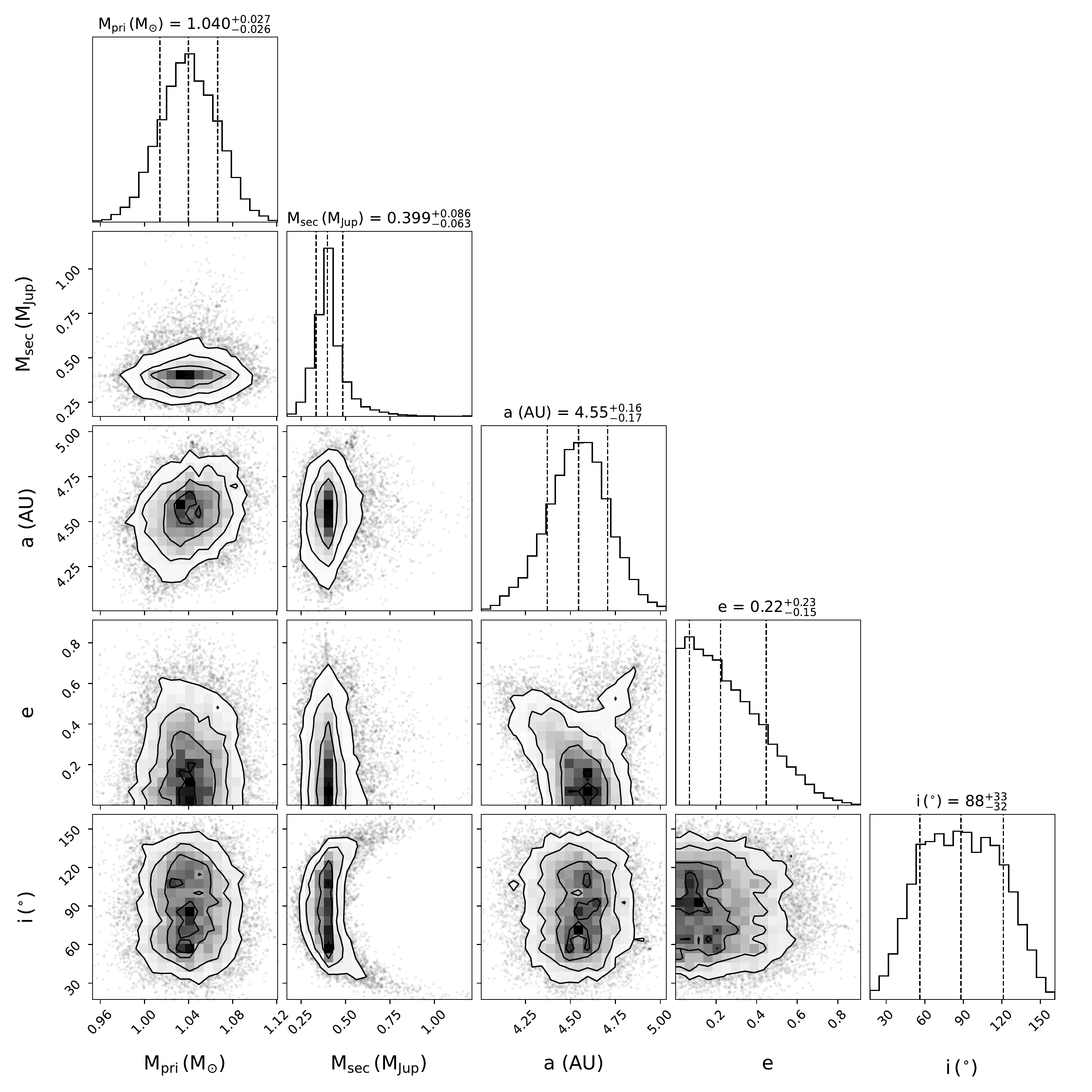}
    \caption{Corner plot with the \texttt{orvara}-derived primary star mass, companion mass, semimajor axis, eccentricity, and inclination of HD74698 c. The inclination is not well-constrained at $90^{\circ} \pm 33^{\circ}$; the true mass is equal to the found minimal mass. The fit also includes HD74698 b.}
\end{figure}

\begin{figure}[H]
    \centering
    \includegraphics[width=.55\textwidth]{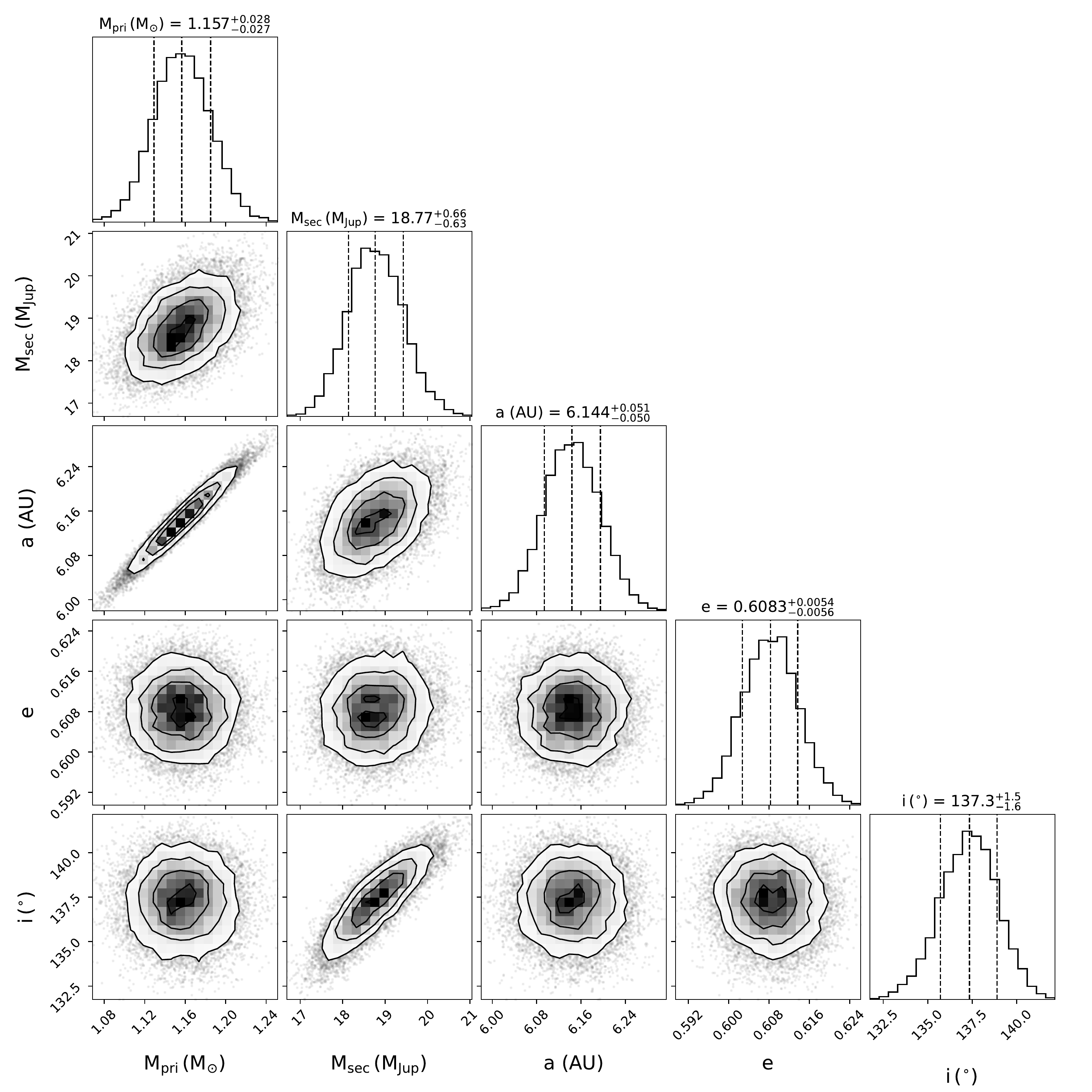}
    \caption{Corner plot with the \texttt{orvara}-derived primary star mass, companion mass, semimajor axis, eccentricity, and inclination of HD62364 b. The mass and inclination are well-constrained. }
\end{figure}

\begin{figure}[H]
    \centering
    \includegraphics[width=.55\textwidth]{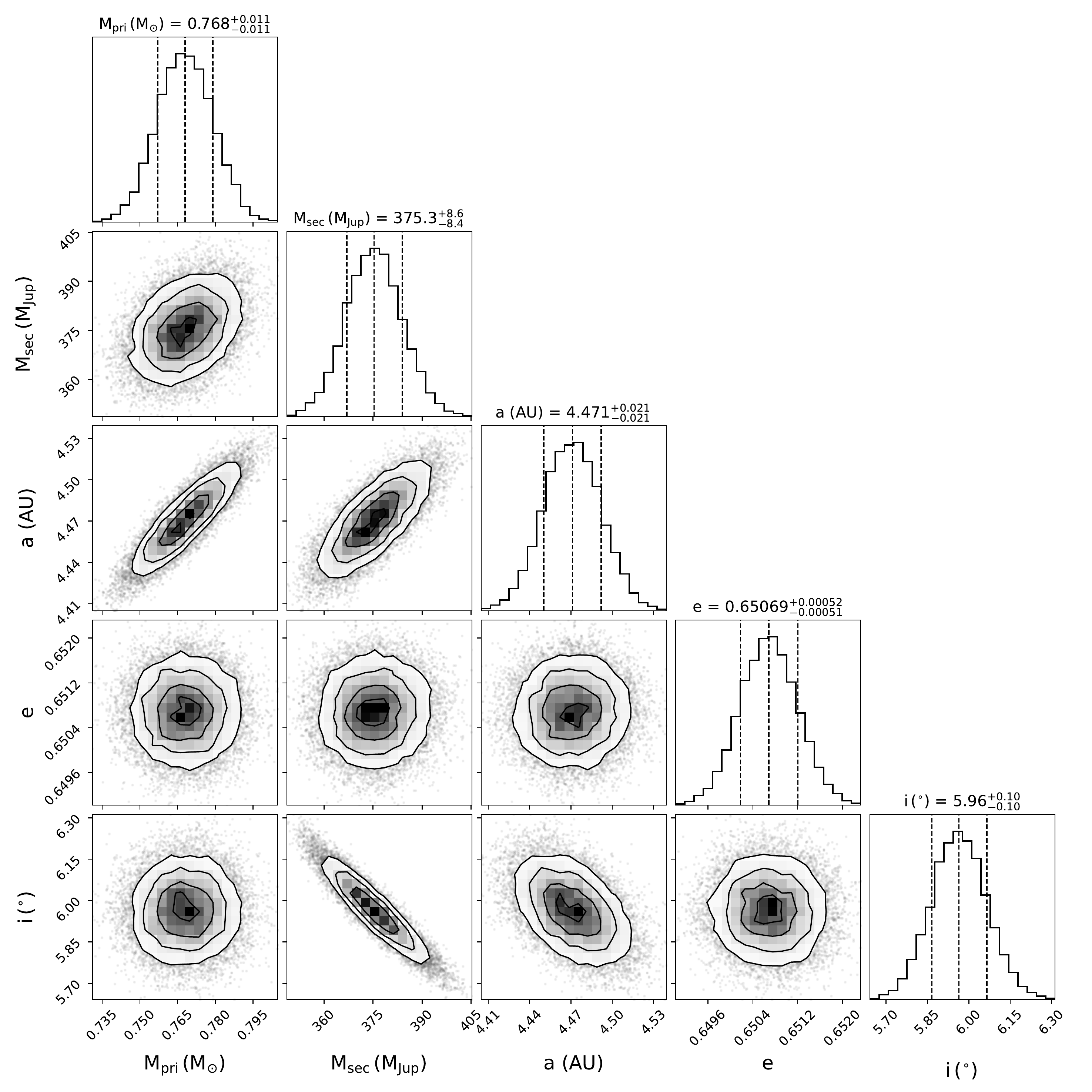}
    \caption{Corner plot with the \texttt{orvara}-derived primary star mass, companion mass, semimajor axis, eccentricity, and inclination of HD56380B. The mass and inclination are well-constrained.}
\end{figure}

\begin{figure}[H]
    \centering
    \includegraphics[width=.55\textwidth]{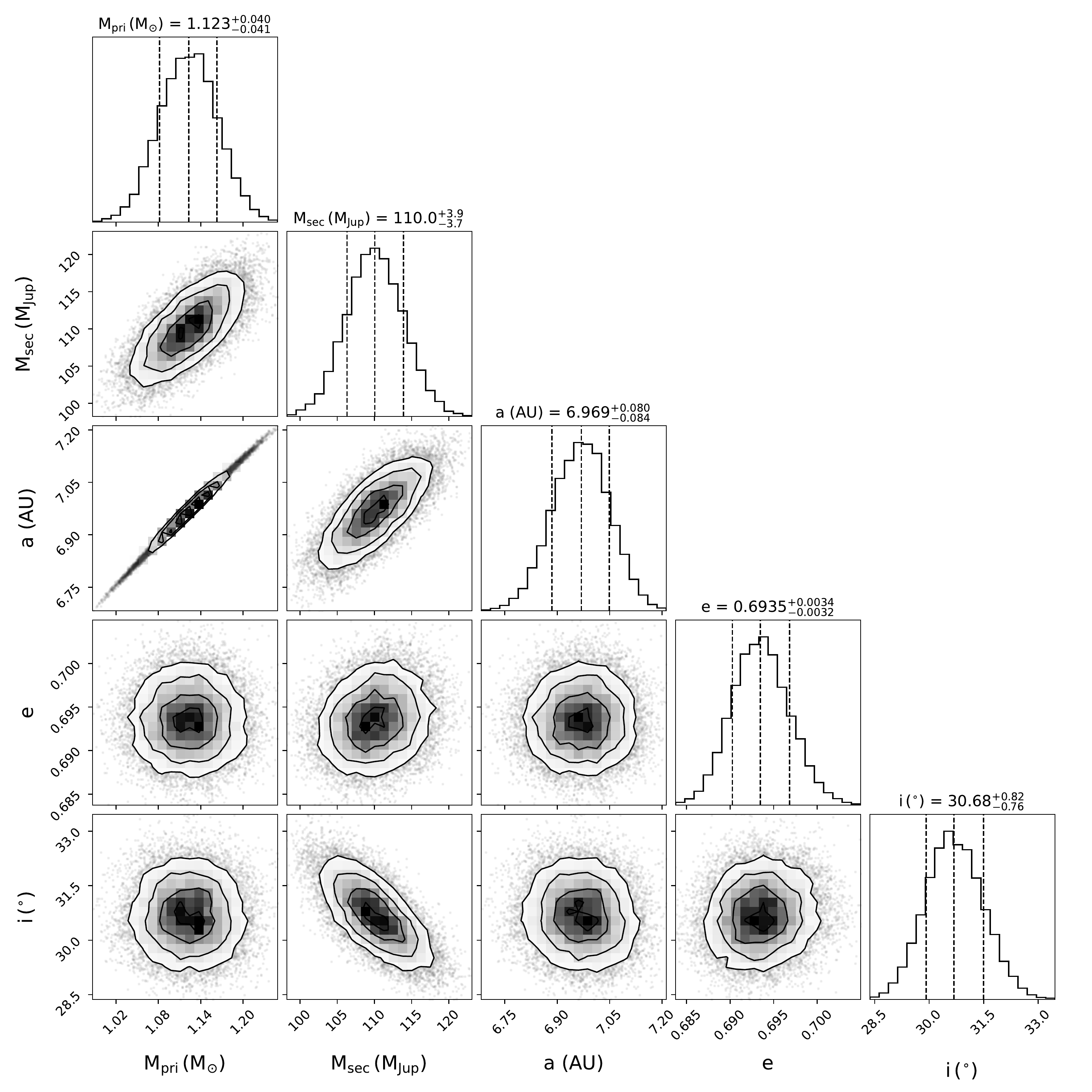}
    \caption{Corner plot with the \texttt{orvara}-derived primary star mass, companion mass, semimajor axis, eccentricity, and inclination of HD221638B. The mass and inclination are well-constrained.}
\end{figure}

\begin{figure}[H]
    \centering
    \includegraphics[width=.55\textwidth]{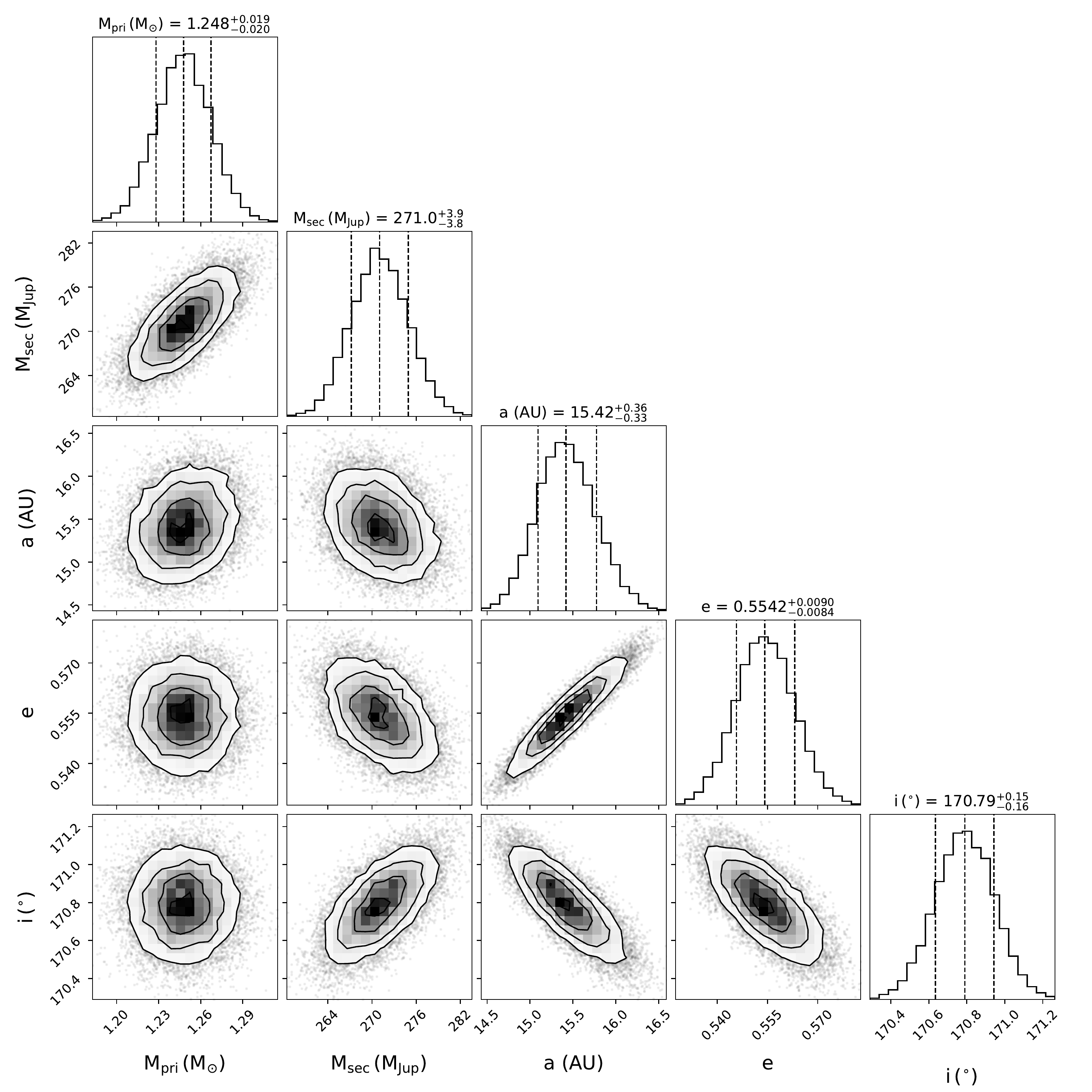}
    \caption{Corner plot with the \texttt{orvara}-derived primary star mass, companion mass, semimajor axis, eccentricity, and inclination of HD33473C. The mass and inclination are well-constrained.}
\end{figure}

\begin{figure}[H]
    \centering
    \includegraphics[width=.55\textwidth]{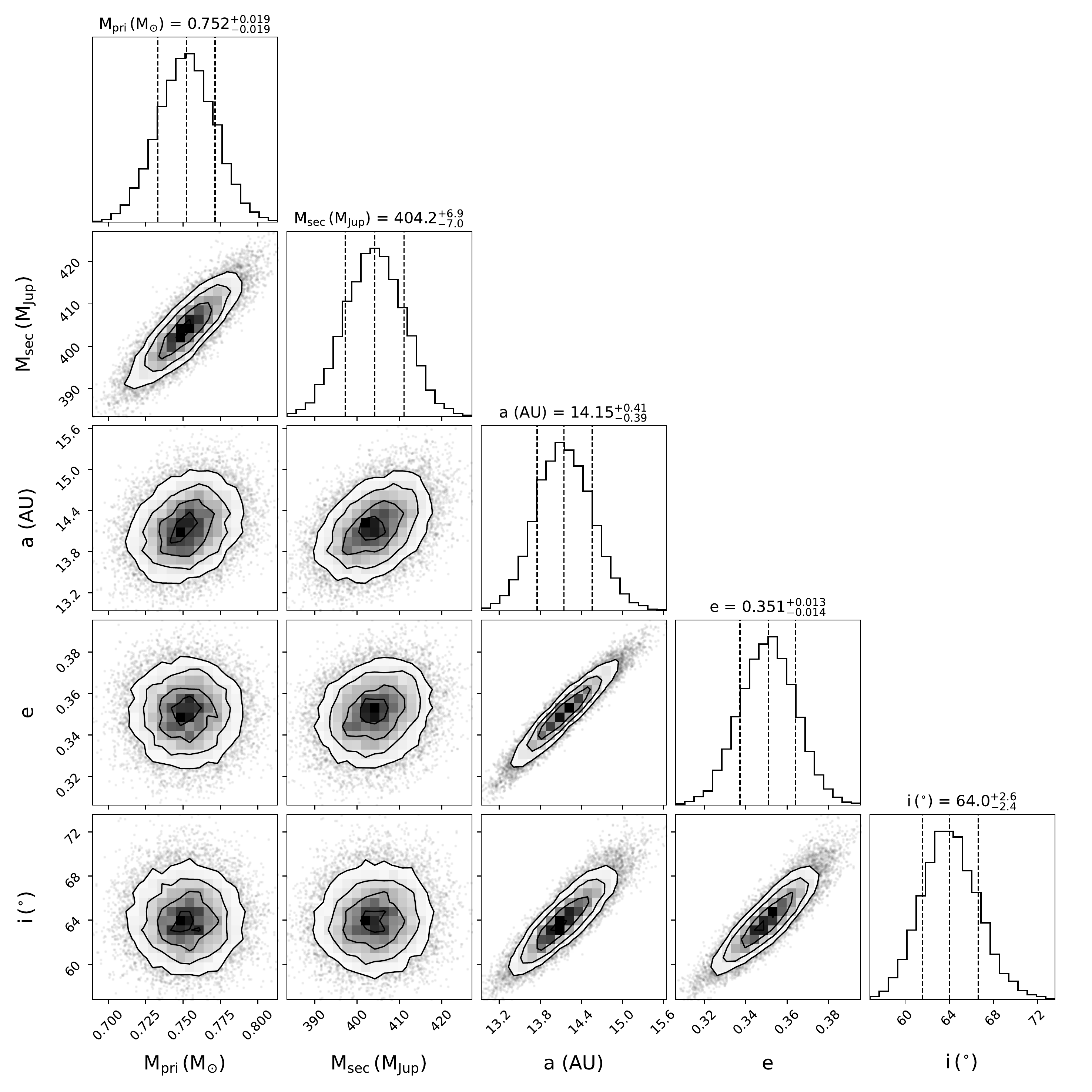}
    \caption{Corner plot with the \texttt{orvara}-derived primary star mass, companion mass, semimajor axis, eccentricity, and inclination of HD11938B. The mass and inclination are well-constrained.}
\end{figure}

\begin{figure}[H]
    \centering 
    \includegraphics[width=.55\textwidth]{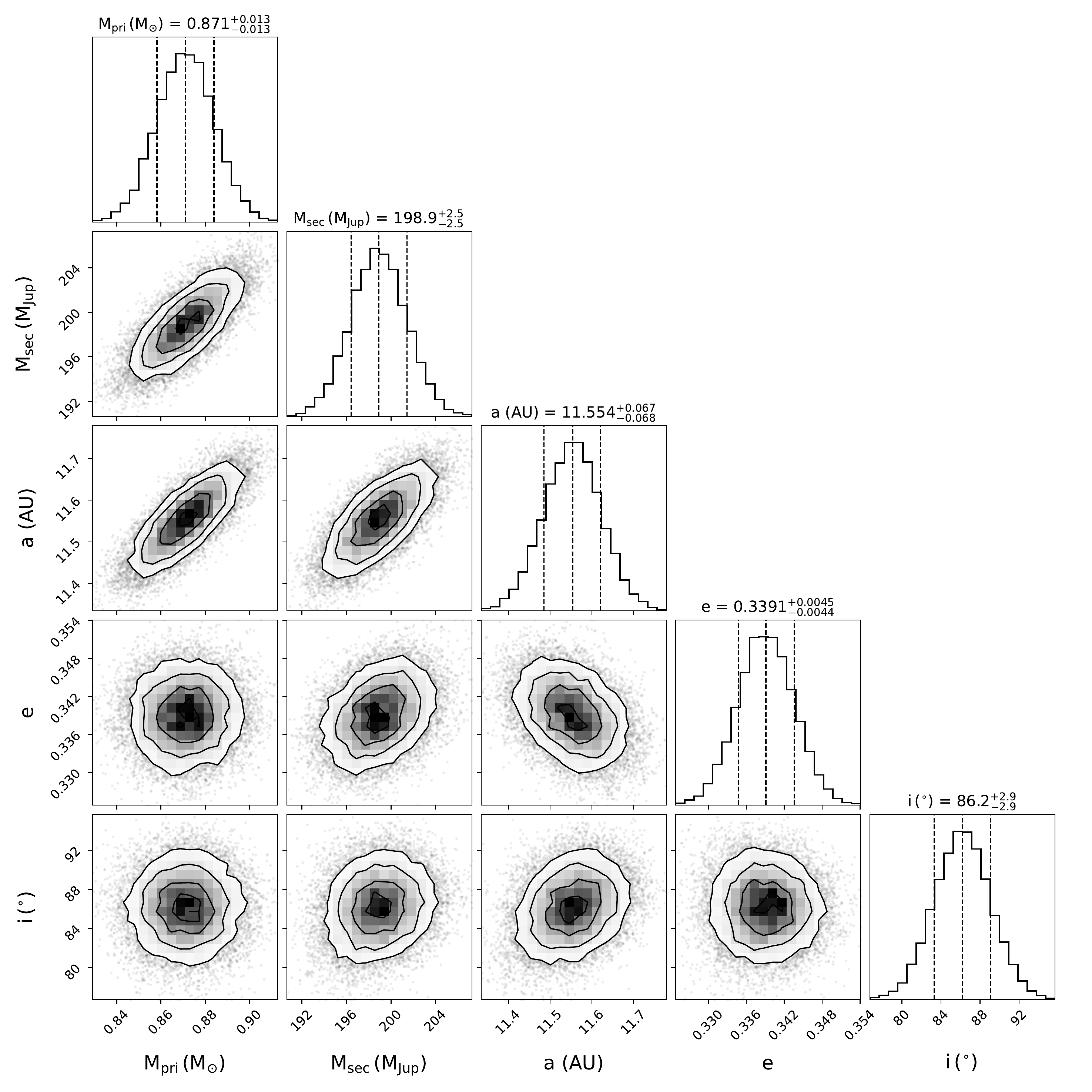}
    \caption{Corner plot with the \texttt{orvara}-derived primary star mass, companion mass, semimajor axis, eccentricity, and inclination of HD61383B. The mass and inclination are well-constrained.}
\end{figure}

\FloatBarrier
\section{Orvara proper motion plots}
\label{app:orvara-res2}
\begin{figure}[H]
    \begin{subfigure}[b]{0.49\textwidth}
    \centering
    \includegraphics[width=\textwidth]{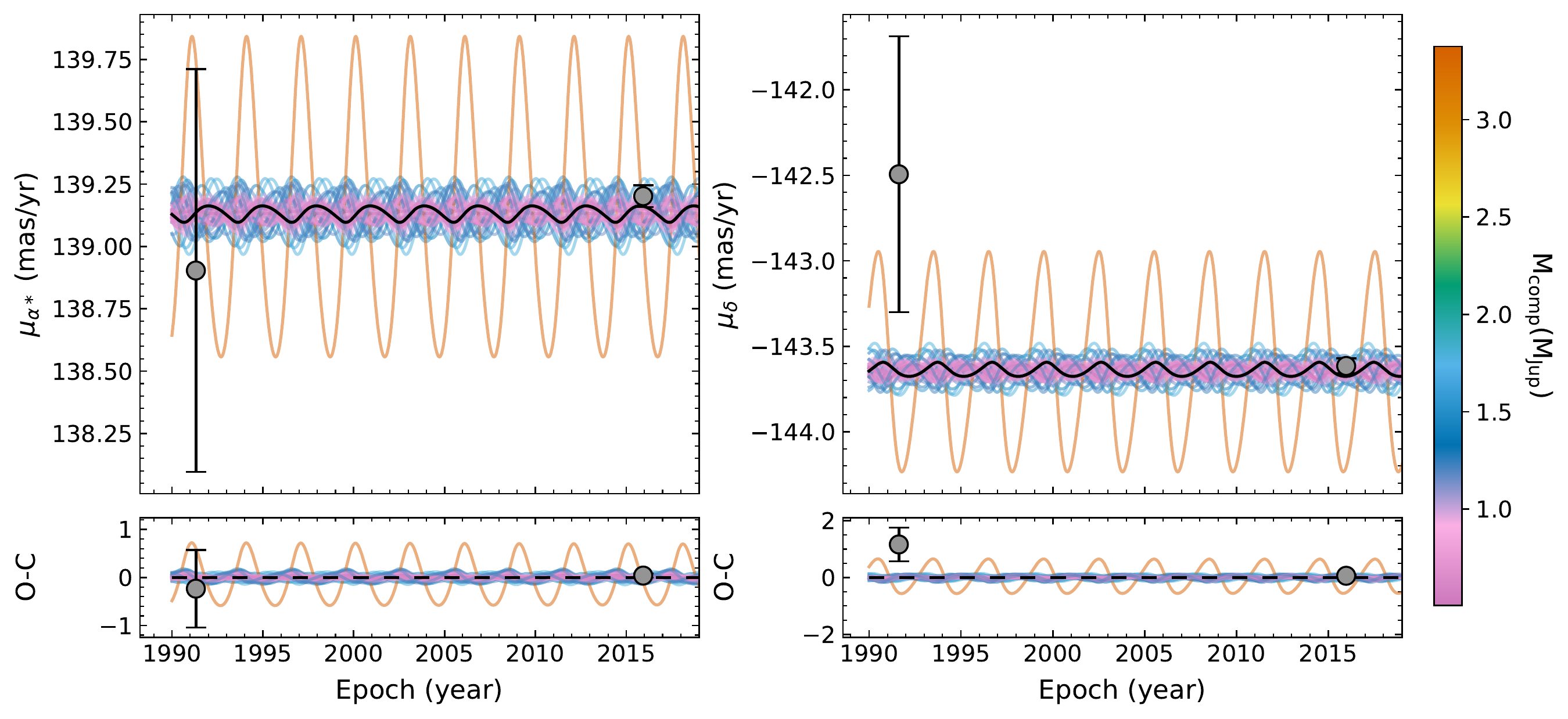}
    \caption{HD3964}
    \end{subfigure}
    \begin{subfigure}[b]{0.49\textwidth}
        \includegraphics[width=\textwidth]{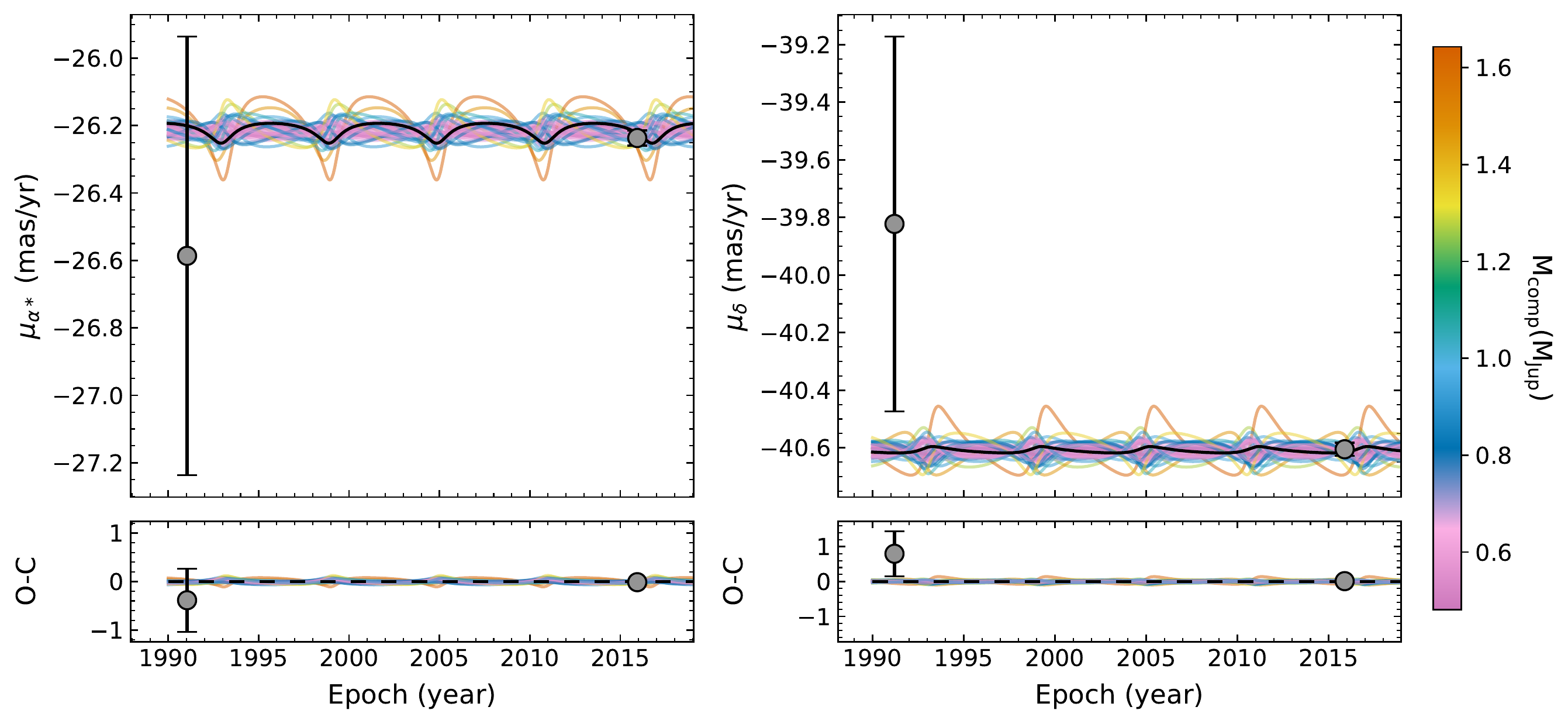}
    \caption{HD94771}
    \end{subfigure}

    \begin{subfigure}[b]{0.49\textwidth}
        \includegraphics[width=\textwidth]{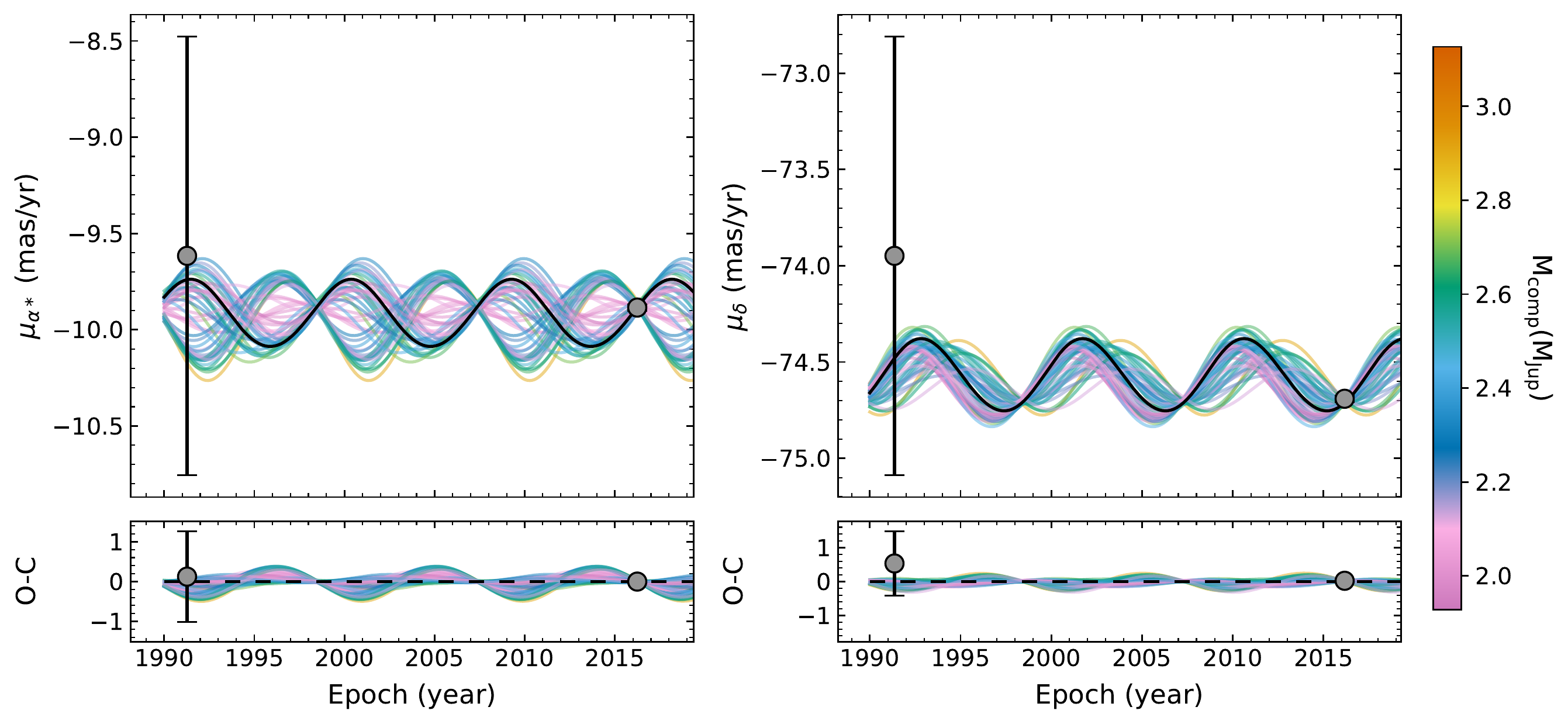}
    \caption{HIP54597}        
    \end{subfigure}
    \begin{subfigure}[b]{0.49\textwidth}
        \includegraphics[width=\textwidth]{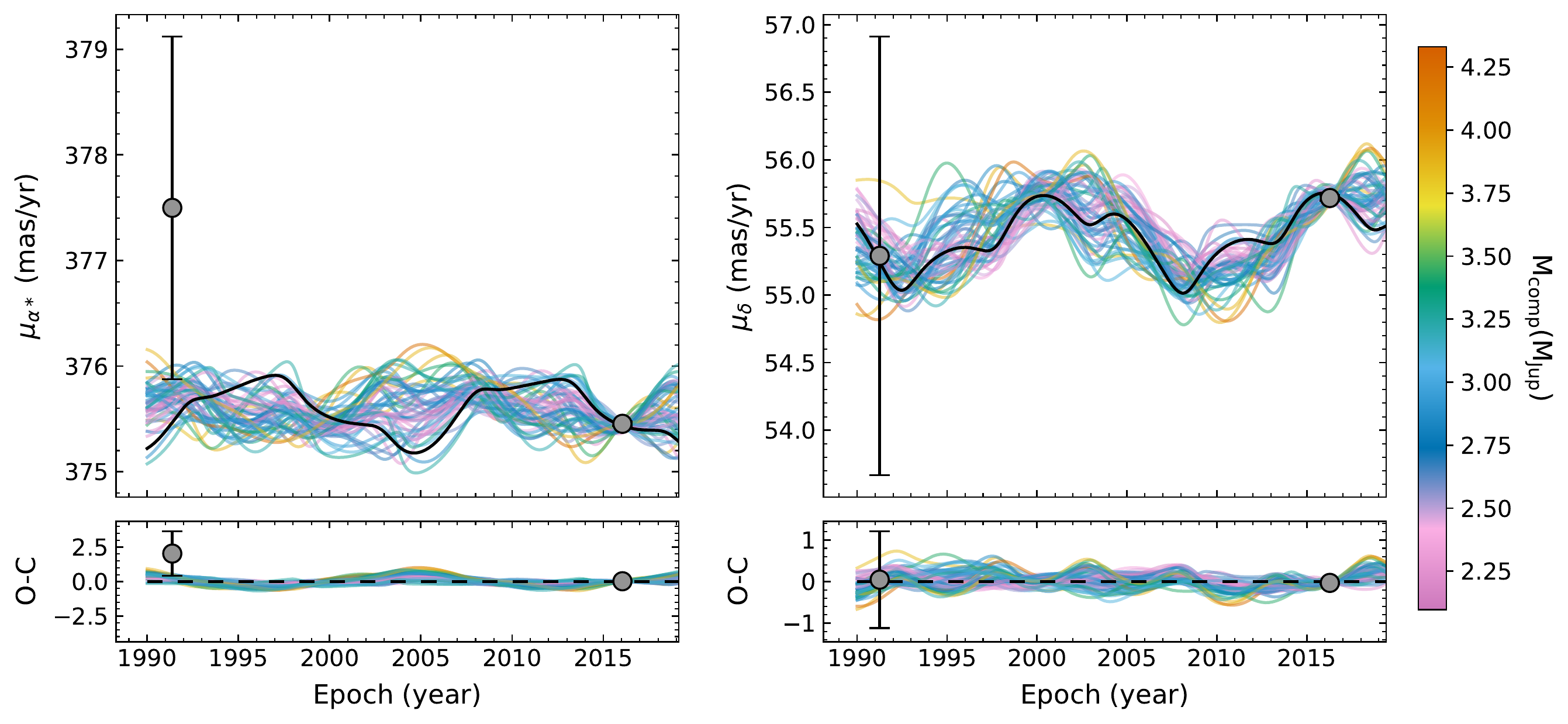}
    \caption{BD-210397}
    \end{subfigure}
    \caption{Proper motion measurements from Hipparcos and Gaia (grey markers) and fitted proper motions in right ascension and declination. The black line is the orbit with the highest likelihood and the 50 coloured lines are randomly drawn orbits from the posterior distribution, where the colour corresponds to the mass of the companion. The bottom figure shows the difference between observed and calculated astrometry.}
\end{figure}

\addtocounter{figure}{-1}
\begin{figure}[H]
    \begin{subfigure}[b]{0.49\textwidth}
        \addtocounter{subfigure}{4}
        \includegraphics[width=\linewidth]{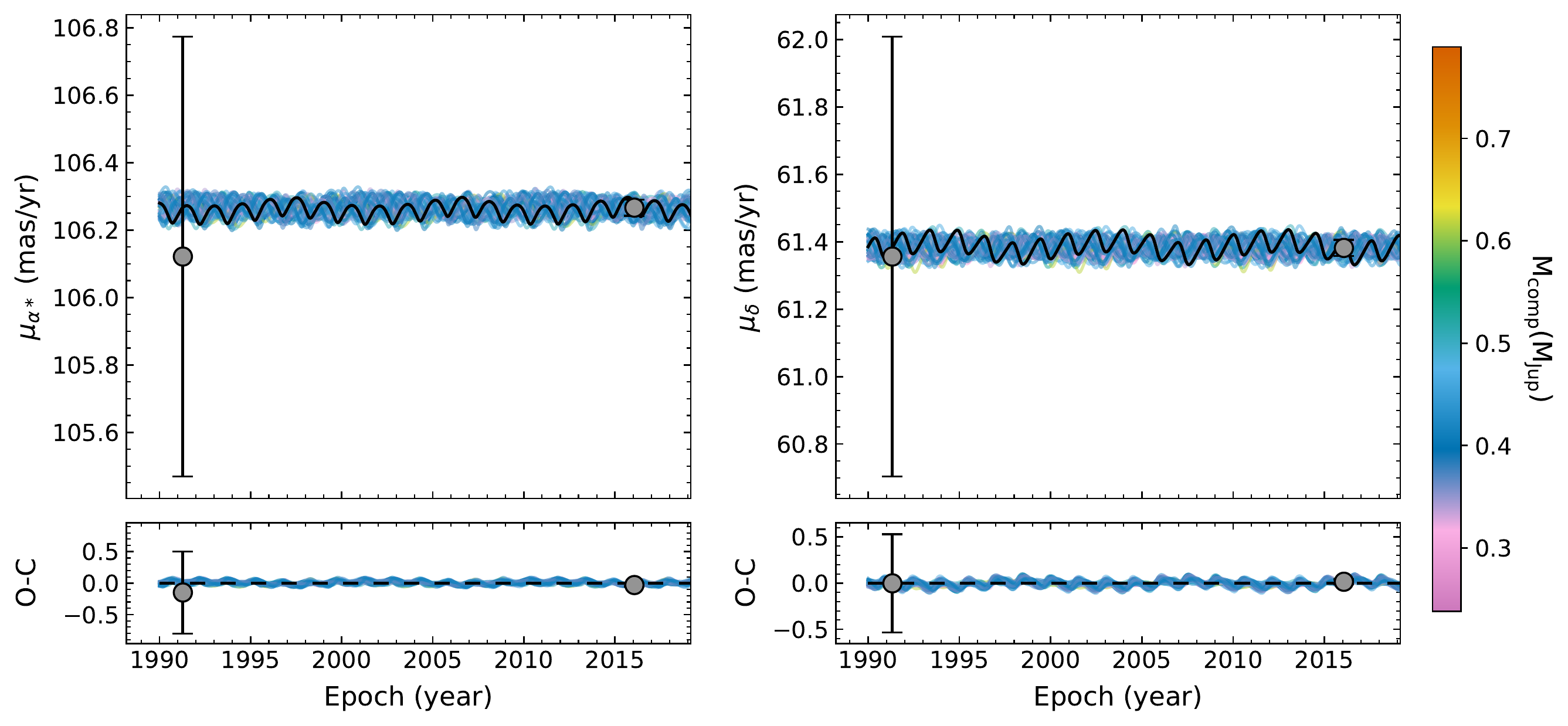}
    \caption{HD74698}
    \end{subfigure} 
    \begin{subfigure}[b]{0.49\textwidth}
        \includegraphics[width=\linewidth]{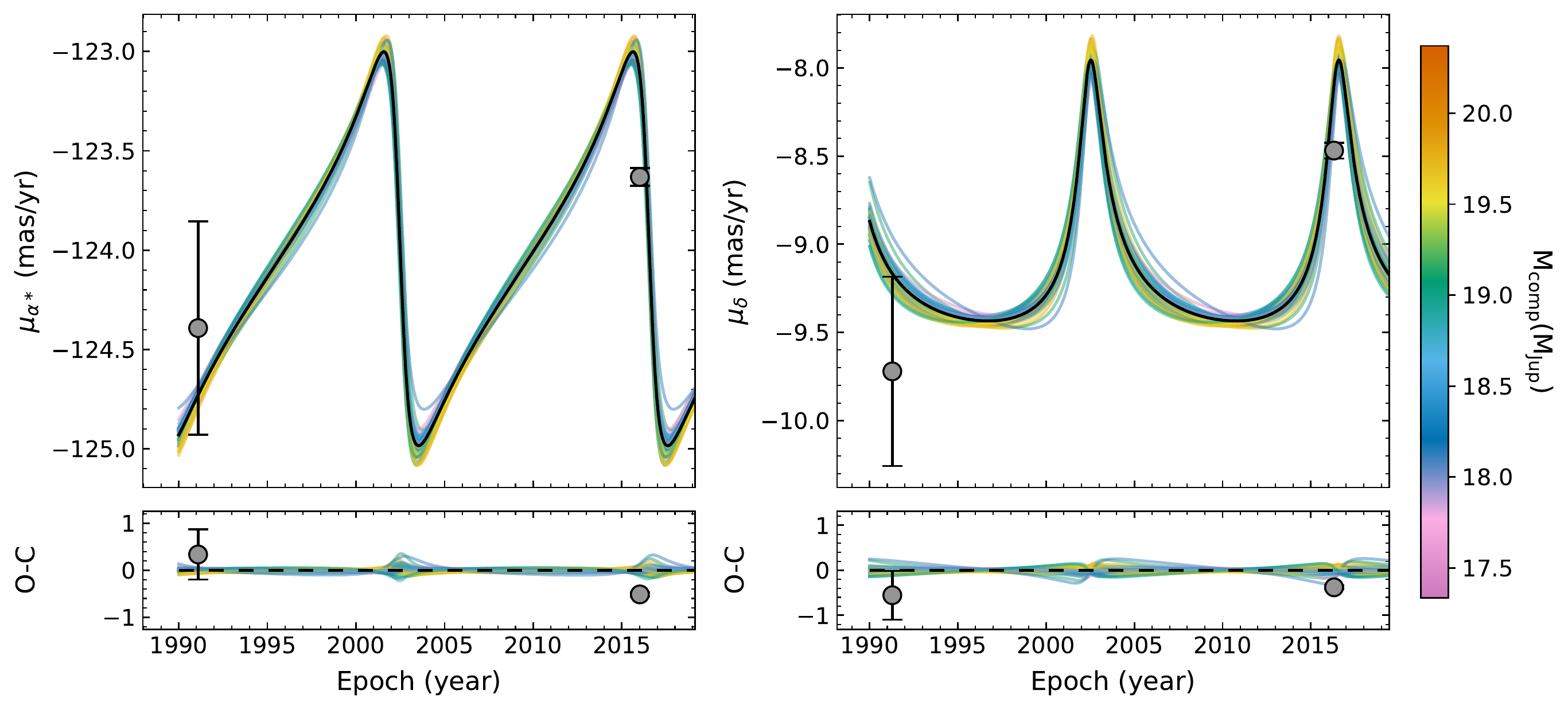}
        \caption{HD62364}
    \end{subfigure}

    \begin{subfigure}[b]{0.49\textwidth}
        \includegraphics[width=\linewidth]{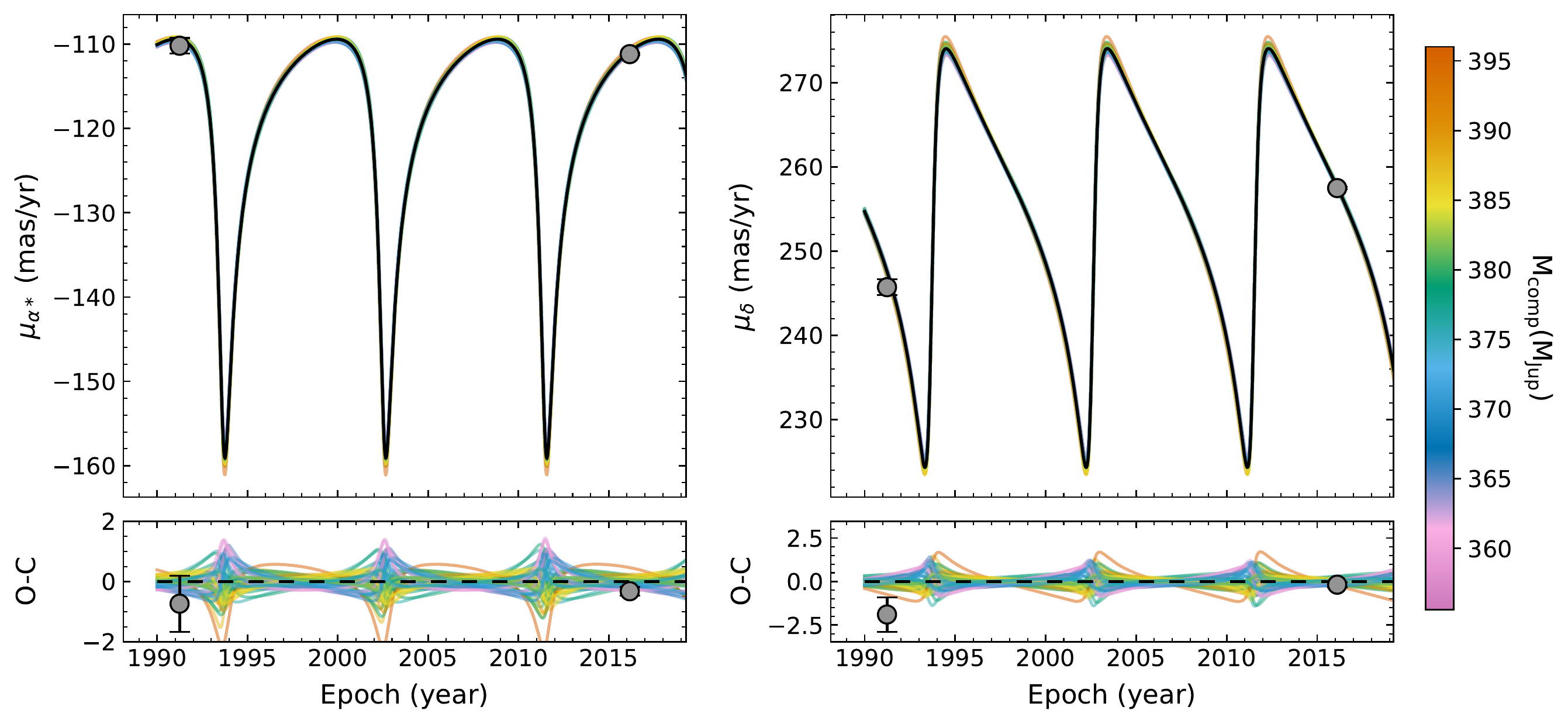}
        \caption{HD56380}
    \end{subfigure}
    \begin{subfigure}[b]{0.49\textwidth}
        \includegraphics[width=\linewidth]{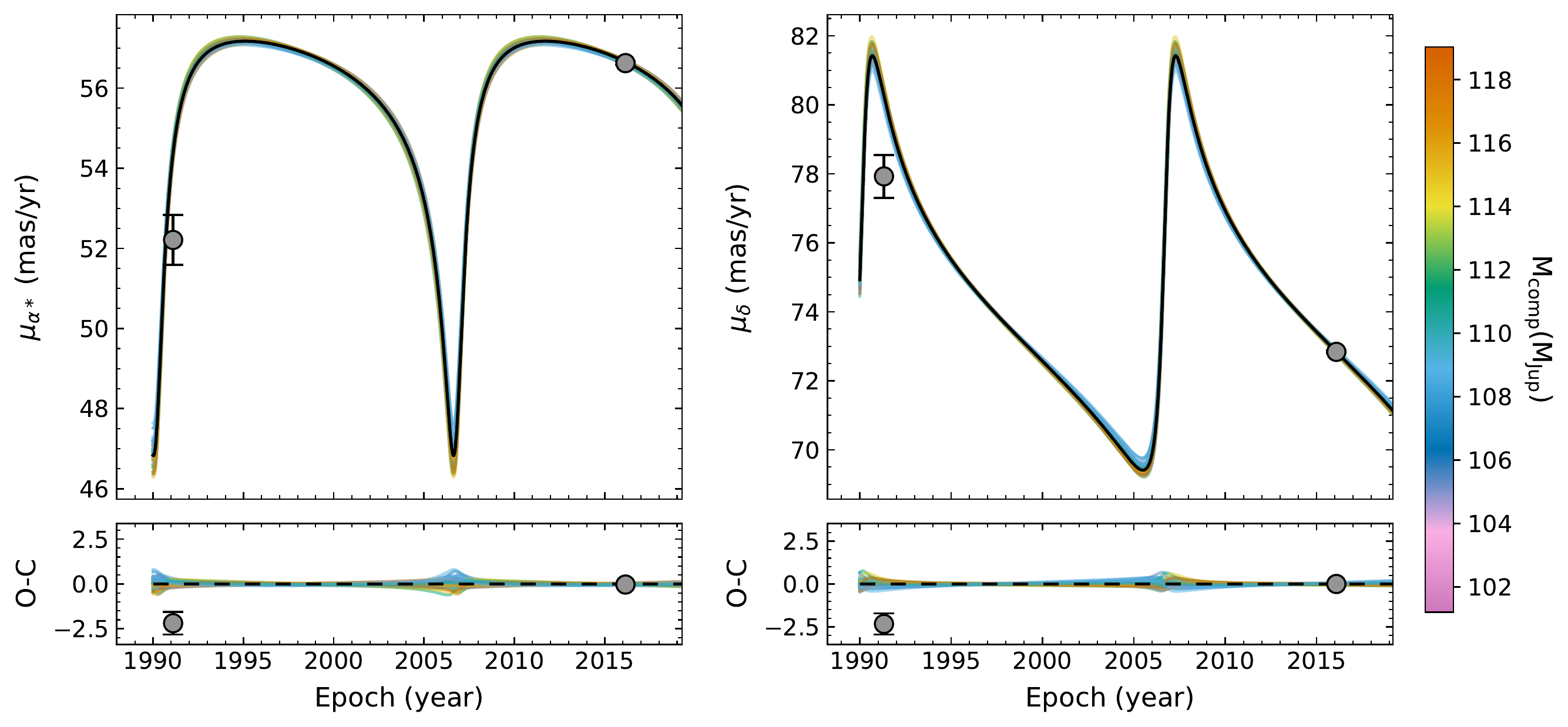}
        \caption{HD221638}
    \end{subfigure}

    \begin{subfigure}[b]{0.49\textwidth}
        \includegraphics[width=\linewidth]{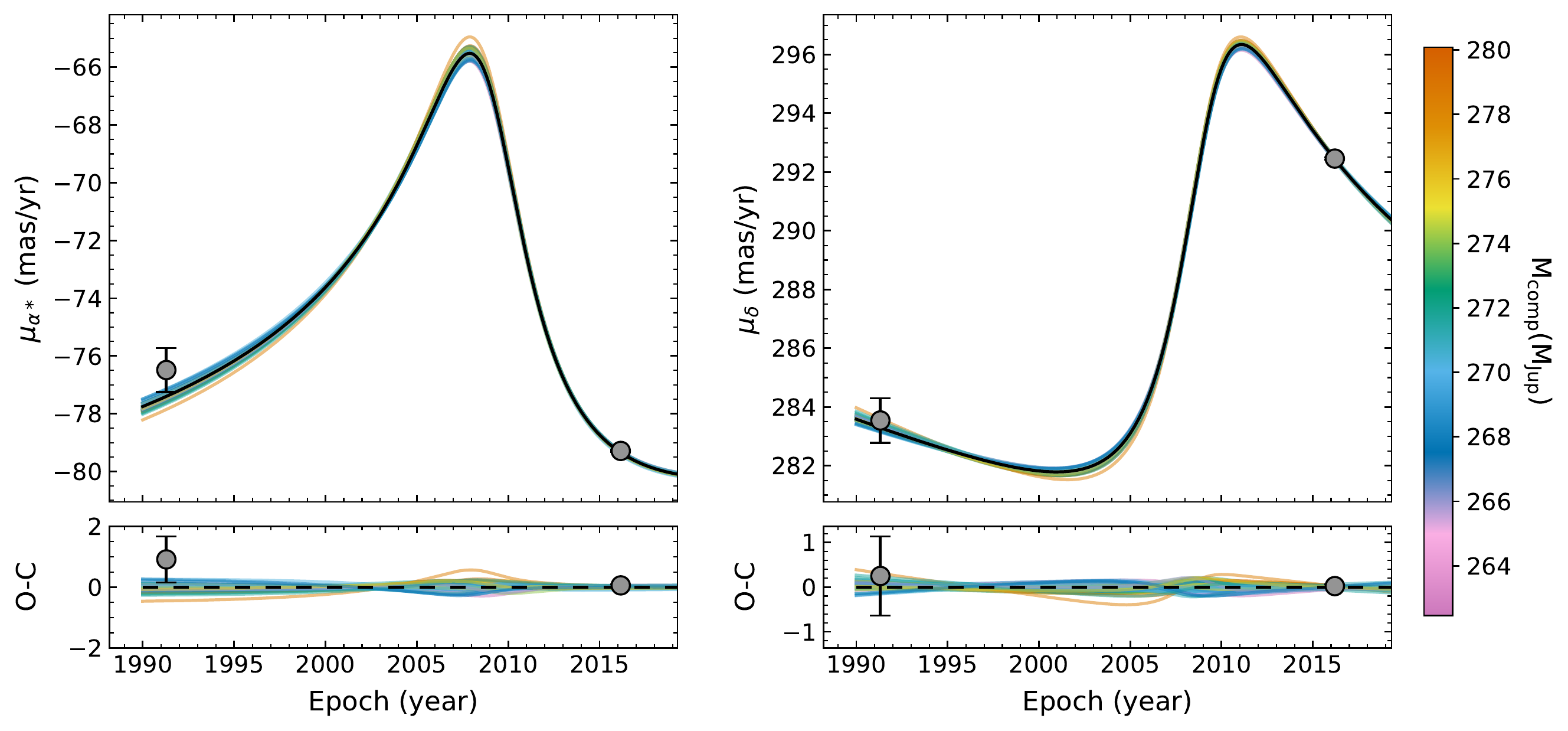}
        \caption{HD33473A}
    \end{subfigure}
    \begin{subfigure}[b]{0.49\textwidth}
        \includegraphics[width=\linewidth]{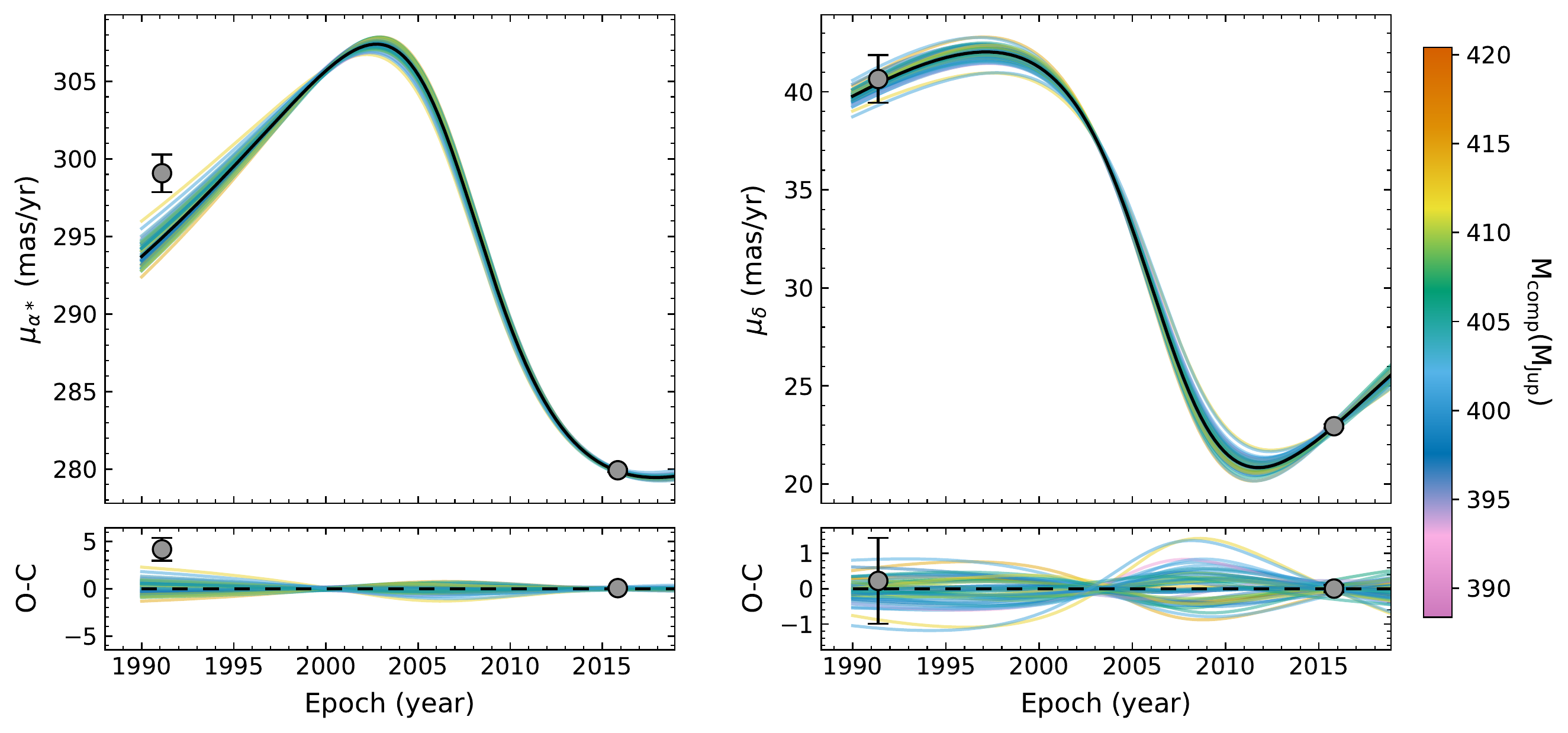}
        \caption{HD11938}
    \end{subfigure}

    \begin{subfigure}[b]{0.49\textwidth}
        \includegraphics[width=\linewidth]{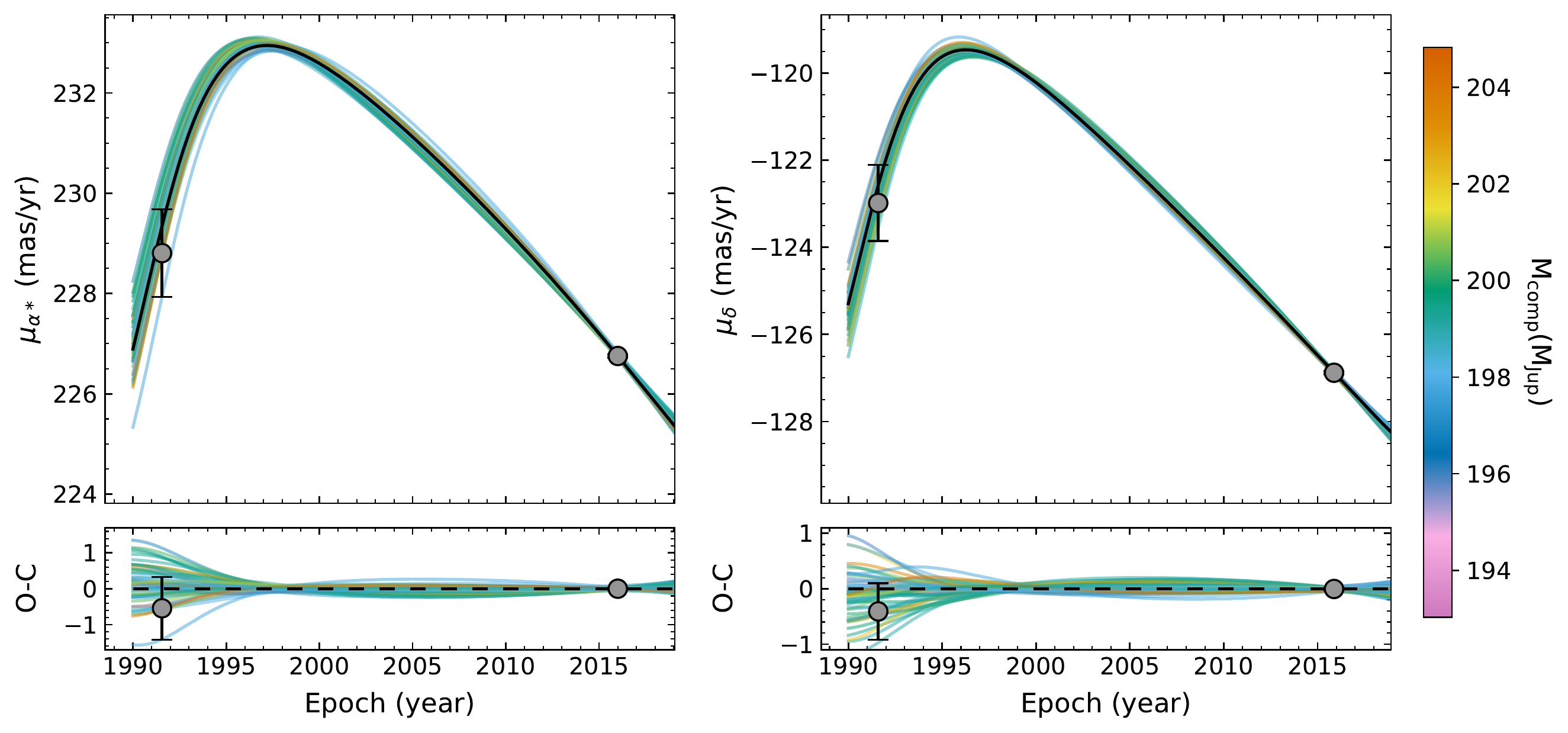}
        \caption{HD61383}
    \end{subfigure}
    \caption{continued.}
\end{figure}

\end{document}